\documentclass[floatfix,showpacs,rmp,aps,twocolumn,superscriptaddress]{revtex4-1}

\usepackage{lastpage}
\usepackage{multirow}
\usepackage{graphicx,epsfig}
\usepackage{amsmath}
\usepackage{amssymb}
\usepackage{minitoc}  
\usepackage{amscd}
\usepackage{dcolumn}
\usepackage{bm, rotating}
\usepackage[normalem]{ulem} 
\usepackage{color} 
\input{Qcircuit}

\DeclareMathSymbol{\N}{\mathbin}{AMSb}{"4E}
\DeclareMathSymbol{\Z}{\mathbin}{AMSb}{"5A}
\DeclareMathSymbol{\R}{\mathbin}{AMSb}{"52}
\DeclareMathSymbol{\Q}{\mathbin}{AMSb}{"51}
\DeclareMathSymbol{\I}{\mathbin}{AMSb}{"49}
\DeclareMathSymbol{\C}{\mathbin}{AMSb}{"43}
\begin{document}

\title{Quantum Error Correction for Beginners}

\author{Simon J. Devitt}\email{electronic address: devitt@nii.ac.jp}
\affiliation{National Institute of Informatics 2-1-2 Hitotsubashi Chiyoda-ku Tokyo 101-8340.  Japan}
 \author{William J. Munro}
 \affiliation{NTT Basic Research Laboratories NTT Corporation 3-1 Morinosato-Wakamiya, Atsugi Kanagawa 243-0198. Japan}
\author{Kae Nemoto}\affiliation{National Institute of Informatics 2-1-2 Hitotsubashi Chiyoda-ku Tokyo 101-8340.  Japan}

\date{\today}

\begin{abstract}
Quantum error correction (QEC) and fault-tolerant quantum computation represent one of the most vital 
theoretical aspect of quantum information processing.  
It was well known from the early developments of this exciting field that the fragility of coherent quantum systems would be a 
catastrophic obstacle to the development of large scale quantum computers.  The introduction of quantum error correction in 
1995 showed that active techniques could be employed to mitigate this fatal problem.  However, quantum error 
correction and fault-tolerant computation is now a much larger field and many new codes, techniques, and 
methodologies have been developed to implement error correction for large scale quantum algorithms.  In response, we have attempted to summarize the basic aspects of quantum error correction and fault-tolerance, not as a 
detailed guide, but rather as a basic introduction.  
This development 
in this area has been so pronounced that many in the field of quantum information, specifically researchers who are 
new to quantum information or people focused on the many other important 
issues in quantum computation, have found it difficult to keep up with the general formalisms and methodologies employed in this area.  Rather than introducing these concepts from a rigorous mathematical and 
computer science framework, we instead examine error correction and fault-tolerance largely through detailed examples, 
which are more relevant to experimentalists today and in the near future.  


\end{abstract}


\maketitle

\tableofcontents

\section{Introduction}\label{sec:sec:QECIntro}
The micro-computer revolution of the late 20th century has arguably been of greater impact to 
the world than any other technological revolution in history.  The advent of transistors, integrated 
circuits, and the modern microprocessor has spawned literally hundreds of devices from 
pocket calculators to the iPod, all now integrated through an extensive worldwide communications 
system.  However, as we enter the 21st century ever increasing computational power is driving us 
quickly to the realm where quantum physics dominates.  The component size of 
individual transistors on modern microprocessors are becoming so small that quantum effects will soon  
begin to dominate.  Unfortunately, 
quantum mechanical behaviour will tend to result in unpredictable and 
unwanted operation in classical microprocessor designs.  We therefore have two choices: keep trying to suppressing quantum effects 
in classically fabricated electronics or move to 
the field of quantum information processing (QIP) where we exploit them.  This leads to 
a paradigm shift in the way we view and process information and has commanded considerable interest from 
physicists, engineers, computer scientists and mathematicians.  The counter-intuitive and strange 
rules of quantum physics offers enormous possibilities for information processing and the development 
of a large scale quantum computer is the holy grail for many researchers worldwide. 

While the advent of Shor's algorithm~\cite{S94} 
certainly spawned great interest in quantum information processing, 
demonstrating that quantum algorithms could be far more 
efficient than those used in classical computing, there was a great deal of debate surrounding the 
practicality of building a large scale, controllable, quantum system.  
It was well known even before the introduction of 
quantum information that coherent quantum states were extremely fragile and many believed that 
to maintain large, multi-qubit, coherent quantum states for a long enough time to 
complete {\em any} quantum algorithm was unrealistic~\cite{U95}.  Additionally, 
classical error correction techniques are intrinsically based on a digital framework, unlike quantum 
computing where  the {\em readout} of qubits is digital but actual manipulations are 
analogue.  Therefore, how can 
we adapt the vast knowledge gained from classical coding theory to the 
quantum regime?   

Starting in 1995, several papers appeared, in rapid succession, proposing codes which were 
appropriate to perform error correction on quantum data~\cite{S95,S96,CS96,LMPZ96,BDSW96}.  
This was a key theoretical development needed 
to convince the general community that quantum computation was indeed a possibility.  
Since this initial introduction, the progress in this field has been extensive.  

Initial research on error correction focused heavily on developing 
quantum codes~\cite{S96++,S96+++,G96,PVK96,KL97,K96}, introducing a more rigorous theoretical framework 
for the structure, properties and operation of 
QEC~\cite{BDSW96,KL00,CRSS98,G97+,CG97,KLV99}.  Additionally, the introduction of 
concepts such as fault-tolerant quantum computation~\cite{S96+,DS96,G98,KLZ98,G00+}  
leading directly to the threshold theorem for concatenated QEC~\cite{KLZ96,AB97}.
In more recent years, QEC protocols have been developed for various systems, such as,
continuous variables~\cite{LS98,B98,L08,ATKYBLF08}, ion-traps and other systems containing motional 
degrees of freedom~\cite{ST98}, adiabatic computation~\cite{JFS06} and
globally controlled quantum computers~\cite{BBK03}.  Work also continues on not only developing 
better (and in some ways, more technologically useful) protocols such as subsystem codes~\cite{KLP05,KLPL06,B06} and topological codes~\cite{K97,DKLP02,BD06,BD07,BD+07,RHG07,FSG08}. Advanced 
techniques to analyse fault-tolerant thresholds~\cite{AGP08,PR11,BD08,BAO12} have also been developed to implement error correction in a fault-tolerant 
manner with concatenated codes~\cite{S97,S02,DA07,K05} and topological codes~\cite{BD07,KBAD10,B11}.  

Along with QEC, other methods of protecting quantum information were also developed.  These other techniques 
would technically be placed in a separate category of error suppression rather than error correction.  
The distinction between error suppression and error correction is also known as passive and active QEC, the latter 
terminology is much more prevalent in the theoretical community.  The most well 
known techniques of error suppression are protocols such as decoherence free subspaces 
(DFS)~\cite{PSE96,DG97,LCW98,DG98,ZR97,ZR97+,DG98+,LW03}.  
DFS protocols are passive, encoding information in a subspace which is 
invariant under the dynamics that generate errors.  Active methods for error suppression include applying repeated rotations 
to qubits such that these rotations, combined with the natural dynamics that induce errors partially cancel.  
These methods were initially referred to as Bang-Bang control~\cite{VL98,VT98,Z99} and developed primarily for NMR 
systems.  In recent years these techniques have been further generalized and optimized for 
use in essentially all quantum information processing systems~\cite{VKL99,FLP04,VK03,VK05,KL05,U07}.  These more advanced techniques 
are now routinely categorized as dynamical decoupling protocols.  As with QEC, this field is vast, incorporating well established 
ideas from quantum control to create specially designed control sequences to suppress errors before more resource intensive 
encoding is applied.  

This review deals exclusively with the concepts of QEC and fault-tolerant 
quantum computation.  Many papers have reviewed error correction and 
fault-tolerance~\cite{G97+,NC00,G00,KLABVZ02,S03+,G09}, however to cater for 
a large audience, we attempt to describe QEC and fault-tolerance in a more basic manner, largely through examples.  Instead of providing a 
more rigorous review of error correction, we instead focus on more practical 
issues involved when working with these ideas.    
For those who have recently begun investigating quantum information processing or those who are focused on other important theoretical 
and/or experimental aspects related to quantum computing, searching through this enormous collection of work is daunting especially 
if a basic working knowledge of QEC is all that is required.  This review of the basic 
aspects of QEC and fault-tolerance 
is designed to allow those with little knowledge of the field to quickly become accustomed to the various techniques and tricks that are commonly 
used.  The examples in this review are selected due to their relative 
simplicity.  For individuals seriously interested in examining new ideas for QEC 
and fault-tolerant computation we strongly advise that the references cited throughout are 
consulted.  

We begin the discussion in section~\ref{sec:prelim} where we describe some preliminary concepts on the required properties of any 
quantum error correcting protocol.  In section~\ref{sec:error} we review some basic noise models from the context of 
how they influence quantum algorithms.  Section~\ref{sec:sec:3qubit}
introduces quantum error correction through the traditional 
example of the 3-qubit code, illustrating the circuits used for encoding and correction and why the principle of 
redundant encoding suppresses the failure of encoded qubits.  Sections~\ref{sec:9qubit} and~\ref{sec:detection} introduce the 
first examples of complete quantum codes, Shor's 9-qubit code and a 4-qubit error detection code. 
Sections~\ref{sec:sec:QEC} and~\ref{sec:sec:QEC2} introduces the 
stabiliser formalism, 
demonstrating how QEC circuits are synthesized once the structure of the code is known.  
In section~\ref{sec:sec:decoherence} we briefly return to noise models and relate the 
abstract analysis of QEC, where errors are assumed to be discrete and probabilistic, to some of the 
physical mechanisms which can cause errors.  Sections~\ref{sec:Fault-tolerance} and~\ref{sec:operations} 
introduce the concept of fault-tolerant error correction, the 
threshold theorem and how logical gate operations can be applied directly to encoded data.  
We then move onto circuit synthesis in section~\ref{sec:FTcircuit}, presenting a 
fault-tolerant circuit design for state 
preparation using the 7-qubit Steane code as a representative example.  
Finally, in section~\ref{sec:modern}, we review specific codes for qubit loss and 
examine two more modern techniques for error correction.  We briefly examine 
quantum subsystem codes~\cite{KLP05,KLPL06,B06} and topological surface codes~\cite{DKLP02,FSG08} due to both their 
theoretical elegance and their increasing relevance in quantum architecture designs~\cite{IFIITB02,DFSG08,SJ08,MF10}.  

\section{Preliminaries} 
\label{sec:prelim}
Before discussing quantum errors and how they can be corrected, 
we first introduce the basics of qubits and quantum gates.  We assume a basic 
working knowledge with quantum information~\cite{EJ96,NC00} 
and this brief discussion is used simply to define 
our notation for the remainder of this review.  

\subsection{Basic structure of the qubit}
The fundamental unit of quantum information is the qubit.  Unlike the classical bit, the qubit can exist in 
coherent superpositions of its two states, denoted as $\ket{0}$ and $\ket{1}$.  These basis states 
can be, for example, photonic polarization, atomic spin states, electronic states of an ion or charge states of 
superconducting systems.  An arbitrary state of an individual qubit, $\ket{\phi}$, can 
be expressed as,
\begin{equation}
\ket{\phi} = \alpha\ket{0} + \beta\ket{1}
\end{equation}
where $\ket{0}$ and $\ket{1}$ are the two orthonormal basis states of the qubit and 
 $|\alpha|^2+|\beta|^2 = 1$.  Quantum gate operations are 
represented by unitary operations acting on the Hilbert space of a collection of qubits.  Unlike 
classical information processing, conservation of probability for quantum states require that all 
operations be reversible and hence unitary.

When describing a quantum gate on an individual qubit, any dynamical operation, $G$, is 
a member of the unitary group $U(2)$, which consists of all $2\times 2$ matrices where 
$G^{\dagger} = G^{-1}$.  Up to a global (and unphysical) phase factor, any single qubit operation 
can be expressed as a linear combination of the generators of $SU(2)$ as,
\begin{equation}
G = c_I \sigma_I + c_x \sigma_x + c_y\sigma_y + c_z\sigma_z
\end{equation}
where,
\begin{equation}
\sigma_x = \begin{pmatrix} 0 & 1 \\ 1 & 0 \end{pmatrix},  \quad
\sigma_y = \begin{pmatrix} 0 & -i \\ i & 0 \end{pmatrix}, \quad 
\sigma_z = \begin{pmatrix} 1 & 0 \\ 0 & -1 \end{pmatrix},
\end{equation}
are the Pauli matrices, $\sigma_I$ is the $2\times 2$ identity matrix,  
$c_I$ is a real number, and $(c_x,c_y,c_z)$ are complex numbers satisfying,
\begin{equation}
\begin{aligned}
 &|c_I|^2+|c_x|^2+|c_y|^2+|c_z|^2 = 1, \\
 &\Re(c_Ic_x^*)+\Im(c_yc_z^*) = 0, \\
&\Re(c_Ic_y^*)+\Im(c_xc_z^*) = 0, \\
 &\Re(c_Ic_z^*)+\Im(c_xc_y^*) = 0.
  \end{aligned} 
 \end{equation}
Here $\Re$ and $\Im$ are the real and imaginary components respectively.
\subsection{Some general requirements of quantum error correction}
Although the field of QEC is largely based on classical coding theory, there are several issues that need to 
be considered when transferring classical error correction techniques to the quantum regime.  

First, coding based on data-copying, which 
is extensively used in classical error correction cannot be used 
due to the no-cloning theorem of quantum 
mechanics~\cite{WZ82}.  This result implies that there exists no transformation resulting in the 
following mapping,
\begin{equation}
U(\ket{\phi} \otimes \ket{\psi}) = \ket{\phi} \otimes \ket{\phi} \quad \forall \ket{\phi}.
\end{equation}
i.e. it is impossible to perfectly copy an unknown quantum state.  This means that quantum data cannot be 
protected from errors by simply making multiple copies.  
Secondly, direct measurement cannot be used to effectively protect against errors, since this will 
act to destroy any quantum superposition that is being used for computation.  Error correction protocols 
must therefore 
be employed, which can detect and correct errors without determining {\em any} information regarding the 
qubit state.  
Unlike classical information, qubits are susceptible to both traditional bit errors 
$\ket{0} \leftrightarrow \ket{1}$, 
{\em and} also phase errors $\ket{0} \leftrightarrow \ket{0}$, $\ket{1} \leftrightarrow -\ket{1}$.  
Hence any error correction procedure needs to be able to 
simultaneously correct for both.  Finally, errors in quantum information are intrinsically continuous (i.e. 
qubits do not experience full bit or phase flips but rather an angular shift of the qubit state by any angle).  
This issue will be discussed in more depth in Section \ref{sec:sec:decoherence}

At its most basic level, QEC utilizes the idea of redundant encoding. 
This is where the total size of the Hilbert space is expanded beyond what is needed to store a single 
qubit of information.  This way, errors on individual qubits are mapped to large set of mutually orthogonal 
subspaces, the size of which is determined by the number of qubits utilized in the code.  Finally, the 
error correction protocol cannot allow us to gain information regarding 
the coefficients, $\alpha$ and $\beta$ of the encoded state; as doing so would collapse the system.

\section{Quantum errors: Cause and effect}\label{sec:error}
Before we begin discussing the details of QEC, we first examine some of the 
common sources of errors in quantum information processing and contextualize what they imply 
for computation.  We will consider several important sources of errors and how they influence two 
trivial, single qubit quantum algorithms.

The first algorithm is a computation consisting of a single qubit, initialized in the $\ket{0}$ state 
undergoing $N$ identity operations.  If this algorithm is performed perfectly the final state is,
\begin{equation}
\ket{\psi}_{\text{final}} = \prod_i^N I_i\ket{0} = \ket{0},
\end{equation}
where $I \equiv \sigma_I$ is the identity gate.  Measurement 
of the qubit in the $\ket{0}$, $\ket{1}$ basis 
will consequently yield the result 0 with a probability of unity.  

The second illustration is an algorithm of three gates,
\begin{equation}
\begin{aligned}
\ket{\psi}_{\text{final}} = HIH\ket{0} &=HI \frac{1}{\sqrt{2}}(\ket{0} + \ket{1}) \\
&= H \frac{1}{\sqrt{2}}(\ket{0}+\ket{1})  \\
&= \ket{0}.
\end{aligned}
\end{equation}
This algorithm will {\em ideally} 
yield a result $\ket{0}$ with probability of 1 when measured in the $\ket{0}$, $\ket{1}$ basis.
This algorithm implements two $H \equiv $ Hadamard operations separated by a wait stage, 
represented by the Identity gate.  

We examine, independently, several 
common sources of error from the effect they have on these two quantum algorithms.  Hopefully, 
this introductory section will show that while quantum errors are complicated physical effects, 
in QIP the relevant measure is the theoretical success probability of a given quantum algorithm. 

Errors that exist on any quantum system are highly dependent on the specific 
physical mechanisms that govern the system.  The situations that follow are only 
basic illustrative examples which can often arise in discussions related to errors and 
QEC.  It is important to stress that the types of errors and the way in which they are analyzed 
needs to be considered in the context of the physical system under consideration.

\subsection{Coherent quantum errors: Gates which are incorrectly applied}
The first possible source of error is coherent, systematic control errors.  This type of 
error is typically associated with incorrect knowledge of the system dynamics.  Namely, otherwise 
perfect control assuming a system governed by a Hamiltonian $H'$ when the actual system 
dynamics is governed by $H$.  As qubit systems must be characterized prior to being used \cite{PCZ97,CSDWOH05}, 
finite experimental resolution with these processes will lead to coherent control errors.  As these errors 
are coherent inaccuracies applied to the qubit, they can be mitigated using coherent techniques such as composite pulse sequences \cite{L79,L86,BHC04,J09}.
Other control errors 
(e.g. arising from fluctuations in control fields) are not coherent processes and are dealt with in a 
similar way to environmental coupling described further on.

As this source of error is coherent and systematic it 
causes you to apply the same, undesired gate operation and can be analyzed without moving to the density matrix formalism.  
With respect to the first of our algorithms, we are able to model this several different ways.  To 
keep things simple, we assume that incorrect characterization of the control dynamics leads to a 
gate which is not $\sigma_I$, but instead introduces a small rotation around the 
$X$-axis of the Bloch sphere.  This results in the state,
\begin{equation}
\ket{\psi}_{\text{final}} = \prod^N e^{i\epsilon \sigma_x}\ket{0} = \cos(N\epsilon)\ket{0} + i\sin(N\epsilon)\ket{1}.
\end{equation}
We now measure the system in the $\ket{0}$,  $\ket{1}$ basis.  
Due to these errors, the probability of measuring the system in the $\ket{0}$ or $\ket{1}$ state is,
\begin{equation}
\begin{aligned}
&P(\ket{0}) = \cos^2(N\epsilon) \approx 1- (N\epsilon)^2, \\
&P(\ket{1}) = \sin^2(N\epsilon) \approx (N\epsilon)^2.
\end{aligned}
\end{equation}
Hence, the probability of error in this trivial quantum algorithm is given by $p_{\text{error}} \approx (N\epsilon)^2$, which will be 
small given that $N\epsilon \ll 1$.  The systematic error in this system is quadratic in both the small systematic over rotation  
and the total number of applied identity operations.  This is expected as this rotational error always acts 
in the same direction every time it is applied.

\subsection{Environmental decoherence}
Environmental decoherence is another important source of errors in quantum systems.  We will take a simple 
decoherence model and examine how our second algorithm behaves.  
Later in section~\ref{sec:sec:decoherence} we will illustrate a more 
complicated decoherence model that arises from standard mechanisms.  

Consider a very simple environment, which is another two level quantum system.  This environment has two basis states, $\ket{e_0}$ 
and $\ket{e_1}$, that satisfy the completeness relations,
\begin{equation}
\bra{e_i}e_j\rangle = \delta_{ij}, \quad \ket{e_0}\bra{e_0} + \ket{e_1}\bra{e_1} = I.
\end{equation}
We will also assume that the environment couples to the qubit in a specific way.  When the qubit is in the $\ket{1}$ state, the coupling 
flips the environmental state, while if the qubit is in the $\ket{0}$ state nothing happens to the environment.
Finally, this model assumes the system/environment interaction only occurs during the 
wait stage of the algorithm (while the Identity gate is being applied).  As with 
the first algorithm we should measure the state $\ket{0}$ with probability one.  The reason for 
utilizing the second of our algorithms in this example 
is that this specific decoherence model acts to reduce coherence between the $\ket{0}$ and $\ket{1}$ 
states.  Therefore, we require a coherent superposition to observe any effect from the environmental coupling. 
We assume the environment is initialized in the state, 
$\ket{E} = \ket{e_0}$, and then coupled to the system,
\begin{equation}
HI H\ket{0}\ket{E} = \frac{1}{2}(\ket{0}+\ket{1})\ket{e_0} + \frac{1}{2}(\ket{0}-\ket{1})\ket{e_1}
\end{equation}
As we are considering environmental decoherence, pure states will be transformed into classical mixtures, hence we now move 
into the density matrix representation for the state $HI H\ket{0}\ket{E}$,
\begin{equation}
\begin{aligned}
\rho_{f} &= \frac{1}{4}( \ket{0}\bra{0} + \ket{0}\bra{1}+\ket{1}\bra{0}+\ket{1}\bra{1})\ket{e_0}\bra{e_0}\\
&+\frac{1}{4}( \ket{0}\bra{0} - \ket{0}\bra{1}-\ket{1}\bra{0}+\ket{1}\bra{1})\ket{e_1}\bra{e_1} \\
&+ \frac{1}{4}( \ket{0}\bra{0} - \ket{0}\bra{1}+\ket{1}\bra{0}-\ket{1}\bra{1})\ket{e_0}\bra{e_1}\\
&+\frac{1}{4}( \ket{0}\bra{0} + \ket{0}\bra{1}-\ket{1}\bra{0}-\ket{1}\bra{1})\ket{e_1}\bra{e_0}.
\end{aligned}
\end{equation}
As we have no access to the environmental degrees of freedom, we trace over this part of the system, giving,
\begin{equation}
\begin{aligned}
\text{Tr}_{E}(\rho_f) &= \frac{1}{4}( \ket{0}\bra{0} + \ket{0}\bra{1}+\ket{1}\bra{0}+\ket{1}\bra{1})\\
&+\frac{1}{4}( \ket{0}\bra{0} - \ket{0}\bra{1}-\ket{1}\bra{0}+\ket{1}\bra{1}) \\
&= \frac{1}{2}(\ket{0}\bra{0}+\ket{1}\bra{1}).
\end{aligned}
\end{equation}
Measurement of the system will consequently return $\ket{0}$ 50\% of the time and $\ket{1}$ 50\% of the time.  This final state 
is a complete mixture of the two qubit states and is consequently a classical system.  The coupling to the environment removed all the 
coherences between the $\ket{0}$ and $\ket{1}$ states and the second Hadamard transform, 
intended to rotate $(\ket{0}+\ket{1})/\sqrt{2} \rightarrow \ket{0}$, has no effect on the qubit state.  

As we assumed that the system/environment 
coupling during the wait stage causes the environmental degree of freedom to ``flip" (when the qubit is in the $\ket{1}$ state) this decoherence model implicitly incorporates a temporal effect.  
The temporal interval of the identity gate in the above algorithm is long enough to enact the controlled-flip operation.  If we 
assumed a controlled rotation that is not a full flip on the environment, the final mixture will not be 50/50.  Instead there 
would be a residual coherence between the qubit states and an increased probability of our algorithm returning a 
$\ket{0}$.  Section~\ref{sec:sec:decoherence} 
revisits the decoherence model and illustrates how time-dependence is explicitly incorporated.  

\subsection{Simple models of loss, leakage, measurement and initialization}
\label{sec:lossleakage}

Other sources of error such as qubit initialization, measurement errors, qubit loss and qubit leakage can be modeled in 
a very similar manner.  As with coherent control errors and environmental decoherence, the specifics of other error 
channels can be modeled in a coherent or incoherent manner, depending on the physical mechanisms. 
The examples shown below are commonplace within QEC analysis 
and adequately apply to a large number of systems. 

Measurement errors can be modeled in the same way as environmental decoherence, acting incoherently on the system.  
Measurement errors can be described in two slightly different ways.  The first it to use the 
following positive operator value measures (POVM's),
\begin{equation}
\begin{aligned}
F_0 = (1-p_M)\ket{0}\bra{0} + p_M\ket{1}\bra{1}, \\
F_1 = (1-p_M)\ket{1}\bra{1} + p_M\ket{0}\bra{0},
\end{aligned}
\label{eq:POVM}
\end{equation}
where $p_M$ is the probability of measurement error.  These are not measurement projectors as $F_0^2 \neq 
F_0$ and $F_1^2 \neq F_1$.  The second method is to apply the following mapping to the qubit,
\begin{equation}
\rho \rightarrow \rho' = (1-p_M)\rho + p_M X\rho X
\end{equation}
followed by a perfect measurement in the $(\ket{0}, \ket{1})$ basis.  
These two models give exactly the same probabilities for each outcome.  Defining $A_0 = \ket{0}\bra{0}$ and $A_1 = \ket{1}\bra{1}$ as the  measurement projectors onto the ($\ket{0}, \ket{1}$) basis we find,
\begin{equation}
\begin{aligned}
&\text{Tr}(F_0\rho) = (1-p_M)\text{Tr}(A_0\rho) + p_M\text{Tr}(A_1\rho), \\
&\text{Tr}(F_1\rho) = (1-p_M)\text{Tr}(A_1\rho) + p_M\text{Tr}(A_0\rho), 
\end{aligned}
\end{equation}
\begin{equation}
\begin{aligned}
\text{Tr}(A_0\rho') &= (1-p_M)\text{Tr}(A_0\rho) + p_M\text{Tr}(XA_0X\rho) \\
&= (1-p_M)\text{Tr}(A_0\rho) + p_M\text{Tr}(A_1\rho), \\
\text{Tr}(A_1\rho') &= (1-p_M)\text{Tr}(A_1\rho) + p_M\text{Tr}(XA_1X\rho) \\
&= (1-p_M)\text{Tr}(A_1\rho) + p_M\text{Tr}(A_0\rho).
\end{aligned}
\end{equation}
Hence, either method will result in the same probabilities for a given value of $p_M$.  
The difference between these two models is the state the measured qubit is projected to.  

When using the POVM's in Eq.(\ref{eq:POVM}), the collapsed state of the qubit is given by,
\begin{equation}
\begin{aligned}
\rho \rightarrow \frac{M_i\rho M_i^{\dagger}}{\text{Tr}(F_i\rho)} \quad i = 0,1,
\end{aligned}
\end{equation}
where,
\begin{equation}
\begin{aligned}
M_0 = \sqrt{1-p_M}\ket{0}\bra{0} + \sqrt{p_M}\ket{1}\bra{1}, \\
M_1 = \sqrt{1-p_M}\ket{1}\bra{1} + \sqrt{p_M}\ket{0}\bra{0}.
\end{aligned}
\end{equation}
Therefore the resulting state, after measurement, will {\em not} be initialized in a known state.  If the second model is used, then the system is 
initialized to either $\ket{0}$ or $\ket{1}$ depending on which projector is used.  To see this explicitly, consider the 
state,
\begin{equation}
\frac{1}{\sqrt{2}}\left(\ket{0} + \ket{1}\right).
\end{equation}
Using both models, the probability of measuring $\ket{0}$ is $1/2$  (as the above state is 
invariant under a bit-flip error).  If the POVM model is used (using the $F_0$ and $M_0$ operators), the state after measurement is,
\begin{equation}
\sqrt{(1-p_m)}\ket{0} + \sqrt{p_m}\ket{1}.
\end{equation}
The collapsed state, after measurement, is a superposition of $\ket{0}$ and $\ket{1}$, with amplitudes related 
to $p_m$.  If we use the second model (using the projector $A_0$), the state after measurement is simply $\ket{0}$.
As measurement is generally followed by either discarding the qubit or reinitializing it in a known state, both 
models can generally be used interchangeably with no adverse consequences.

Qubit loss is modeled in a slightly different manner.  When a qubit is lost, it is effectively removed from the system.  
This channel can be modeled by simply tracing out the lost qubit, $\text{Tr}_i(\rho)$, where $i$ is the index of the lost qubit.  This reduces 
the dimensionality of the qubit space by a factor of two.

The loss of the physical object means that it cannot be directly measured or coupled to any other ancillary system.  
Even though this model of qubit loss (with 
regards to the information stored on the qubit) is equivalent to incoherent 
processes such as environmental 
coupling, correcting this type of error requires additional machinery on top of standard 
QEC protocols.   Standard correction 
protocols can protect against the loss of information on a qubit, provided that the physical object still exists in the 
computer.  Therefore, an initial non-demolition detection must be employed (which 
determines if the qubit is actually present without performing a projective measurement on the computational 
state) before standard correction can correct the error.  Provided that these non-demolition protocols are employed, loss 
actually becomes a preferred channel in error correction.  If a loss event is detected and the quit is replaced, this heralds the qubit experiencing the 
error.  This additional information can be used within error correction protocols to increase performance.  

Initialization of the qubit can be modeled either using a coherent systematic or incoherent process.  If an incoherent model is 
employed, initialization can be modeled essentially the same way as imperfect measurement.  If we have a probability $p_I$ of 
initialization error (where $p_I$ encapsulates the internal mechanisms that introduce errors) and we initialize the qubit 
in the $\ket{0}$ state, the initial state of the system is given by the mixture,
\begin{equation}
\rho_i = (1-p_I)\ket{0}\bra{0} + p_I\ket{1}\bra{1}.
\end{equation}
In contrast, we could consider an initialization model which is achieved via a coherent unitary operation where the 
target is the desired initial state.  In this case, the initial state is pure, but contains a non-zero amplitude 
of the undesired target, for example,
\begin{equation}
\ket{\psi} = \alpha\ket{0} + \beta\ket{1},
\end{equation}
where $|\alpha|^2+|\beta|^2 = 1$ and $|\beta|^2 \ll 1$.  The interpretation of these two types of initialization models is identical 
to the coherent and incoherent models presented.  Again, the effect of these types of errors relates 
to the probabilities of measuring the system in an erred state.  

\begin{figure}[ht]
\begin{center}
\includegraphics[width=0.55\textwidth]{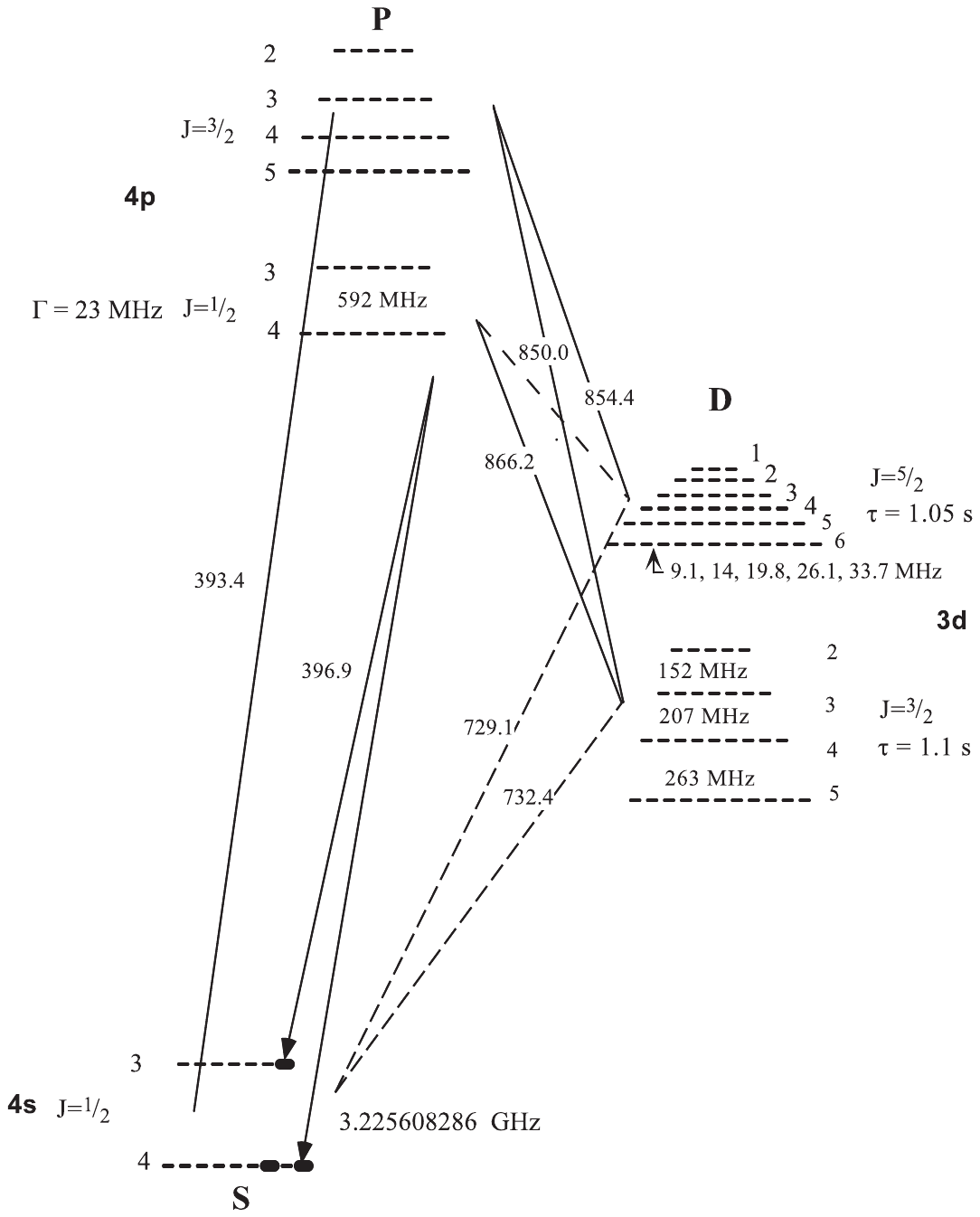}
\caption{(from Ref.~\cite{S97+}) Energy level structure for the $^{43}$Ca$^+$ investigated by the Oxford ion-trapping group.  
The structure of this ion is clearly not a 2-level quantum system.  Hence leakage into non-qubit states is an important factor to 
consider.}
\label{fig:calcium}
\end{center}
\end{figure}

One final type of error that we can briefly mention is qubit leakage.  Qubit leakage manifests due to the fact 
that most systems utilized for qubits do not consist of just 2 levels.  For example, Fig~\ref{fig:calcium} (from Ref.~\cite{S97+}) 
illustrates the energy level structure for a $^{43}$Ca$^+$ ion utilized for ion trap quantum computing.  
The qubits in this system are 
defined with two electronic states,  however the system itself contains many more levels 
(including some which are used for qubit readout and initialization).  
As with systematic errors, leakage can occur when improper control is applied to such a system.  In the case of ion-traps, 
qubit transitions are performed by focusing finely tuned lasers resonant on the relevant transitions.  If the laser frequency 
fluctuates or additional levels are not sufficiently detuned from the qubit resonance, the following transformation could 
occur,
\begin{equation}
U\ket{0} = \alpha\ket{0} + \beta\ket{1} + \gamma\ket{2},
\label{eq:leak}
\end{equation}
where the state $\ket{2}$ is a third level, which is now populated due to improper control.  The actual effect of this type of 
error can manifest in several different ways.  
As quantum circuits and algorithms are fundamentally designed assuming the 
computational array is a  of 2-level systems, the transformation shown 
in Eq.~(\ref{eq:leak}) (which in this case is operating over 
a 3-level space) will naturally induce unwanted dynamics.  Another important implication of applying non-qubit operations 
is how these levels interact with the environment and hence how decoherence effects the system.  
For example, in the above case, the unwanted level, $\ket{2}$, may be extremely short lived leading to an 
emission of a photon and the system relaxing back to the ground state.  For these reasons, leakage is one of 
the most problematic error channels to correct using QEC.  In general, leakage induced errors need to be 
corrected in a similar manner to loss, which can also be thought of as a 
leakage process into a vacuum state.  Non-demolition techniques are first required to determine if the 
system is still confined to the qubit subspace~\cite{P98,GBP97,VWW05} after which standard correction 
protocols can be utilized.  As with qubit loss, if these additional protocols are employed, leakage errors 
are heralded and the performance of the resulting error correction will increase.  

Another well known method is to use a complicated pulse 
sequence which acts to re-focus an improperly confined operation back to the 
qubit subspace~\cite{WBL02,BLWZ05}.  This can be advantageous as it does not require additional qubit resources.  

Leakage effects arise from multiple sources, and the best method for correction is often heavily dependent on the 
error mechanisms.  One 
example is when internal fluctuations 
in the system change otherwise perfect qubit dynamics.  Unfortunately this leakage source generally cannot be predicted and/or characterized 
and so dedicated leakage protection will need to be employed.  A second source is imprecise 
fabrication of an otherwise perfect qubit system.  This source of leakage can, in principal, 
be engineered away, thereby eliminating the need for specialized leakage correction. 
In the context of mass manufacturing of qubit systems, leakage would ideally be 
quantified immediately after the fabrication of a device~\cite{DSOCH07} to eliminate these systematic leakage errors.  
If a particular system is found to be improperly confined to the qubit subspace, it would simply be discarded.  Employing characterization at this stage could potentially eliminate the need to 
implement a large amount of leakage protection in the computer, shortening gate times and ultimately reducing error rates in the computer.

In this section we have presented a very basic set of examples to explain 
some of the ideas of quantum errors and how they effect the success of a quantum 
algorithm.  Section~\ref{sec:sec:decoherence} will return to a more realistic set 
of error models.  For those interested in a more complete treatment of quantum errors 
we encourage readers to refer to~\cite{G97+,NC00,G00,KLABVZ02,S03+,G09}.  

\section{The 3-qubit code: a good starting point for quantum error correction}  
\label{sec:sec:3qubit}
The 3-qubit bit-flip code is traditionally used as a basic introduction to the concept of quantum error correction.  
It should be emphasized that the 3-qubit code {\em does not} represent a full quantum code.  This is 
due to the fact that the code cannot simultaneously correct for both bit and phase flips 
[Sec.~\ref{sec:sec:decoherence}].  
This code is a repetition code extended by 
Shor~\cite{S95} to construct the 9-qubit quantum code, demonstrating that QEC was possible.

\begin{figure}[ht]
\begin{center}
\includegraphics[width=0.28\textwidth]{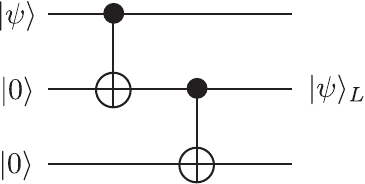}
\caption{Quantum Circuit to prepare the $\ket{0}_L$ state for the 3-qubit code where an arbitrary 
single qubit state, $\ket{\psi}$ is coupled to two freshly initialized ancilla qubits via 
CNOT gates to prepare $\ket{\psi}_L$.}
\label{fig:3qubit}
\end{center}
\end{figure}

The 3-qubit code encodes a single logical qubit into three physical qubits with the property that 
it can correct for a single, $\sigma_x$, bit-flip error.  The two logical basis states 
$\ket{0}_L$ and $\ket{1}_L$ are defined as,
\begin{equation}
\ket{0}_L = \ket{000}, \quad \quad \ket{1}_L = \ket{111},
\end{equation}
such that an arbitrary single qubit state $\ket{\psi} = \alpha\ket{0} + \beta\ket{1}$ is mapped to,
\begin{equation}
\begin{aligned}
\alpha\ket{0} + \beta\ket{1} &\rightarrow \alpha\ket{0}_L + \beta\ket{1}_L \\
&= \alpha\ket{000} + 
\beta\ket{111} \\
&= \ket{\psi}_L.
\end{aligned}
\end{equation}
Fig.~\ref{fig:3qubit} illustrates the quantum circuit required to encode a single logical qubit via the initialization of 
two ancilla qubits and two CNOT gates.  
The reason that this code is able to correct for a single bit-flip error is the binary distance between the two 
codeword states.  Notice that three individual bit flips are required to take 
$\ket{0}_L \leftrightarrow \ket{1}_L$, hence 
if we assume $\ket{\psi} = \ket{0}_L$, a single bit flip on any qubit leaves the final state closer to $\ket{0}_L$ 
than $\ket{1}_L$.  The distance between two codeword states, $d$, defines the number of errors that 
can be corrected, $t$, as, $t = \lfloor(d-1)/2\rfloor$.  In this case, $d=3$, hence $t=1$.  

\begin{figure}[ht]
\begin{center}
\includegraphics[width=0.5\textwidth]{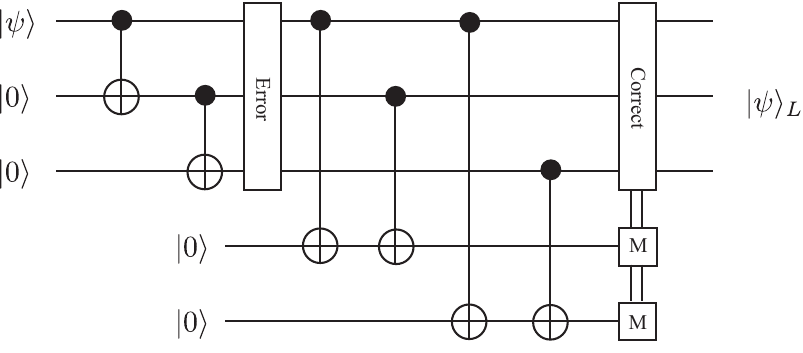}
\caption{Circuit required to encode and correct for a single $\sigma_x$-error.  We assume that after encoding a single 
bit-flip occurs on one of the three qubits (or no error occurs).  Two initialized ancilla are then 
coupled to the data block 
which only checks the parity between qubits.  These ancilla are then measured, with the measurement result 
indicating where (or if) an error has occurred, without directly measuring any of the data qubits.  Using this 
{\em syndrome} information, the error can be corrected with a classically controlled $\sigma_x$ gate. }
\label{fig:3qubit2}
\end{center}
\end{figure}

How are we able to correct errors using this code without directly measuring or obtaining information 
about the logical state?  
Two additional ancilla qubits are introduced, which are used to extract {\em syndrome} information (information 
regarding possible errors) from the data block 
without discriminating the exact state of any qubit.  The encoding and correction circuit is illustrated in Fig.~\ref{fig:3qubit2}.   Correction 
proceeds by introducing two ancilla qubits and performing a sequence of CNOT gates, which checks the 
parity of the three qubit data block.
For the sake of simplicity we assume that all gate operations are perfect and the only place where the 
qubits are susceptible to error is the region between encoding and correction (As illustrated in 
Fig~\ref{fig:3qubit2}).  We will return to 
this issue in section~\ref{sec:Fault-tolerance} when we discuss fault-tolerance.  We also assume that 
at most a single, complete bit flip error occurs on one of the three data qubits. Table~\ref{tab:errors} summarizes the state of the whole system, for each possible error, just prior to measurement.
\begin{table}[ht!]
\begin{center}
\vspace*{4pt}   
\begin{tabular}{c|c}
Error Location & Final State, $\ket{\text{data}}\ket{\text{ancilla}}$ \\
\hline
No Error & $\alpha\ket{000}\ket{00} + \beta\ket{111}\ket{00}$ \\
Qubit 1 & $\alpha\ket{100}\ket{11} + \beta\ket{011}\ket{11}$ \\
Qubit 2 & $\alpha\ket{010}\ket{10} + \beta\ket{101}\ket{10}$ \\
Qubit 3 & $\alpha\ket{001}\ket{01} + \beta\ket{110}\ket{01}$ \\
\end{tabular}
\caption{Final state of the five qubit system prior to the syndrome measurement for no error or a single 
$X$ error on one of the qubits.  The last two qubits represent the state of the ancilla.  Note that each possible 
error will result in a unique measurement result (syndrome) 
of the ancilla qubits.  This allows for a $\sigma_x$ correction gate to be applied 
to the data block which is classically controlled from the syndrome result.} 
\label{tab:errors}
\end{center}
\end{table} 

For each possible situation, either no error or a single bit-flip error, the ancilla qubits are flipped to a unique state based on 
the parity of the data block.  These qubits are then measured to obtain the classical {\em syndrome} result.  The 
result of the measurement will then dictate if a correction gate needs to be applied.  This is illustrated in 
Table.~\ref{tab:3qubit}.
\begin{table}[ht!]
\begin{center}
\vspace*{4pt}   
\begin{tabular}{c|c|c}
Ancilla Measurement& Collapsed State & Consequence \\
\hline
$00$ & $\alpha\ket{000} + \beta\ket{111}$ & No Error\\
$01$ & $\alpha\ket{001} + \beta\ket{110}$ & $\sigma_x$ on Qubit 3\\
$10$ & $\alpha\ket{010} + \beta\ket{101}$ & $\sigma_x$ on Qubit 2\\
$11$ & $\alpha\ket{100} + \beta\ket{011}$ & $\sigma_x$ on Qubit 1\\
\end{tabular}
\caption{Ancilla measurements for a single $\sigma_x$ error with the 3-qubit code.  Each of the four possible results correspond to 
either no error or a bit flip on one of the three qubits.}
\label{tab:3qubit}
\end{center}
\end{table} 

Provided that only a single error has occurred, the data block is restored.  At no point during 
correction do we gain any information regarding the coefficients $\alpha$ and $\beta$, hence superpositions 
of the computational state will remain in-tact during correction.  

This code will only work if a maximum of one error occurs.  The 3-qubit code is a $d=3$ code, hence if $t > 1$, then 
the resulting state becomes ``closer" to the wrong logical state.  This can be seen in Table.~\ref{tab:errors3} when tracking multiple errors through 
the correction circuit.
\begin{table}[ht!]
\begin{center}
\vspace*{4pt}   
\begin{tabular}{c|c|c}
Error Location & Final State, $\ket{\text{data}}\ket{\text{ancilla}}$ & Assumed Error\\
\hline
Qubit 1 \& 2 & $\alpha\ket{110}\ket{01} + \beta\ket{001}\ket{01}$  &$\sigma_x$ on Qubit 3\\
Qubit 2 \& 3 & $\alpha\ket{011}\ket{11} + \beta\ket{100}\ket{11}$ &$\sigma_x$ on Qubit 1\\
Qubit 1 \& 3 & $\alpha\ket{101}\ket{10} + \beta\ket{010}\ket{10}$ &$\sigma_x$ on Qubit 2\\
Qubit 1, 2 \& 3 & $\alpha\ket{111}\ket{00} + \beta\ket{000}\ket{00}$ & No Error
\end{tabular}
\caption{Ambiguity of syndrome results when multiple errors occur.} 
\label{tab:errors3}
\end{center}
\end{table} 
In each case, we 
assume that the total number of errors is $> d/2$, therefore our correction 
takes us to the wrong logical state.   
In every case where $t > 1$ our mis-correction induces a logical bit flip, causing the code to fail.

To be absolutely clear on how QEC acts to restore the system and protect against errors let us now consider a different and 
slightly more physically realistic example.  We will 
assume that the errors acting on the qubits are coherent rotations of the form 
$U = \exp (i\epsilon \sigma_x)$ on each qubit, with $\epsilon \ll 1$.  We 
choose coherent rotations so that we can remain in the state vector representation.  This is not a necessary requirement, however more 
general incoherent mappings would require us to move to density matrices.  

We assume that each qubit experiences the same error, hence the error operator acting on the state is,
\begin{equation}
\begin{aligned}
\ket{\psi}_E = E&\ket{\psi}_L,\\
E = U^{\otimes 3} &= (\cos(\epsilon)\sigma_I + i\sin(\epsilon)\sigma_x)^{\otimes 3} \\
&= c_0\sigma_I\sigma_I\sigma_I\\
&+ c_1 (\sigma_x\sigma_I\sigma_I+\sigma_I\sigma_x\sigma_I+\sigma_I\sigma_I\sigma_x) \\
&+ c_2 (\sigma_x\sigma_x\sigma_I+\sigma_I\sigma_x\sigma_x+\sigma_x\sigma_I\sigma_x) \\
&+ c_3 \sigma_x\sigma_x\sigma_x.
\end{aligned}
\end{equation}
where,
\begin{equation}
\begin{aligned}
&c_0 = \cos^3(\epsilon), \\
&c_1 = i\cos^2(\epsilon)\sin(\epsilon),  \\
&c_2 = -\cos(\epsilon)\sin^2(\epsilon),\\
 &c_3 = -i\sin^3(\epsilon).
\end{aligned}
\end{equation}
Now let's examine 
the transformation that occurs when we run the error correction circuit in Fig.~\ref{fig:3qubit2}.  This is represented via the unitary transformation, 
$U_{\text{QEC}}$, over {\em both} the data and ancilla qubits,
\begin{equation}
\begin{aligned}
U_{\text{QEC}} (E\ket{\psi}_L\ket{00}) &= 
c_0\ket{\psi}_L\ket{00} \\
&+c_1 \sigma_x\sigma_I\sigma_I\ket{\psi}_L\ket{11} \\
&+ c_1\sigma_I\sigma_x\sigma_I\ket{\psi}_L\ket{10} \\
&+ c_1\sigma_I\sigma_I\sigma_x\ket{\psi}_L\ket{01} \\
&+c_2 \sigma_x\sigma_x\sigma_I \ket{\psi}_L\ket{10} \\
&+ c_2 \sigma_I\sigma_x\sigma_x\ket{\psi}_L\ket{11}\\ 
&+c_2\sigma_x\sigma_I\sigma_x\ket{\psi}_L\ket{01} \\
&+c_3\sigma_x\sigma_x\sigma_x\ket{\psi}_L\ket{00}
\end{aligned}
\end{equation}
Once again, the ancilla block is measured and the appropriate correction operator is applied, yielding the results (up to renormalisation) 
shown in Table.~\ref{tab:3qubit2}.  
\begin{table}[ht!]
\begin{center}
\vspace*{4pt}   
\begin{tabular}{c|c}
Ancilla Measurement& Collapsed State (with correction) \\
\hline
$00$ & $c_0\ket{\psi}_L + c_3\sigma_x\sigma_x\sigma_x\ket{\psi}_L$ \\
$01$ & $c_1\ket{\psi}_L + c_2\sigma_x\sigma_x\sigma_x\ket{\psi}_L$\\
$10$ & $c_1\ket{\psi}_L + c_2\sigma_x\sigma_x\sigma_x\ket{\psi}_L$\\
$11$ & $c_1\ket{\psi}_L + c_2\sigma_x\sigma_x\sigma_x\ket{\psi}_L$\\
\end{tabular}
\caption{Resulting logical state after a round of error correction with the 3-qubit code.}
\label{tab:3qubit2}
\end{center}
\end{table} 
In each case, after correction (based on the syndrome result), we are left with approximately the same state, a superposition 
of a ``clean state" with the logically flipped state, $\sigma_x\sigma_x\sigma_x\ket{\psi}$.  
Let us now examine in detail the amplitudes now associated with each term in these states.  
If we consider the unitary $U$ acting on a single, unencoded qubit, the rotation takes the 
state $\ket{\psi}$ to,
\begin{equation}
U\ket{\psi} = \cos(\epsilon)\ket{\psi} + i\sin(\epsilon)\sigma_x\ket{\psi}, 
\end{equation}
Consequently, the fidelity of the single qubit state is,
\begin{equation}
F_{\text{unencoded}} = |\bra{\psi}U\ket{\psi}|^2 = \cos^2{\epsilon} \approx 1-\epsilon^2
\end{equation}
In contrast, the worst case fidelity\footnote{worst case fidelity assumes that the state $\ket{\psi}_L$ is orthogonal to the state $\sigma_x\sigma_x\sigma_x\ket{\psi}_L$.} of the encoded qubit state after a cycle of error correction is,
\begin{equation}
\begin{aligned}
F_{\text{no error detected}} &= \frac{|c_0|^2}{|c_0|^2+|c_3|^2} \\
&= \frac{\cos^6(\epsilon)}{\cos^6(\epsilon)+\sin^6(\epsilon)} \\ &\approx 1-\epsilon^6,
\end{aligned}
\end{equation}
with probability $1-3\epsilon^2+O(\epsilon^4)$ and
\begin{equation}
\begin{aligned}
F_{\text{error detected}} &= \frac{|c_1|^2}{|c_1|^2+|c_2|^2} \\ &= \frac{\cos^4(\epsilon)\sin^2(\epsilon)}{\cos^4(\epsilon)\sin^2(\epsilon)+\sin^4(\epsilon)\cos^2(\epsilon)} \\ &\approx 1-\epsilon^2,
\end{aligned}
\end{equation}
with probability $3\epsilon^2 + O(\epsilon^4)$.
This is the general idea of how QEC suppresses errors at the logical level.  During a round of error correction, if no error is detected, the error on the resulting state is suppressed from $O(\epsilon^2)$ to 
$O(\epsilon^6)$.  If a single error is detected, the fidelity of the resulting state remains the same.  
As the 3-qubit code is a single error correcting code, if one error has already been corrected 
then the failure rate of the logical qubit is conditional on experiencing one further error 
(which will be proportional to $\epsilon^2$).  As $\epsilon \ll 1$ 
the majority of correction cycles will detect no error and the fidelity of the resulting 
encoded state is higher than when 
unencoded.  

It should be stressed that {\bf no error correction scheme will, in general, fully restore a corrupted state to the original logical state}.  The resulting state will generally
contain a superposition or a mixture 
of a clean state and a logically erred state depending on whether the error process is coherent or incoherent. 
The point is that the fidelity of the corrupted states, at the logical level, 
is greater than the corresponding fidelity for unencoded qubits.  Consequently the probability of measuring the correct result at the 
end of a specific algorithm increases when the system is encoded.  The example shown here is somewhat unphysical.  i.e. it 
assumes perfect gate operations, errors that only consists of $\sigma_x$-rotations at a specific point in the circuit.  In the coming 
sections we will introduce the concepts necessary when relaxing these assumptions.

\section{The 9-qubit code:  The first full quantum code}
\label{sec:9qubit}

\begin{figure}[ht]
\begin{center}
\includegraphics[width=0.4\textwidth]{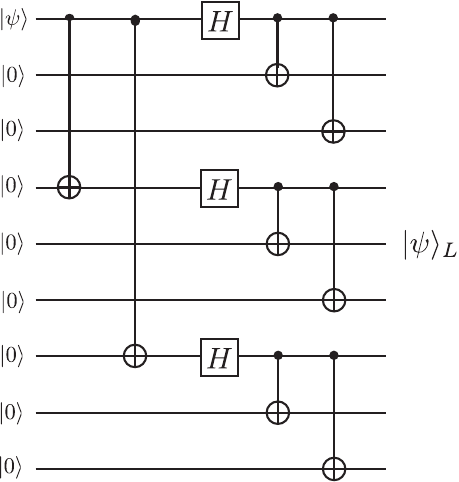}
\caption{Circuit required to encode a single qubit with Shor's nine qubit code.}
\label{fig:9encode}
\end{center}
\end{figure}

\begin{figure*}[ht]
\begin{center}
\includegraphics[width=0.8\textwidth]{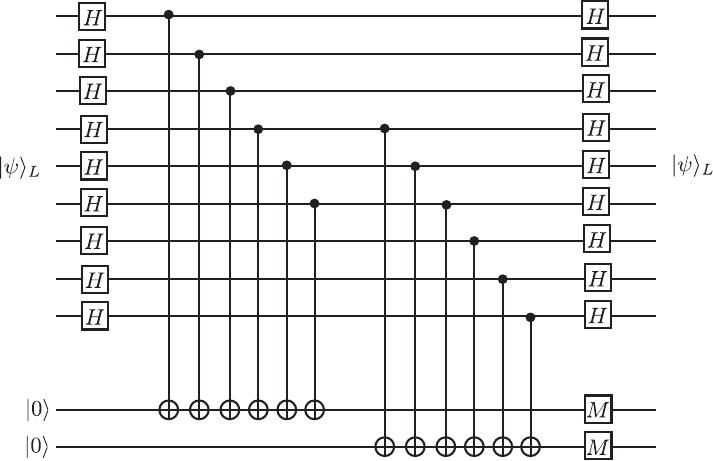}
\caption{Circuit required to perform $Z$-error correction for the 9-qubit code.}
\label{fig:9qubit2}
\end{center}
\end{figure*}

The nine qubit error correcting code was first developed by Shor~\cite{S95} in 1995 and is based largely on the 3-qubit repetition code.  The Shor code is a degenerate\footnote{Degenerate quantum codes are ones where different types of errors have the same effect on the codestates.}
single error correcting code, able to correct a logical qubit from one bit-flip, one phase-flip or 
one of each, on any of the nine physical qubits.   This code is therefore
sufficient to correct for an arbitrary single qubit error [Sec.~\ref{sec:sec:decoherence}].  
The two basis states for the code are,
\begin{equation}
\begin{aligned}
\ket{0}_L = \frac{1}{\sqrt{8}}(\ket{000}+\ket{111})(\ket{000}+\ket{111})(\ket{000}+\ket{111}) \\
\ket{1}_L = \frac{1}{\sqrt{8}}(\ket{000}-\ket{111})(\ket{000}-\ket{111})(\ket{000}-\ket{111}) \\
\end{aligned}
\end{equation}
and the circuit to perform the encoding is shown in Fig.~\ref{fig:9encode}.  
Correction for $X$ errors, for each block of three qubits encoded to $(\ket{000}\pm \ket{111})/\sqrt{2}$ 
is identical to the three qubit code shown earlier.  By performing the correction circuit shown in Fig.~\ref{fig:3qubit2} for each block of three
qubits, single $\sigma_x \equiv X$ errors can be detected and corrected.  Phase errors 
($\sigma_z \equiv Z$) are corrected by examining the sign differences between the 
three blocks.  The circuit shown in Fig.~\ref{fig:9qubit2} achieves this.
The first set of six CNOT gates compares the sign of blocks one and two and the second set of CNOT gates compares the 
sign for blocks two and three.  Note that a phase flip on {\em any} one qubit in a block of three has the same effect, this is why the 9-qubit 
code is referred to as a degenerate code.  In other error correcting codes, such as the 5- or 7-qubit codes~\cite{S96,LMPZ96}, there is a one-to-one 
mapping between correctable errors and unique states.  In degenerate codes such as the 9-qubit code, the mapping is not unique.  Hence 
provided we know in which block the error occurs it does not matter which qubit we apply the correction operator to.  

As the 9-qubit code can correct for a single $X$ error in any one block of three and a single phase error on any of the nine qubits, this 
code is a full quantum error correcting code.  Even if a bit {\em and} phase error occurs on the same qubit, the $X$ correction circuit will detect and correct for 
bit flips, while the $Z$ correction circuit will detect and correct for phase flips.  
As mentioned, the $X$ error correction does 
have the ability to correct for up to three individual bit flips (provided each bit flip occurs in a different block of three).  However, in general, 
the 9-qubit code is only a single error correcting code as it cannot handle multiple errors if they occur in certain locations.

The 9-qubit code is in fact related to a useful 
class of error correcting codes known as Bacon-Shor codes~\cite{B06}.  
These codes have the property that certain subgroups of error operators do not corrupt the logical space.  
For example, In the 9-qubit code, specific pairs of phase errors do not corrupt the logical states.
Bacon-Shor codes are very nice codes from a computer architectural point of view.  Error correction circuits and 
gates are generally simpler, allowing for circuit structures more amenable to the physical restrictions of 
a computer architecture~\cite{AC07}.  Additionally, Bacon-Shor 
codes correcting a larger number of errors have a similar structure.  Therefore,  
are able to perform dynamical switching between codes, in a fault-tolerant manner.  This allows us to adapt the amount of error correction 
to better reflect the noise present at a physical level~\cite{SEDH07}.  We will return and revisit these codes later in section~\ref{sec:subsystem}.

\section{Quantum error detection}
\label{sec:detection}
So far we have focused on the ability not only to detect errors, but also to correct them.  Another approach is to not enforce the correction requirement.  Post-selected quantum computation, developed by Knill~\cite{K05} demonstrated that large scale quantum 
computing could be achieved with much higher noise rates when error detection is employed instead of more costly correction protocols.  
The basic idea in post-selected schemes is to encode a large number of 
ancilla qubits with error detecting circuits.  Sets of encoded qubits which 
pass error detection are selected and further utilized as encoded ancillas for error correction.  
In general, error detection is faster and requires fewer qubits than performing active error correction.  By producing and verifying large numbers of encoded ancillas which are post-selected 
after verification, error correction can be performed without data qubits waiting as long for appropriate ancilla 
to be prepared, decreasing the number of errors that need to be corrected.  One of the downsides 
to these types of schemes is that although they lead to large tolerable error rates, the resource requirements are much higher. 

\begin{figure*}[ht]
\begin{center}
\includegraphics[width=0.8\textwidth]{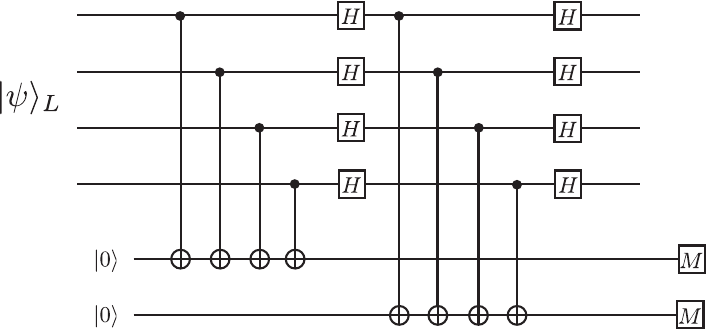}
\caption{Circuit required to detect errors in the 4-qubit error detection code.  If both ancilla measurements return $\ket{0}$, then 
the code state is error free.  If either measurement returns $\ket{1}$, an error has occurred.  Unlike the 9-qubit code, the detection of 
an error does not give sufficient information to correct the state.}
\label{fig:4qubit}
\end{center}
\end{figure*}

The simplest error detecting circuit is the 4-qubit code~\cite{GBP97}.  This encodes two logical qubits on to 
four physical qubits with the ability to detect a single 
error on either of the two logical qubits.  The four basis states for the code are,
\begin{equation}
\begin{aligned}
&\ket{00} = \frac{1}{\sqrt{2}}(\ket{0000}+\ket{1111}), \\ 
&\ket{01} = \frac{1}{\sqrt{2}}(\ket{1100}+\ket{0011}), \\
&\ket{10} = \frac{1}{\sqrt{2}}(\ket{1010}+\ket{0101}), \\
&\ket{11} = \frac{1}{\sqrt{2}}(\ket{0110}+\ket{1001}).
\end{aligned}
\end{equation}
Fig.~\ref{fig:4qubit} illustrates the error detection circuit that can be utilized to detect a single bit and/or phase flip on one of
these encoded qubits.
If a single bit and/or phase flip occurs on one of the four qubits then the ancilla qubits 
will be measured in the state $\ket{1}$.  For example, let us consider the cases when a single bit flip occurs on one of 
the four qubits.  
The state of the system, just prior to the measurement of the ancilla, is given in Table.~\ref{tab:errors2}.
\begin{table}[ht!]
\begin{center}
\vspace*{4pt}   
\begin{tabular}{c|c}
Error Location & Final State, $\ket{\text{data}}\ket{\text{ancilla}}$ \\
\hline
No Error & $\ket{\psi}_L\ket{00}$ \\
Qubit 1 & $X_1\ket{\psi}_L\ket{10}$ \\
Qubit 2 & $X_2\ket{\psi}_L\ket{10}$ \\
Qubit 3 & $X_3\ket{\psi}_L\ket{10}$ \\
Qubit 4 & $X_4\ket{\psi}_L\ket{10}$ \\
\end{tabular}
\caption{Qubit and ancilla state, just prior to measurement for the 4-qubit error detection code when a single bit-flip has occurred on at most one 
of the four qubits.} 
\label{tab:errors2}
\end{center}
\end{table} 
Regardless of the location of the bit flip, the ancilla system is measured in the state $\ket{10}$.  Similarly if one considers a single 
phase error on any of the four qubits the ancilla measurement will return $\ket{01}$.  In both cases no information is obtained regarding {\em where} 
the error has occurred, hence it is not possible to correct the state.  Instead the circuit can 
must reset and re-run. 

\section{Stabiliser formalism}
\label{sec:sec:QEC}
So far we have described error correcting codes from the state vector representation of their encoded states.  
This is a rather inefficient method for describing the codes as the state representations and circuits will differ from code to code.  
Consequently we would like a representation that has a generalised method for error correction and 
circuit construction, regardless of the code used.  The majority of error correcting codes that are 
used within the literature are members of a class known as stabilizer codes.  Stabilizer codes are very 
useful to work with.  The general formalism applies broadly and there exists general rules to 
construct preparation circuits, correction circuits and fault-tolerant logical gate operations once the 
stabilizer structure of the code is known.  

The stabilizer formalism was first introduced by Daniel Gottesman~\cite{G97+} and uses the Heisenberg 
representation for quantum mechanics.   Describing quantum states in terms of operators rather than the 
states.  A state $\ket{\psi}$ is defined to be stabilized by some operator, $K$, if it is a 
$+1$ eigenstate of $K$, i.e.
\begin{equation}
K\ket{\psi} = \ket{\psi}.
\end{equation}
For example, the single qubit state $\ket{0}$ is stabilized by the operator $K = \sigma_z$, i.e.
\begin{equation}
\sigma_z\ket{0} = \ket{0}.
\end{equation}
We can definine multi-qubit states in the same way, by examining some of the group 
properties of multi-qubit operators.  

Within the group of all possible, 
single qubit operators, there exists a subgroup, denoted the Pauli group, 
$\mathcal{P}$, which contains the following elements,
\begin{equation}
\mathcal{P} = \{\pm \sigma_I, \pm i \sigma_I, \pm \sigma_x, 
\pm i \sigma_x,\pm \sigma_y, \pm i \sigma_y,\pm \sigma_z, \pm i \sigma_z\}.
\end{equation}
By utilizing the commutation and anti-commutation 
rules for the Pauli set, $\{\sigma_i \}_{i=x,y,z}$,
\begin{equation}
[\sigma_i,\sigma_j] = 2i\epsilon_{ijk}\sigma_k, \quad \quad \{\sigma_i,\sigma_j\} = 2\delta_{ij},
\end{equation}
where,
\begin{equation}
\epsilon_{ijk} = \Bigg \{
\begin{array}{l}
+1\text{ for } (i,j,k) \in \{(1,2,3), (2,3,1), (3,1,2)\}\\
-1 \text{ for } (i,j,k) \in \{(1,3,2), (3,2,1), (2,1,3)\}\\
0 \text{    for } i=j, j=k, \text{ or } k=i
\end{array}
\end{equation}
and
\begin{equation}
\delta_{ij} = \Bigg \{
\begin{array}{cr}
1\text{ for } i = j\\
0 \text{ for } i \neq j
\end{array}
\end{equation}
it is easy to show that $\mathcal{P}$ forms a 
group under matrix multiplication. 

The Pauli group extends over $N$-qubits by simply taking the $N$ fold tensor product of all elements 
of $\mathcal{P}$, denoted as,
\begin{equation}
\begin{aligned}
\mathcal{P}_N &= \mathcal{P}^{\otimes N}. \\
\end{aligned}
\end{equation}
An $N$-qubit stabilizer state, $\ket{\psi}_N$ is then defined by the $N$ generators of an Abelian (all elements commute)
subgroup, $\mathcal{G}$, of the $N$-qubit Pauli group,
\begin{equation}
\begin{aligned}
\mathcal{G}  = 
&\{\; K^i \;|\; K^i\ket{\psi} = \ket{\psi}, \; [K^i,K^j] = 0, \; \forall \;(i,j)\} \subset \mathcal{P}_N.
\label{eq:stabdef}
\end{aligned}
\end{equation}
The state $\ket{\psi}_N$ can be equivalently defined either through the 
state vector representation {\em or} by specifying the generators of the 
stabilizer group, $\mathcal{G}$. As each 
stabilized state, $\ket{\psi}$, satisfies, $K\ket{\psi} = \ket{\psi}$, and each stabilizer is 
Hermitian, each stabilizer squares to the 
identity, $K\cdot K = I$.

Many extremely useful multi-qubit states are stabilizer states, including two-qubit Bell states, 
Greenberger-Horne-Zeilinger (GHZ) states~\cite{GHZ89,GHSZ90}, Cluster states~\cite{BR01,RB01} 
and codeword states for QEC.  As an example, consider a three qubit GHZ state, 
\begin{equation}
\ket{\text{GHZ}}_3 = \frac{\ket{000} + \ket{111}}{\sqrt{2}}.
\end{equation}
This state can be expressed via any three linearly independent generators 
of the $\ket{\text{GHZ}}_3$ stabilizer group,
\begin{equation}
\begin{aligned}
K^1 &= \sigma_x\otimes \sigma_x \otimes \sigma_x \equiv XXX, \\
K^2 &= \sigma_z\otimes \sigma_z \otimes \sigma_I \equiv ZZI, \\
K^3 &= \sigma_I \otimes \sigma_z \otimes \sigma_z \equiv IZZ, 
\end{aligned}
\end{equation}
where the right-hand side of each equation is the short-hand representation of stabilizers.  
Similarly, the four orthogonal Bell states,
\begin{equation}
\begin{aligned}
\ket{\Phi^{\pm}} &= \frac{\ket{00} \pm \ket{11}}{\sqrt{2}}, \\
\ket{\Psi^{\pm}} &= \frac{\ket{01} \pm \ket{10}}{\sqrt{2}},
\end{aligned}
\end{equation}
are stabilized by the operators, $K^1 = (-1)^aXX$, and $K^2 = (-1)^b ZZ$, where $[a,b] \in \{0,1\}$.  Each of the 
four Bell states correspond to the four $\pm 1$ eigenstate combinations of these two operators, 
\begin{equation}
\begin{aligned}
&\Phi^+ \equiv  \begin{pmatrix} K^1 = XX \\ K^2 = ZZ \end{pmatrix} \quad \Phi^- \equiv  \begin{pmatrix} K^1 = -XX \\ K^2 = ZZ \end{pmatrix} \\
&\Psi^+ \equiv  \begin{pmatrix} K^1 = XX \\ K^2 = -ZZ \end{pmatrix} \quad \Psi^- \equiv  \begin{pmatrix} K^1 = -XX \\ K^2 = -ZZ \end{pmatrix}
\end{aligned}
\end{equation}

\section{Quantum error correction with stabiliser codes}\label{sec:sec:QEC2}
The use of the stabilizer formalism to describe quantum error correction codes is extremely useful since it 
allows easy synthesis of correction circuits and also allows for quick determination of what logical 
operations can be applied directly on encoded data.  
The link between stabilizer codes and stabilizer states comes about by defining a relevant coding {\em subspace}
within the larger Hilbert space of a multi-qubit system.  

To illustrate this reduction let us examine a simple two qubit example. 
A 2-qubit system has a Hilbert space dimension of 
four, however if we require that this two qubit state is stabilized by the $XX$ operator, then there are only two orthogonal basis states 
which satisfies this condition,
\begin{equation}
\ket{0}_L \equiv \frac{1}{\sqrt{2}} \left( \ket{01}+\ket{10}\right), \quad \ket{1}_L \equiv \frac{1}{\sqrt{2}}\left( \ket{00}+\ket{11}\right),
\end{equation}
which can be used to define a effective logical 
qubit.  Hence by using stabilizers we can reduce the size of the Hilbert space 
for a multi-qubit system to an effective single qubit system.  In the context of QEC, the stabilizers that are used to define the logical 
subspace are utilized to detect and correct for errors.  A stabilizer code is therefore a subspace defined via stabilizer 
operators for a multi-qubit system.

Returning to QEC, the example we focus on is
the most well known quantum code; the 7-qubit Steane code first proposed in 1996~\cite{S96}.  
The 7-qubit code is 
defined as a $[[n,k,d]] = [[7,1,3]]$ 
quantum code, where $n=7$ physical qubits encode $k=1$ logical qubit with a distance between basis states $d=3$, 
correcting $t = \lfloor(d-1)/2\rfloor=1$ error.  
As the code contains a single logical qubit, the code must contain two valid code states:  
$\ket{0}_L$ and $\ket{1}_L$ basis states which are, in state vector notation,
\begin{widetext}
\begin{equation}
\begin{aligned}
|0\rangle_L = \frac{1}{\sqrt{8}}(&|0000000\rangle + |1010101\rangle + |0110011\rangle + |1100110\rangle + 
|0001111\rangle + |1011010\rangle + |0111100\rangle + |1101001\rangle),\\
|1\rangle_L = \frac{1}{\sqrt{8}}(&|1111111\rangle + |0101010\rangle + |1001100\rangle + |0011001\rangle + 
|1110000\rangle + |0100101\rangle + |1000011\rangle + |0010110\rangle).
\label{eq:log}
\end{aligned}
\end{equation}
\end{widetext}
For a general 7-qubit state, the total dimension of the Hilbert space is $2^7$.  For a single logically encoded qubit, we 
must restrict this to a $2$-dimensional subspace spanned by the 
states in Eq.~(\ref{eq:log}).  This reduction can be clearly seen via stabilizers.  

The stabilizer set for the 7-qubit code is fully specified by the six operators,
\begin{equation}
\begin{aligned}
&K^1 = IIIXXXX, \quad \quad K^2 = XIXIXIX,\\ 
&K^3 = IXXIIXX, \quad \quad K^4 = IIIZZZZ \\
&K^5 = ZIZIZIZ, \quad \quad K^6 = IZZIIZZ.
\end{aligned}
\label{eq:stab7}
\end{equation}
Each of the valid codestates for the 7-qubit code are stabilized by these operators.  As the dimensionality of a 7-qubit system is 
$2^7$ and there are six stabilizers, the total dimension of this subspace defined 
by the stabilizer group is $2^{7-6} = 2$.  The stabilizer set now defines an effective 2-dimensional subspace, 
and thus a single encoded qubit\footnote{In general, for an $n$ qubit system, the total dimensionality of 
the Hilbert space is $2^n$, if a stabilizer set is defined over this system containing $k$ multiplicativity independent generators, then 
the dimension of the subspace is $2^{n-k}$ and therefore the stabilizer set defines a subspace containing 
$n-k$ logical qubits.}.

The final operator fixes the encoded state to one of the two codewords.  For the Steane code this operator is 
$\bar{Z} = ZZZZZZZ=Z^{\otimes 7}$, where
$\bar{Z}\ket{0}_L = \ket{0}_L$ and $\bar{Z}\ket{1}_L = -\ket{1}_L$.   Notice that these generators for the stabilizer set separate into elements consisting solely of $X$ or $Z$ operators.  This defines the 
code as a Calderbank-Shor-Steane (CSS) code.  CSS codes are useful since they allow for the straightforward 
application of several logical gate operations directly to the encoded data  [Section~\ref{sec:operations}].

Although the 7-qubit code is the most well-known stabilizer code, there are other stabilizer codes 
which encode multiple logical qubits and correct for more errors~\cite{G97+}.  
Despite these advantages, larger codes  
generally require more complicated error correction circuits.  As the complexity of 
the error correction circuits increase, adapting such codes to physical computer architectures becomes more 
difficult.  
Tables~\ref{tab:9qubit} 
and~\ref{tab:5qubit} show the stabilizer structure of two other well-known codes.  The 9-qubit 
code~\cite{S95} examined earlier, and the 5-qubit code~\cite{LMPZ96} which 
represents the smallest possible quantum code that corrects for a single error.  
\begin{table}[ht]
\begin{center}
\vspace*{4pt}   
\begin{tabular}{c|c|c|c|c|c|c|c|c|c}
$K^1$ & $Z$&$Z$&$I$&$I$&$I$&$I$&$I$&$I$&$I$ \\
$K^2$ & $Z$&$I$&$Z$&$I$&$I$&$I$&$I$&$I$&$I$ \\
$K^3$ & $I$&$I$&$I$&$Z$&$Z$&$I$&$I$&$I$&$I$ \\
$K^4$ & $I$&$I$&$I$&$Z$&$I$&$Z$&$I$&$I$&$I$ \\
$K^5$ & $I$&$I$&$I$&$I$&$I$&$I$&$Z$&$Z$&$I$ \\
$K^6$ & $I$&$I$&$I$&$I$&$I$&$I$&$Z$&$I$&$Z$ \\
$K^7$ & $X$&$X$&$X$&$X$&$X$&$X$&$I$&$I$&$I$ \\
$K^8$ & $X$&$X$&$X$&$I$&$I$&$I$&$X$&$X$&$X$ \\
\end{tabular}
\caption{The eight stabilizers for the 9-qubit Shor code, encoding nine physical qubits into one logical 
qubit to correct for a single $X$ and/or $Z$ error. } 
\label{tab:9qubit}
\end{center}
\end{table} 
\begin{table}[ht]
\begin{center}
\vspace*{4pt}   
\begin{tabular}{c|c|c|c|c|c}
$K^1$ & $X$&$Z$&$Z$&$X$&$I$ \\
$K^2$ & $I$&$X$&$Z$&$Z$&$X$ \\
$K^3$ & $X$&$I$&$X$&$Z$&$Z$ \\
$K^4$ & $Z$&$X$&$I$&$X$&$Z$ \\
\end{tabular}
\caption{The four stabilizers for the [[5,1,3]] quantum code, encoding five physical qubits into one logical qubit
to correct for a single $X$, $Y$ or $Z$ error.  Unlike the 7- and 9-qubit codes, the [[5,1,3]] code is a non-CSS 
code, since the stabilizer set does not separate into $X$ and $Z$ sectors.  Additionally, this code can only correct 
a single error where as the 7- and 9-qubit codes can correct for a maximum of 2 errors (a single $X$ and $Z$ error 
provided they occur on different qubits)} 
\label{tab:5qubit}
\end{center}
\end{table} 

\subsection{State preparation}
Using the stabilizer structure for QEC codes, the logical state preparation and error correcting procedure is 
straightforward.  
Recall that valid codeword states are defined as simultaneous $+1$ eigenstates of each generator 
of the stabilizer group.  In order to prepare 
a logical state from an arbitrary input, we need to project qubits into 
eigenstates of each of these operators.

\begin{figure}[ht]
\begin{center}
\includegraphics[width=0.45\textwidth]{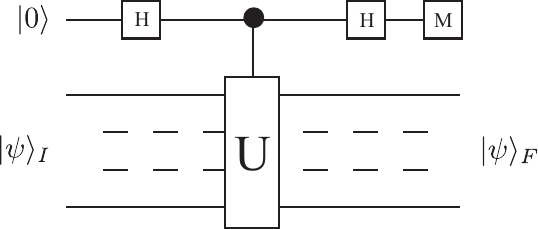}
\caption{Quantum Circuit required to project an arbitrary state, $\ket{\psi}_I$ into a $\pm 1$ eigenstate 
of the Hermitian operator, $U = U^{\dagger}$.  
The measurement result of the ancilla determines which eigenstate $\ket{\psi}_I$ is projected to and since the operator, $U$ 
is Hermitian and unitary can only have the two eigenvalues $\pm 1$.}
\label{fig:opmeas}
\end{center}
\end{figure}

Consider the circuit shown in Fig.~\ref{fig:opmeas}.
For an arbitrary input state, $\ket{\psi}_I$, an ancilla which is initialized in the $\ket{0}$ state is used as a control qubit 
for a unitary and Hermitian operation ($U^{\dagger} = U$, $U^2 = I$) on $\ket{\psi}_I$.  
After the second Hadamard gate is performed, the state of the system is,
\begin{equation}
\ket{\psi}_F = \frac{1}{2} ( \ket{\psi}_I + U\ket{\psi}_I)\ket{0} + \frac{1}{2}(\ket{\psi}_I - U\ket{\psi}_I)\ket{1}.
\end{equation}
The ancilla qubit is then measured in the computational basis.  If the result is $\ket{0}$, the input state is projected to 
(neglecting normalisation),
\begin{equation}
\ket{\psi}_F = \ket{\psi}_I+U\ket{\psi}_I.
\end{equation}
Since $U$ is unitary and Hermitian, $U\ket{\psi}_F=\ket{\psi}_F$, hence $\ket{\psi}_F$ is a $+1$ eigenstate of $U$.
If the ancilla is measured to be $\ket{1}$, then the input is projected to the state,
\begin{equation}
\ket{\psi}_F = \ket{\psi}_I-U\ket{\psi}_I,
\end{equation}
which is the $-1$ eigenstate of $U$.  Therefore, provided $U$ is Hermitian, the general circuit of 
Fig.~\ref{fig:opmeas} 
will project an arbitrary input state to a $\pm 1$ eigenstate of $U$.\footnote{An operator, $U$, which is both Hermitian and unitary can only have eigenvalues of $\pm1$}  This procedure is well known and is 
referred to as either a ``parity" or ``operator" measurement~\cite{NC00}.  

\begin{figure}[ht]
\begin{center}
\includegraphics[width=0.45\textwidth]{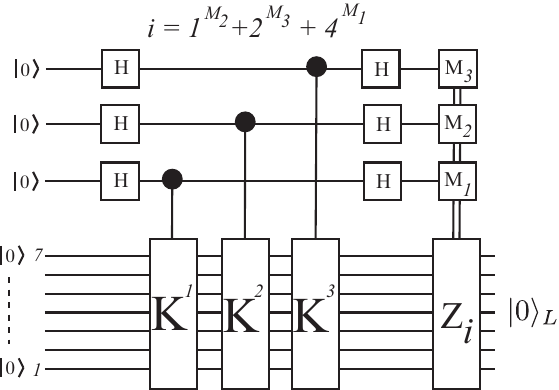}
\caption{Quantum circuit to prepare the $[[7,1,3]]$ logical $\ket{0}$ state.  The input state $\ket{0}^{\otimes 7}$ is 
projected into an eigenstate of each of the $X$ stabilizers shown in Eq. (\ref{eq:stab7}).  After each ancilla measurement 
the classical results are used to apply a single qubit $Z$ gate to qubit $i = 1^{M_2}+2^{M_3}+4^{M_1}$ which 
converts the state from a $-1$ eigenstates of $(K^1,K^2,K^3)$ to $+1$ eigenstates.}
\label{fig:7qubitprep}
\end{center}
\end{figure}

From this construction it should be clear how QEC state preparation proceeds.  Taking the $[[7,1,3]]$ 
code as an example (the following technique is not unique or optimal for 
preparing and correcting the 7-qubit code \cite{S97,S02} but once this idea is understood, readers 
should have little difficultly understanding other correction techniques), 7-qubits are first initialized in the 
state $\ket{0}^{\otimes 7}$.  The circuit shown in Fig.~\ref{fig:opmeas} is applied three times with 
$U = K^1,K^2,K^3$, projecting the input state into a simultaneous $\pm 1$ eigenstate of each $X$ generator of the stabilizer group
for the $[[7,1,3]]$ code.  The result of each operator measurement is then used to classically control 
a single qubit $Z$ gate which is applied to one of the seven qubits at the end of the preparation.  This single $Z$ 
gate converts any $-1$ projected eigenstates into $+1$ eigenstates.  
Notice that the final three stabilizers 
do not need to be measured due to the input state, $\ket{0}^{\otimes 7}$, already being a $+1$ eigenstate 
of $K^4,K^5$ and $K^6$.  Additionally as the state $\ket{0}^{\otimes 7}$ is also a $+1$ eigenstate of $K_7$, the initial state will 
be $\ket{0}_L$ state.  Fig.~\ref{fig:7qubitprep} illustrates the final circuit, where instead 
of one ancilla, three are utilized to speed up the state preparation by 
performing each operator measurement in parallel.  

As a quick aside, let us detail exactly how the relevant logical basis states can be derived from the 
stabilizer structure of the code by utilizing this preparation procedure.  
We will use the stabilizer set shown in Table~\ref{tab:5qubit} 
to calculate the $\ket{0}_L$ state for the 5-qubit code as we have not yet explicitly shown the state vectors 
for the two logical state.  The four stabilizer generators are given by,
\begin{equation}
\begin{aligned}
&K^1 = XZZXI, \quad \quad K^2 = IXZZX,\\
&K^3 = XIXZZ, \quad \quad K^4 = ZXIXZ.
\end{aligned}
\end{equation}
In an identical way to the 7-qubit code, projecting an arbitrary state into a $+1$ eigenstate of these operators define 
the two logical basis states, $\ket{0}_L$ and $\ket{1}_L$.   The logical operator, $\bar{Z} = ZZZZZ$, then fixes the 
state to either $\ket{0}_L$ or $\ket{1}_L$.  Therefore, calculating $\ket{0}_L$ from some initial un-encoded state 
requires us to project the initial state into a $+1$ eigenstate of these operators.  If we take the initial, un-encoded 
state as $\ket{00000}$, then it is already a $+1$ eigenstate of $\bar{Z}$.  Therefore, to find $\ket{0}_L$ we 
simply calculate,
\begin{equation}
\begin{aligned}
\ket{0}_L&=\prod_{i=1}^4 (I^{\otimes 5} + K^i)\ket{00000},
\end{aligned}
\end{equation} 
up to normalization.  Expanding out this product, we find,
\begin{equation}
\begin{aligned}
\ket{0}_L = \frac{1}{4}( &\ket{00000}+\ket{01010}+\ket{10100}-\ket{11110}+\\
&\ket{01001}-\ket{00011}-\ket{11101}-\ket{10111}+\\
&\ket{10010}-\ket{11000}-\ket{00110}-\ket{01100}-\\
&\ket{11011}-\ket{10001}-\ket{01111}+\ket{00101}).
\end{aligned}
\end{equation}
Note, that the above state vector does not match up with those given in~\cite{LMPZ96}\footnote{The stabilizer formalism was introduced after the 5-qubit code.}.  However, these vectors are 
equivalent up to local rotations on each qubit.  

\subsection{Error correction}
Error correction using stabilizer codes is an extension of the state preparation.  Consider an arbitrary single 
qubit state that has been encoded as, 
\begin{equation}
\alpha\ket{0} + \beta\ket{1} \rightarrow \alpha\ket{0}_L + \beta\ket{1}_L = \ket{\psi}_L.  
\end{equation}
Now assume that an error occurs on one (or multiple) qubits which is described via the operator $E$, where $E$ is 
a combination of $X$ and/or $Z$ errors over the $N$ physical qubits of the logical state (and therefore 
an element of the $N$-qubit Pauli group, $\mathcal{P}_N$).  
By definition of stabilizer 
codes, $K^i\ket{\psi}_L = \ket{\psi}_L$, $i \in [1,..,N-k]$, for a code encoding $k$ logical qubits.  
Hence the erred state, 
$E\ket{\psi}_L$, satisfies,
\begin{equation}
K^iE\ket{\psi}_L = (-1)^m EK^i\ket{\psi}_L = (-1)^m E\ket{\psi}_L.
\end{equation}
where $m$ is defined as $m=0$, if $[E,K^i]=0$ and $m=1$, if $\{E,K^i\} = 0$ ($E$ and $K$ are Pauli group operators 
and all Pauli group operators either commute or anti-commute).  Therefore, if the error operator commutes 
with the stabilizer, the state remains a $+1$ eigenstate of $K^i$, if the error operator anti-commutes with the 
stabilizer then the logical state is flipped to a $-1$ eigenstate of $K^i$.  

The general procedure for error correction is identical to state preparation.  Each of the code 
stabilizers is sequentially measured.  Since a error free state is already a $+1$ eigenstate of all the 
stabilizers, errors which anti-commute with any of the stabilizers describing the code will flip the relevant 
eigenstate and consequently measuring the parity of these stabilizers will return a result of $\ket{1}$.
Taking the $[[7,1,3]]$ code as an example, if the error operator is 
$E = X_i$ (where $i = 1,...,7$), 
representing a bit-flip on any {\em one} of the 7 physical qubits, then regardless of the location, $E$ will 
anti-commute with a unique combination of $K^4,K^5$ and $K^6$.  Hence the classical results of measuring these 
three operators will indicate if and where a single $X$ error has occurred.  Similarly, if $E=Z_i$, then the 
error operator will anti-commute with a unique combination of, $K^1,K^2$ and $K^3$.  Consequently, the first three 
stabilizers for the $[[7,1,3]]$ code correspond to $Z$ error correction 
while the second three stabilizers correspond to $X$ error correction.  Note that correction for Pauli $Y$ errors 
is achieved by correcting in the $X$ and $Z$ sector since a $Y$ error on a single qubit is 
equivalent to both an $X$ and $Z$ error on the same qubit, i.e. $Y = iXZ$.  

\begin{figure*}[ht]
\begin{center}
\includegraphics[width=\textwidth]{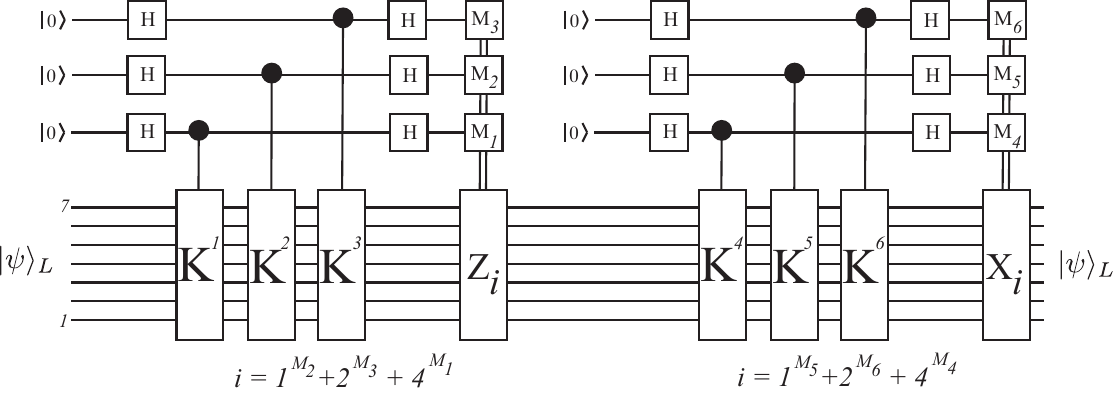}
\caption{Quantum circuit to to correct for a single $X$ and/or $Z$ error using the $[[7,1,3]]$ code.  Each of the 
six stabilizers are measured, with the first three detecting and correcting for $Z$ errors, while the last three 
detect and correct for $X$ errors.}
\label{fig:correct}
\end{center}
\end{figure*}

The circuit shown in Fig.~\ref{fig:correct} illustrates 
the circuit for full error correction with the $[[7,1,3]]$ code.  As you can see it is simply an extension of the 
preparation circuit shown in Fig.~\ref{fig:7qubitprep}, where all six stabilizers are measured across the data block.  
Even though we have specifically used the $[[7,1,3]]$ code as an example, this procedure for error correction 
and state preparation will work for all stabilizer codes (although other correction procedures 
are possible~\cite{S97,S02,K05}).

\section{Digitization of quantum errors}\label{sec:sec:decoherence}
Up until now we have remained fairly abstract regarding the analysis of quantum errors.  Specifically, 
we have examined QEC from the standpoint of a discrete set of Pauli errors occurring at certain locations 
within a larger quantum circuit.  In this section we examine how this analysis of errors relates to 
some of the 
more realistic processes such as environmental decoherence and systematic gate errors. 

Digitization of quantum noise is often assumed when people examine the stability of quantum circuit 
design or attempt to calculate thresholds for concatenated error correction [Section. \ref{sec:threshold}].  
However, the 
link between discrete Pauli errors to more general, continuous noise only makes sense when 
we consider the stabilizer nature of the correction procedure.  Recall from section~\ref{sec:sec:QEC} that 
correction is performed by re-projecting a potentially corrupt data block into $+1$ eigenstates of the code
stabilizers.  A general continuous 
mapping from a ``clean" codeword state to a corrupt one will not be eigenstates of the code stabilizers.   
Instead they will be in superpositions of eigenstates.  
Measuring the parity of each of the stabilizers acts to digitize the quantum noise.   
We will first introduce how a coherent systematic error, caused by imperfect implementation of quantum 
gates, are digitized during correction, after which we will briefly discuss environmental decoherence 
from the standpoint of a Markovian decoherence model.  

\subsection{Systematic gate errors} 
We have already shown a primitive example of how systematic gate errors are digitized into a discrete set of 
Pauli operators in Sec.~\ref{sec:error}.  However, in that case we only considered a very restrictive type of 
error, namely the coherent operator $U=\exp(i\epsilon X)$.  We can easily extend this analysis to 
cover all types of systematic gate errors.
Consider an $N$ qubit unitary operation, $U_N$, which is valid on encoded data.  Assume that $U_N$ is 
applied inaccurately such that the resultant operation is actually $\mathcal{U}_N$.  Given a general 
encoded state $\ket{\psi}_N$, the final state can be expressed as,
\begin{equation}
\mathcal{U}_N \ket{\psi}_L = U_E U_N \ket{\psi}_L = \sum_j \alpha_j E_j \ket{\psi_2}_L,
\end{equation}
where $\ket{\psi_2}_L = U_N\ket{\psi}_L$ is the perfectly applied $N$ 
qubit gate, (the stabilizer group for $\ket{\psi'}_L$ 
remains invariant under the operation $U_N$ [see Sec.~\ref{sec:operations}]) and 
$U_E$ is a coherent error operator which is expanded in terms of the $N$ qubit Pauli Group, 
$E_j \in P_N$.  Now append two ancilla blocks, $\ket{A_0}^X$ and $\ket{A_0}^Z$.  These 
are generally initialized to the state that corresponds to no detected errors (but can be 
other initial states up to a redefinition of the measurement results)
and are used for $X$ and $Z$ correction.  We then run the 
syndrome extraction procedure, which we represent by the unitary operator, $U_{\text{QEC}}$. 
It will be assumed that $\ket{\psi}_L$ is encoded with a 
QEC code which can correct for a single error (both $X$ and/or $Z$), and the error operators $E_j$ 
are a maximum of weight one (i.e. $E_j$ contains at most one non-identity term e.g. 
$E_1 = X_1\otimes I^{\otimes (N-1)}$) \footnote{This assumption is for demonstration purposes.  In reality, all qubits will 
experience errors and hence $E_j$ can be of higher weight (up to a weight $N$ operator on an $N$ qubit system).  The ability 
of the error correction code to correct for higher weight errors depend on how all these $E_j$ map the ancilla states under 
$U_{\text{QEC}}.$},  
hence there is a one-to-one mapping between the error operators, $E_j$, and the orthogonal basis states of the ancilla blocks,
\begin{equation}
\begin{aligned}
&U_{\text{QEC}}U_N'\ket{\psi}_L\ket{A_0}^X\ket{A_0}^Z \\
&= U_{\text{QEC}}
\sum_j \alpha_jE_j\ket{\psi'}_L \ket{A_0}^X\ket{A_0}^Z \\
&= \sum_j \alpha_j E_j \ket{\psi'}_L\ket{A_j}^X\ket{A_j}^Z. 
\label{eq:ancilla}
\end{aligned}
\end{equation}
The above assumes a non-degenerate quantum code\footnote{A degenerate quantum code is one where multiple unique errors can 
map to the same state, in the case of Eq. (\ref{eq:ancilla}) this would mean two operators $E_j$ and $E_j'$, under the unitary 
$U_{\text{QEC}}$ map 
$\ket{A_0}^X$ and $\ket{A_0}^Z$ to the 
same ancilla state}.  The ancilla blocks are then measured, projecting the data blocks into the state $E_j\ket{\psi'}_L$ with 
probability $|\alpha_j|^2$.  After measurement the correction $E_j^{\dagger}$ is applied based on the syndrome
result.  As the error operation $E_j$ is simply an element of $\mathcal{P}_N$, correcting for $X$ and $Z$ 
independently is sufficient to correct for all error operators (as $Y$ errors are corrected when a bit and phase 
error is detected and corrected on the same qubit).

For well designed gates, very small systematic inaccuracies lead to the expansion co-efficient 
$\alpha_0 \approx 1$, with all other coefficients, $\alpha_{j\neq 0} \ll 1$.  Hence during correction 
there will be a very high probability that no error is detected.  This is the digitization effect of QEC.  
Since codeword states are eigenstates of the stabilizers, re-projecting the state 
when each stabilizer is measured forces any continuous noise operator to collapse.  
The strength of the error is then related to the probability that the data block 
collabses to a specific Pauli error acting on the codestate.
The physical mechanisms which are used to construct quantum gates is what determines the 
form of $U_E$ and hence the form of the error operators, $E_j$ and their respective coefficients, 
$\alpha_j$.  
\subsection{Environmental decoherence}
A complete analysis of environmental decoherence in relation to quantum information is a lengthy topic.  Instead 
of a detailed review, we will instead present a simplified example to highlight how QEC 
relates to environmental effects.

The Lindblad formalism~\cite{G91,NC00,DWM03} 
provides an elegant method for analyzing the effect of decoherence on open
quantum systems.  This model does have several assumptions, most notably that the environmental 
bath couples weakly to the system (Born approximation), the system and environment are initially in some separable state 
and that each qubit experiences 
temporally un-correlated noise (Markovian approximation).  While these assumptions are utilized for a variety of 
systems~\cite{BHPC03,BM03,BKD04}, it is known that they may not 
hold in some cases~\cite{HMCS00,MCMS05,APNYT04,ALKH02}.  This is particularly important in superconducting systems, 
where decoherence can be caused by small numbers of fluctuating charges which induces coloured noise therefore violating 
the assumption of non-markovian dynamics.  In this case 
more specific decoherence models need to be considered.  

Using this 
formalism, the evolution of the density matrix can be written as,
\begin{equation}
\partial_t \rho = -\frac{i}{\hbar} [H,\rho] + \sum_k \Gamma_k \mathcal{L}_k[\rho],
\end{equation}
where $H$ is the Hamiltonian, representing coherent, dynamical evolution of the system and 
$\mathcal{L}_k[\rho]=([L_k,\rho L_k^{\dagger}]+[L_k\rho, L_k^{\dagger}])/2$ represents the 
incoherent evolution.  The operators $L_k$ are known as the Lindblad quantum jump operators and 
are used to model specific decoherence channels, with each operator associated with some rate 
$\Gamma_k \geq 0$.  This differential equation is known as the quantum Louiville equation or more 
generally, the density matrix master equation.   

To link Markovian decoherence to QEC, consider a special set of decoherence 
channels which represent a single qubit undergoing dephasing, spontaneous emission and spontaneous 
absorption.  This helps to simplify the 
calculation.  Dephasing of a single qubit is modeled by the Lindblad operator $L_1 = Z$ while spontaneous 
emission/absorption are modeled by the operators $L_2 = \ket{0}\bra{1}$ and $L_3 = \ket{1}\bra{0}$ 
respectively.  For the sake of simplicity, we assume that absorption/emission occur at the same rate, $\Gamma$.
Although this is unphysical, it does simplify the calculation significantly for this example.  The density matrix evolution is given by,
\begin{equation}
\partial_t \rho = -\frac{i}{\hbar}[H,\rho] + \Gamma_Z (Z\rho Z - \rho) + \frac{\Gamma}{2}(X\rho X + Y\rho Y -2\rho).
\label{eq:diff}
\end{equation}
If we assumed that the qubit is not undergoing any coherent evolution ($H = 0$), 
i.e. a memory stage within a quantum algorithm, 
then Eq.~(\ref{eq:diff}) can be solved by re-expressing the density matrix in the Bloch formalism.  
Set $\rho(t) = I/2 + x(t)X + y(t)Y + z(t)Z$, then Eq. (\ref{eq:diff}) with $H=0$ reduces to $\partial_t S(t) = AS(t)$ with
$S(t) = (x(t),y(t),z(t))^T$ and
\begin{equation} 
A = \begin{pmatrix} -(\Gamma + 2\Gamma_z) & 0 &0 \\ 0 &-(\Gamma + 2\Gamma_z) &0\\ 0 &0 &-2\Gamma \end{pmatrix}.
\end{equation}
This differential equation is easy to solve, leading to,
\begin{equation}
\begin{aligned}
\rho(t) &= [1-p(t)]\rho(0) + p_x(t) X\rho(0) X \\
&+ p_y(t) Y \rho(0) Y + p_z(t) Z \rho(0) Z,
\end{aligned}
\end{equation}
where,
\begin{equation}
\begin{aligned}
p_x(t) = &p_y(t) = \frac{1}{4}(1-e^{-2\Gamma t}), \\
&p_z(t) = \frac{1}{4}(1+e^{-2\Gamma t}-2e^{-(\Gamma +2\Gamma_z)t}), \\
&p(t) = p_x(t) + p_y(t) + p_z(t).
\end{aligned}
\end{equation} 
If this single qubit is part of an encoded data block, then each term represents a single 
error on the qubit experiencing decoherence.  Two blocks of ancilla qubits, 
initialized to the state corresponding to no detected error, are added 
to the system.  The error correction protocol is then run.  
Once the ancilla qubits are measured, the state will collapse to 
no error with probability $1-p(t)$, or a single $X$,$Y$ or $Z$ error,
with probabilities $p_x(t),p_y(t)$ and $p_z(t)$ respectively.   
Although the above example is somewhat artificial, the above should allow the reader to perform their own calculations for expected QEC error 
rates given a more physical complete error models. 

We can also see how temporal effects are incorporated into the error correction model.  The temporal integration window $t$ 
of the master equation will influence how probable an error is detected for a fixed rate $\Gamma$.  The longer 
between correction cycles, the more probable the qubit experiences an error. 
 
\subsection{More general mappings}
Both the systematic gate errors and the errors induced by environmental decoherence illustrate the digitization 
effect of QEC.  However, we can generalize digitization to other mappings of 
the density matrix.  In this case consider a more general Krauss map on a multi-qubit density matrix,
\begin{equation}
\rho \rightarrow \sum_k A_k^{\dagger}\rho A_k,
\end{equation}
where $\sum A_k^{\dagger}A_k = I$.  For the sake of simplicity let us choose a simple mapping where 
$A_1 = (Z_1+iZ_2)/\sqrt{2}$ and $A_k = 0$ for $k\neq 1$.  This mapping essentially represents dephasing on two qubits.  
However, this type of mapping (when considered in the context of error correction) 
represents independent $Z$ errors on either qubit one or two.  

The above discussion assumes that we are 
working with a non-degenerate quantum 
code.  If the code is degenerate, i.e. if the errors $Z_1$ and $Z_2$ have the same effect 
on the codestates then the correction protocol does not simplify this error mapping as described above.  

To illustrate, first expand out the density matrix (neglecting normalization),
\begin{equation}
\rho \rightarrow A_1^{\dagger}\rho A_1 = Z_1\rho Z_1 + Z_2\rho Z_2 - iZ_1\rho Z_2 + iZ_2 \rho Z_1.
\end{equation}
Note that only the first two terms in this expansion, on their own, represent physical mixtures.  
However, the last two off-diagonal terms are removed 
by the process of syndrome extraction and are irrelevant in the context of QEC.
To illustrate, we assume that $\rho$ represents a protected qubit, where $Z_1$ and $Z_2$ 
are {\em physical} errors on qubits comprising the code block.  As we are only considering phase 
errors in this example, we will ignore $X$ correction (but the analysis automatically generalizes if the error 
mapping contains $X$ terms).  A fresh ancilla block, represented by the density matrix $\rho^z_0$ is coupled 
to the system and the unitary $U_{QEC}$ is run,
\begin{equation}
\begin{aligned}
U_{QEC}^{\dagger}\rho'\otimes \rho^z_0 U_{QEC}  &= Z_1\rho Z_1 \otimes \ket{Z_1}\bra{Z_1} \\
&+Z_2\rho Z_2 \otimes \ket{Z_2}\bra{Z_2} \\
&-iZ_1\rho Z_2 \otimes \ket{Z_1}\bra{Z_2} \\
&+iZ_2\rho Z_1 \ket{Z_2}\bra{Z_1},
\end{aligned}
 \end{equation}
where $\ket{Z_1}$ and $\ket{Z_2}$ represent the two orthogonal syndrome states of the ancilla that 
are used to detect phase errors on qubits one and two respectively.   The important part of the above expression is 
that when the syndrome qubits are measured, the state collapses to,
\begin{equation}
\begin{aligned}
&\rho \rightarrow \frac{\bra{Z_1}\rho\ket{Z_1}\ket{Z_1}\bra{Z_1}}{\text{Tr}(\rho \ket{Z_1}\bra{Z_1})} \\ 
\text{or   }  &\rho \rightarrow \frac{\bra{Z_2}\rho\ket{Z_2}\ket{Z_2}\bra{Z_2}}{\text{Tr}(\rho \ket{Z_2}\bra{Z_2})}.
\end{aligned}
\end{equation}
The two cross terms in the above expression are 
never observed.  In this mapping the only two possible states that exist after the measurement of the 
ancilla system are,
\begin{equation}
\begin{aligned}
Z_1\rho Z_1 \otimes \ket{Z_1}\bra{Z_1} \quad \text{with Probability } =\frac{1}{2}, \\
Z_2\rho Z_2 \otimes \ket{Z_2}\bra{Z_2} \quad \text{with Probability } =\frac{1}{2}. 
\end{aligned}
\end{equation}
Therefore, not only are the cross terms eliminated via error correction but the final density matrix again 
collapses to a single error perturbation of ``clean" codeword states with no correlated errors.  

Consequently, it is common in standard QEC analysis to assume that discrete $X$ and/or $Z$ errors are applied 
stochastically to each qubit after each elementary gate operation, measurement,
 initialization and memory step with some probability $p$.  The QEC protocol itself digitizes 
 the continuous noise, either coherent or incoherent errors, into a discrete set of bit and/or phase flips.  The set of discrete errors is determined by the set of errors that are distinguishable by the quantum code used.  The magnitude of the 
 continuous error is translated to the probability of detecting a discrete error.  In this way error correction can be analyzed by 
 assuming perfect gate operations and discrete, probabilistic errors.  The probability of these errors occurring 
 can then be independently calculated via analysis of the physical 
 mechanisms which produce errors.  While a local, stochastic error model is generally the most common, depending on the 
 physical system under consideration and the type of quantum codes being used a 
 more complicated analysis may be needed~\cite{P98}.

\section{Fault-tolerant quantum error correction and the threshold theorem}\label{sec:Fault-tolerance}
Section~\ref{sec:sec:QEC} detailed the protocols required to correct for quantum errors, however 
in the discussion so far, we have implicitly assumed the following,
\begin{enumerate}
\item Errors only occur during ``memory" regions, i.e. when quantum operations or 
error correction are not being performed and ancilla qubits are error free.
\item The quantum gates themselves do not induce any systematic errors within the 
logical data block.
\end{enumerate}
Clearly these are two very unrealistic assumptions.  Fault-tolerant QEC is how we address these two issues.  
As the name suggests, fault-tolerance is a design methodology that allows us to tolerate faults, allowing 
QEC to remain effective.  Combining error correction techniques with fault-tolerance 
allows us to design correction procedures and 
logical gate operations such that they can still function 
when the above assumptions are relaxed.  

\subsection{Error propagation}

Before discussing the nature of fault-tolerant computation, we first examine how errors can propagate in quantum circuits.  
There are essentially two dominant channels that can cause errors to be copied, the first is quantum gates. For obvious reasons, 
gates operating on single qubits do not copy errors, however quantum gates which couple qubits can cause errors to propagate.  

For example, take the two qubit state $E_j\ket{\psi}$, where $E_j  = \{X_1I_2,Y_1I_2,Z_1I_2,I_1X_2,I_1Y_2,I_1Z_2\}$ 
are each single qubit errors (either on qubit one or two respectively).  We now perform a CNOT 
operation, $U= $CNOT, where the control is qubit one.  This gives,
\begin{equation}
UE_j\ket{\psi} = (UE_jU^{\dagger})U\ket{\psi},
\end{equation} 
using the identity $(U^{\dagger}U) = I$.
The effect of the gate is to transform the error $E_j$ to $E_j' = (UE_jU^{\dagger})$.  For the six single 
qubit errors, this gives,
\begin{equation}
\begin{aligned}
&U(X_1I_2)U^{\dagger} = X_1X_2,\\
&U(Y_1I_2)U^{\dagger}  = Y_1X_2,\\
&U(Z_1I_2)U^{\dagger}  = Z_1I_2,\\  
&U(I_1X_2)U^{\dagger}  = I_1X_2,\\
&U(I_1Y_2)U^{\dagger}  = Z_1Y_2,\\
&U(I_1Z_2)U^{\dagger}  = Z_1Z_2.\\
\end{aligned}
\end{equation}
Therefore, the CNOT gate copies $X$ errors from the control qubit to the target and copies $Z$ errors from the target to the 
control.  Error propagation rules for any multi-qubit unitary can also be calculated for an 
arbitrary gate, $U$, and error, $E$.\footnote{If $U$ is not a member of the Clifford group (operators which map Pauli operators to Pauli operators) 
then $E$ will map to a linear combination of Pauli errors.}  

The above example assumes a perfect 2-qubit gate operation, however the gates themselves can also introduce multiple errors when they fail.  
In the case of coherent errors, an arbitrary 2-qubit unitary, $U$, can be written in the form, 
$U_EU_p$, where $U_p$ is the perfectly applied gate and $U_E$ is some coherent 
error operator.  This error operator can be expressed 
as a linear combination of the 2-qubit Pauli group,
\begin{equation}
U_E = \sum_{(i,j)=1}^{4}  d_{i,j} \; \sigma_i\otimes \sigma_j, \quad [\sigma_1,\sigma_2,\sigma_3,\sigma_4] = [I,X,Y,Z]
\end{equation}
Therefore inaccurate application of the gate $U_p$ 
(including complete failure) can introduce a single $X,Y$ or $Z$ error to each of the two qubits 
or it can introduce an error to {\em both} qubits (the details of what kind of errors can be introduced by faulty gates is dependent on the mechanisms causing the failure).    Therefore, not only can gates coupling qubits copy pre-existing 
errors but a single failure event can introduce errors on both qubits involved in the interaction.   
Correlated errors also occur in the density matrix representation when incoherent noise causes gate inaccuracies.
This generalizes for higher order operations, e.g. errors on three qubit gates can introduce errors on all three qubits, etc.  

 \begin{figure}[ht]
\begin{center}
\includegraphics[width=.4\textwidth]{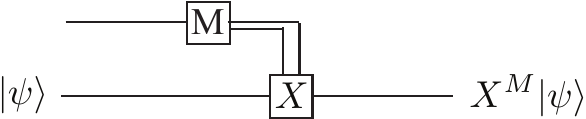}
\caption{Error propagation due to measurement error.  If the measurement result for qubit one is faulty, we inadvertently introduce 
an error to the second qubit. }
\label{fig:measer}
\end{center}
\end{figure}

A second process which can cause errors to cascade is the classical correction of quantum data from a measurement result\footnote{
It should be noted that this process is equivalent to directly coupling qubits with quantum gates and then measuring one of the 
qubits, hence errors will be copied in the same manner as direct coupling.}.  Take 
the circuit shown in Fig. \ref{fig:measer}.  The measurement result on the top qubit is used to determine if the gate, $X$, is applied 
to the second qubit.  In this case the gate is applied if the upper qubit is measured in the state $\ket{1}$.  If we now assume that this 
measurement is faulty, then with some probability the ``classical" information extracted from the measurement device is wrong.  
Therefore the correction that is applied to the second qubit should not have been applied.  Not only has a bit flip error now occurred 
on qubit one (due to the inaccurate measurement result), but we also have accidentally introduced a bit flip error on the second qubit 
due to this inaccurate information.  
 
Hence gates which couple qubits and measurement errors (which subsequently control further quantum gates) can copy errors 
from qubit to qubit.  

\subsection{Concatenation}
\begin{figure*}[ht]
\begin{center}
\includegraphics[width=\textwidth]{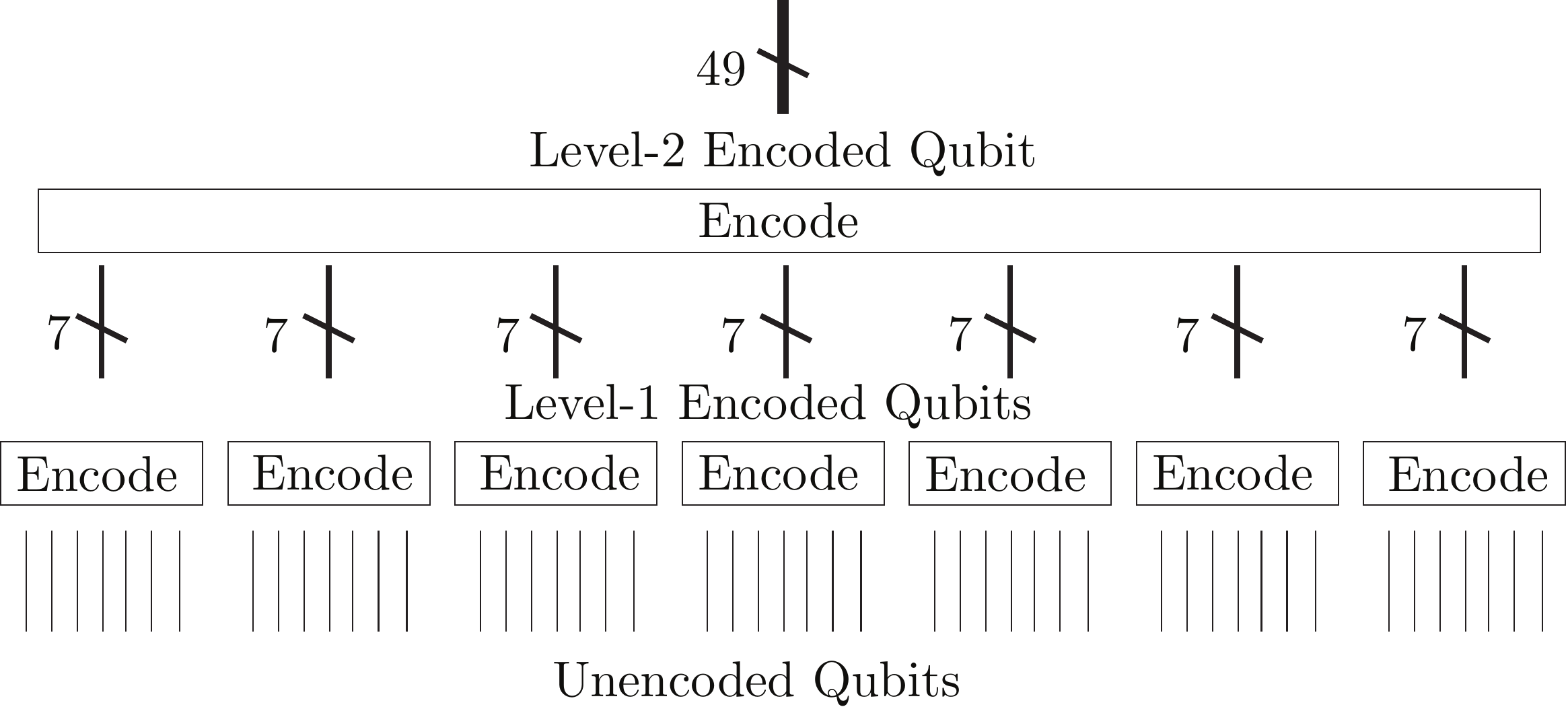}
\caption{Illustration explaining the concept of concatenation with the $[[7,1,3]]$ code.  49 physical qubits are 
first encoded into seven level-1 logical qubits (seven blocks of seven qubits).  These seven logical qubits are 
then encoded into a level-2 logical qubit.  This level-2 qubit can correct unto a maximum of $\frac{3^2-1}{2} =4$ 
errors on the 49 physical qubits forming the encoded block.}
\label{fig:concat}
\end{center}
\end{figure*}

Concatonation is where an group of encoded qubits are further encoded (not necessarily with the same error correction code).  
This forms a second level encoded qubit which Fig. \ref{fig:concat} illustrates for the $[[7,1,3]]$ code.   
Starting with a group of unencoded qubits, we first encode into level-1 logical qubits, each containing a block of seven.  Each of these 
logical qubits are now protected with a distance three quantum code.  If we now take seven of these level-1 logical qubits are 
perform the same encoding operation (using valid encoded gate operations, discussed in the following sections) we form a level-2 
logical qubit.  This is now a block containing 49 physical qubits and has a code distance of $3^2 = 9$.  
A level-2 logical qubit can therefore be faithfully recovered if four or less errors occur on any of the 49 physical qubits.  
This process can be repeated indefinitely, for $g$ levels of concatenation, the encoded state will consist of 
$7^g$ physical qubits and have a code distance of $3^g$, allowing for the correction of $\frac{3^g-1}{2}$ individual errors.  
Hence for a concatenation scheme, both the number of physical qubits required and the number of errors it can correct 
scale exponentially (with the number of qubits growing faster than the number of correctable errors)
 
 \subsection{Fault-tolerance}
The concept of fault-tolerance in computation is not a new idea, it was first developed 
in relation to classical computing~\cite{N55,G83,A87}.  However, in recent years the precise manufacturing 
of digital circuitry has made large-scale error correction and fault-tolerant circuits largely unnecessary.

The basic principle of fault-tolerance is that the circuits used for gate operations and 
error correction procedures should not cause errors to cascade.  As shown in the previous 
section, gates which couple multiple qubits and measurements which are used to control a quantum operation 
can cause errors to be copied.  How do we design operations to avoid cascading errors in quantum algorithms?  

\begin{figure*}[ht]
\begin{center}
\includegraphics[width=0.85\textwidth]{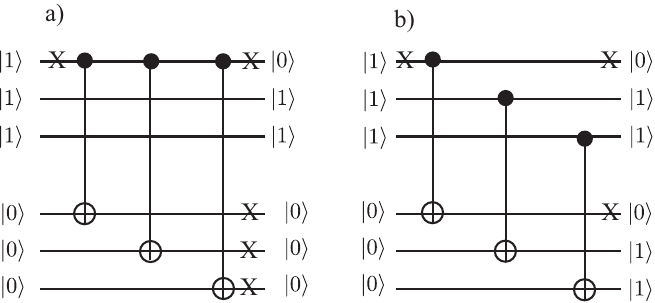}
\caption{Two circuits to implement the transformation 
$\ket{111}\ket{000} \rightarrow \ket{111}\ket{111}$. Subfigure a) shows a version where a single $X$ 
error can cascade into four errors while Subfigure b) shows an equivalent circuit where the error only propagates 
to a second qubit.  }
\label{fig:CNOT}
\end{center}
\end{figure*}

This can be seen more clearly when 
we look at a simple CNOT operation between two qubits [Fig.~\ref{fig:CNOT}].  In this circuit 
we are performing a sequence of three CNOT gates which act to take the state 
$\ket{111}\ket{000} \rightarrow \ket{111}\ket{111}$.  In Fig.~\ref{fig:CNOT}a we consider a single 
$X$ error which occurs on the top most qubit prior to the first CNOT.  This single error 
will cascade through each of the three gates such that the $X$ error has now propagated to 
four qubits.  Fig.~\ref{fig:CNOT}b shows a slightly modified design that implements the 
same operation, but the single $X$ error now only propagates to two of the six qubits.   
If we consider each block of three as a single logical qubit, then the staggered circuit will only induce 
a total of one error in each logical block, given a single $X$ error occurred somewhere during the 
circuit.  Therefore, one of the standard definitions of fault-tolerance is,

{\em fault-tolerant circuit element:  A single error will cause \textbf{at most} one error in the output for each logical qubit block.}

It should be stressed that the idea of fault-tolerance is a discrete definition, either a certain 
quantum operation is fault-tolerant or it is not.  What is defined to be fault-tolerant can change 
depending on the error correction code used.  For example, for a single error correcting code, 
the above definition is the only one available, since any more than one error in a logical qubit 
will result in the error correction procedure failing.  However, if the quantum code employed 
is able to correct multiple errors, then the definition of fault-tolerance can be relaxed, i.e. 
if the code can correct three errors then circuits may be designed such that a single 
failure results in at most three errors in the output (which is then correctable).  In general, for 
an code correcting $t=\lfloor (d-1)/2 \rfloor$ errors, fault-tolerance requires that $\leq t$ errors 
during an operation does not result in $> t$ errors in the output for each logical qubit.  

\subsection{Threshold theorem}
\label{sec:threshold}
The threshold theorem is truly a remarkable result in quantum information and is a consequence 
of fault-tolerant circuit design and the ability to perform dynamical error correction.  
Rather than present a detailed derivation of the theorem for a variety of noise models, we will 
instead take a very simple case where we utilize a quantum code that can only correct for a 
single error, using a model that assumes uncorrelated errors on individual qubits.  For 
more rigorous derivations of the theorem see~\cite{AB97,G97+,A07}.  

Consider a quantum computer where each physical 
qubit experiences either an $X$ and/or $Z$ error independently 
with probability $p$, per gate operation.  
Furthermore, it is assumed that logical gate operations and error 
correction circuits are designed according to the rules of fault-tolerance and that a cycle of 
error correction is performed after each elementary {\em logical} gate operation using a 
code that corrects for a single error ($[[n,k,3]]$ code).  If an error occurs 
during a logical gate operation, then fault-tolerance ensures this error will only propagate 
to at most one error in each block, after which a cycle of error correction will remove the error.  
Hence if the failure probability of un-encoded qubits per time step is $p$, then a single level 
of error correction will ensure that the logical step fails only when two (or more) errors occur.  Hence 
the failure rate of each logical operation, to leading order, is now $p^1_L = cp^2$, where $p^1_L$ is the 
failure rate (per logical gate operation) of a level-1 logical qubit and $c$ is the upper bound for the 
number of possible 2-error combinations 
which can occur at a physical level within the sequence of operations consisting of a  
correction cycle, the logical gate operation and a second correction cycle~\cite{A07}.  
We now concatenate, encoding the computer further, 
such that a level-2 logical qubit is formed.  We assume this is done using the same $[[n,k,3]]$ 
quantum code level-1 encoded 
qubit.  It is assumed that all error correcting procedures and gate operations at level-2 are 
self-similar to the level-1 operations (i.e. the circuit structures for the level-2 encoding are 
identical to the level-1 encoding).  Therefore, if the level-1 failure rate per logical time step is $p^1_L$, 
then by the same argument, the failure rate of a level-2 operation is given by,
$p^2_L = c(p^1_L)^2 = c^3p^4$.  This iterative procedure is then repeated 
up to level-$g$, such that the logical failure rate, per time step, of a level-$g$ encoded qubit is given by,
\begin{equation}
p^g_L = \frac{(cp)^{2^g}}{c}.
\label{eq:threshold}
\end{equation}   
Eq. (\ref{eq:threshold}) implies that for a finite {\em physical} error rate, $p$, per qubit, per time step, 
the failure rate of a level-$g$ encoded qubit can be made arbitrarily small if $cp < 1$.  This inequality defines the threshold.  The physical error rate 
experienced by each qubit per time step must be $p_{th} < 1/c$ to ensure that multiple levels of concatenated 
error correction reduces the failure rate of logical components.

Hence, provided sufficient resources are available, an arbitrarily large quantum circuit can be 
successfully implemented to arbitrary accuracy, once the physical error rate is below threshold.  The 
calculation of thresholds is therefore an extremely important aspect to quantum architecture design.  
Initial estimates at the threshold, which gave $p_{th} \approx 10^{-4}-10^{-6}$~\cite{K97,AB97,G97+}, 
did not sufficiently model the physical architectures.  Recent 
results~\cite{SFH07,SDT07,SBFRYSGF06,MCTBCCC04,BSO05} have been estimated for more 
realistic quantum processor architectures, showing significant differences in threshold when architectural considerations 
are taken into account.  Many of these estimates are summarised in Table. \ref{tab:threshold} with a brief description of the 
constraints considered by the model.  We order the entries by the level of detail presented in each of the papers with regards to 
physical implementation of fault-tolerant error correction protocols on realistic quantum architectures.

\begin{table*}[ht!]
\begin{center}
\vspace*{4pt}   
\begin{tabular}{c|c|c|c|c}
Relevant Work & Code & Threshold & Geometric Constraints & Architectural context. \\
\hline
\cite{AB99} & $[[7,1,3]]$ &  $O(10^{-6})$ &A.I.\footnote{Arbitrary interactions} & None\\
\cite{G97+} & $[[7,1,3]]$ & $O(10^{-4}-10^{-6})$ & A.I. & None \\
\cite{KLZ96} & $[[7,1,3]]$ & $O(10^{-6})$ & A.I. & None. \\
\cite{PR11} & $[[23,1,7]]$ & $O(10^{-3})$ & A.I. & None\\
\cite{K05} & 4- and 6-qubit detection & $O(10^{-2})$ &A.I. & None. \\
\cite{BAO12} & Toric Code & 18\% & 2D NN\footnote{Nearest Neighbour} array, P.B.C.\footnote{Periodic Boundary conditions} & None, Theoretical upper bound\footnote{This result is calculated assuming perfect quantum gates and is known as the code capacity.}\\
\cite{MCTBCCC04} & $[[7,1,3]]$\footnote{Only a preliminary estimate} & $O(10^{-4})$ & A.I. with movement penalty & Specific to Ion-Traps\\
\cite{SFH07} & $[[7,1,3]]$ & $O(10^{-6})$ & Bilinear NN array & Kane P:Si system \cite{K98}\\
\cite{SBFRYSGF06} & $[[7,1,3]]$ & $O(10^{-7})$ & variable width NN array  & General NN systems \\
\cite{SDT07} & $[[7,1,3]]$ & $O(10^{-5})$ & 2D NN arrays & General NN systems \\
\cite{FTYSPW07} & $[[7,1,3]]$ & $O(10^{-6})$ & Bilinear NN array & Superconducting qubits\\
\cite{BSO05} & $[[7,1,3]]$ & $O(10^{-9})$ & A.I. with movement penalty & Ion-Traps\\
\cite{RHG07} & Topological Cluster & $O(10^{-2}-10^{-3})$ & 3D NN array & Photonic Qubits \cite{DFSG08} \\
\cite{WFSH09} & Surface Codes & $O(10^{-2}-10^{-3})$ & 2D NN array & Quantum Dots, Diamond \cite{JMFMKLY10,YJG10}\\
\end{tabular}
\caption{Various threshold estimates, associated geometric constraints and architectural considerations when performing the calculation.  We 
have ordered the table with respect to the level of architectural detail that has been developed compatible with the threshold result.  Note that 
this is not an exhaustive list.} 
\label{tab:threshold}
\end{center}
\end{table*} 

From the above table you can see the differences in threshold that occur once more specific architectural considerations are taken into account.  
This variation is heavily dependant on what error models are assumed in the analysis, if qubit transport can be done directly (such in 
ion traps where the qubits are physically moved throughout the computational array) or via the application of SWAP gates (required 
in condensed matter systems such as P:Si, Quantum Dots or NV defects in Diamond) and what is the structure of the QEC code that is ultimately used.  
In terms of concatenated QEC codes, the majority of analysis was done using the $[[7,1,3]]$ steane code and only the results of Knill \cite{K05} 
show a threshold of the order of 1\%.  However, the results of Knill have not yet been adapted to a physical architecture and it remains 
unclear if the constraints of a 1D, 2D or 3D nearest neighbour architecture will reduce the threshold to that of other concatenated 
codes and/or substantially increase qubit resources to achieve a high threshold.  

Currently, the most detailed architectural designs are based on topological cluster codes \cite{RHG07} and surface codes \cite{K97,DKLP02,FSG08}.  
The general framework 
for this type of error correction is discussed later in Section \ref{sec:topological}.  These codes are promising for large scale computers as they 
exhibit significantly higher thresholds to concatenation and they have an intrinsic structure that is compatible with 2D or 3D nearest Neighbour 
architectures. 
  
\section{Fault-tolerant operations on encoded data}\label{sec:operations}
Sections~\ref{sec:sec:QEC}, \ref{sec:sec:QEC2} and~\ref{sec:Fault-tolerance} showed how fault-tolerant QEC allows for 
any quantum algorithm to be run to arbitrary accuracy.  However, the results of the threshold theorem 
assume that logical operations can be performed directly on the encoded data without the need for 
continual decoding and re-encoding.  Using stabilizer codes, a large class of operations can 
be performed on logical data in an inherently fault-tolerant way. 

If a given logical state, $\ket{\psi}_L$, is stabilized by $K$, and the logical operation $U$ is applied, 
the new state, $U\ket{\psi}_L$ is stabilized by $UKU^{\dagger}$, i.e,
\begin{equation}
UKU^{\dagger}U\ket{\psi}_L = UK\ket{\psi}_L = U\ket{\psi}_L.
\end{equation}
In order for the codeword states to remain valid, the stabilizer set for the code, $\{G_i\} \in \mathcal{G}$, 
must remain fixed through 
every operation.  Hence for $U$ to be a valid operation on the data, 
$UG_iU^{\dagger} = G_j$, $G_j \in \mathcal{G}$ $\forall i$.  As a shorthand notation, we express this 
relationship as $U\mathcal{G}U^{\dagger} = \mathcal{G}$.

\subsection{Single qubit operations}
For an elementary operation $A$, the equivalent logical operator $\bar{A}$ will now be formed via a sequence of 
elementary operations on the physical qubits comprising the code block. 
The logical $\bar{X}$ and $\bar{Z}$ operations on a single encoded qubit are the first examples of 
valid codeword operations.  Taking the $[[7,1,3]]$ code as an example, $\bar{X}$ and $\bar{Z}$ are 
given by,
\begin{equation}
\begin{aligned}
&\bar{X} = XXXXXXX \equiv X^{\otimes 7},\\
&\bar{Z} = ZZZZZZZ \equiv Z^{\otimes 7}.
\end{aligned}
\label{eq:logop}
\end{equation}
Since the single qubit Pauli operators satisfy $XZX = -Z$ and $ZXZ = -X$ then, 
$\bar{X}K^{i}\bar{X} = K^{i}$ and $\bar{Z}K^{i}\bar{Z} = K^{i}$ for each of the 
$[[7,1,3]]$ stabilizers given in Eq.~(\ref{eq:stab7}).  The fact that each stabilizer has a weight of four 
guarantees that $UKU^{\dagger}$ picks up an even number of $-1$ factors.  Since the 
stabilizers remain fixed, the operations are valid.  However, what transformations do Eq.~(\ref{eq:logop}) 
actually perform on encoded data?

\begin{figure*}[ht]
\begin{center}
\includegraphics[width=0.75\textwidth]{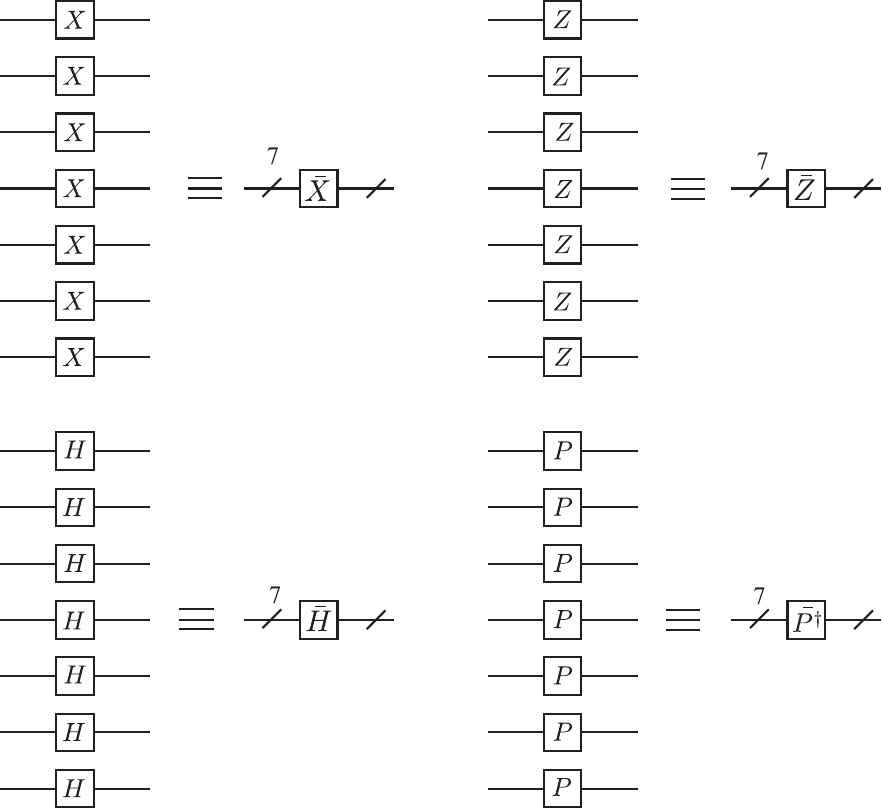}
\caption{Bit-wise application of single qubit gates in the $[[7,1,3]]$ code.  Logical $X$, $Z$ 
$H$ and $P$ gates can trivially be applied by using seven single qubit gates, fault-tolerantly.  
Note that the application of seven $P$ gates results in the logical $\bar{P^{\dagger}}$ being applied 
and vice-versa.}
\label{fig:transversal}
\end{center}
\end{figure*}

For a single qubit, a bit-flip operation $X$ takes $\ket{0} \leftrightarrow \ket{1}$.  Recall that for a single 
qubit $Z\ket{0} = \ket{0}$ and $Z\ket{1} = -\ket{1}$, hence for $\bar{X}$ to actually induce a logical bit-flip 
it must take, $\ket{0}_L \leftrightarrow \ket{1}_L$.  For the $[[7,1,3]]$ code, the final operator which 
fixes the logical state is $K^7 = Z^{\otimes 7}$, 
where $K^7\ket{0}_L = \ket{0}_L$ and $K^7\ket{1}_L = -\ket{1}_L$.  
As $\bar{X}K^7\bar{X} = -K^7$, any state stabilized by $K^7$ becomes stabilized by $-K^7$ (and vice-versa)
after the operation of $\bar{X}$.  Therefore, $\bar{X}$ represents a logical bit flip.  
The same argument can be used for $\bar{Z}$ by considering the stabilizer properties of the states 
$\ket{\pm} = (\ket{0} \pm \ket{1})/\sqrt{2}$.  
Hence, the logical bit- and phase-flip gates can be applied directly to logical data by simply using seven 
single qubit $X$ or $Z$ gates, [Fig.~\ref{fig:transversal}].  

Two other useful gates which can be applied in this manner is the Hadamard rotation and phase gate, 
\begin{equation}
H = \frac{1}{\sqrt{2}} \begin{pmatrix} 1 & 1 \\ 1 & -1 \end{pmatrix}, \quad \quad
P = \begin{pmatrix} 1 & 0 \\ 0 & i \end{pmatrix}.
\end{equation}
These gates are useful since when combined with the two-qubit CNOT gate, they can
generate a subgroup of all multi-qubit gates known as the Clifford 
group (gates which map Pauli group operators back to the Pauli group).   
Again, using the stabilizers of the $[[7,1,3]]$ code and the fact that for single qubits, 
\begin{equation}
\begin{aligned}
HXH = Z, \quad \quad HZH = X, \\
PXP^{\dagger} = iXZ, \quad \quad PZP^{\dagger} = Z,
\end{aligned}
\end{equation}
a seven qubit bit-wise Hadamard gate will switch $X$ with $Z$ and therefore will simply flip 
$\{K^1,K^2,K^3\}$ with $\{K^4,K^5,K^6\}$, and is a valid operation.  The bit-wise application of the 
$P$ gate will leave any $Z$ stabilizer invariant, but takes $X \rightarrow iXZ$.  
This is still valid, since provided there are a multiple 
of four non-identity operators for each generator of the stabilizer group, 
the factors of $i$ will cancel.  Hence seven 
bit-wise $P$ gates is valid for the $[[7,1,3]]$ code.    

What do these logical operators do to the logical state?  For a single qubit, the Hadamard gate 
flips any $Z$ stabilized state to a $X$ stabilized state, i.e $\ket{0,1} \leftrightarrow \ket{+,-}$.  Looking at 
the transformation of $K^7$,  using $\bar{H} = H^{\otimes 7}$ we have, $\bar{H}K^7\bar{H} = X^{\otimes 7}$.  Therefore, the bit-wise 
Hadamard gate, $\bar{H}$, is the logical Hadamard operation.  The single qubit $P$ gate leaves a 
$Z$ stabilized state invariant, while an $X$ eigenstate becomes stabilized by $iXZ$.  
Hence, denoting $\bar{P}^{\dagger} = P^{\otimes 7}$, $\bar{P}(X^{\otimes 7})\bar{P}^{\dagger} = -i(XZ)^{\otimes 7}$ and the bit-wise gate, 
$\bar{P}^{\dagger}$, is the logical $P^{\dagger}$ gate. Similarly, 
bit-wise gate, $\bar{P} = P^{\dagger \otimes 7}$, 
enacts a logical $P$ gate [Fig.~\ref{fig:transversal}].  
Each of these fault-tolerant operations on a logically encoded block are commonly 
referred to as transversal operations.  A transversal operation can be defined as a logical operator which is 
formed by applying the individual physical operators to each qubit in the code block (or between two equivalent physical 
qubits in the case of entangling operations such as the CNOT gate shown below).   The physical operations forming a transversal logical operator need not be the same, for example in Fig.\ref{fig:transversal}, the $\bar{P}^{\dagger}$ gate is formed via 
individual $P$ operations and visa versa.  Hence, valid transversal gates may possibly be constructed from a set of physical 
operations which are unique to each qubit.

\subsection{Two-qubit gate}
\begin{figure}[ht]
\begin{center}
\includegraphics[width=0.45\textwidth]{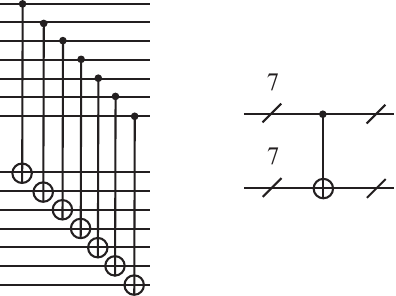}
\caption{Bit-wise application of a CNOT gate between two logical qubits.  Since each CNOT 
only couples corresponding qubits in each block, this operation is inherently fault-tolerant.}
\label{fig:transversal2}
\end{center}
\end{figure}

A two-qubit logical CNOT operation can also be applied in the same way, as a transversal bit-wise operation of individual CNOT 
gates between corresponding physical qubits.  For un-encoded qubits, 
a CNOT operation performs the following mapping on the two qubit stabilizer set,
\begin{equation}
\begin{aligned}
&X\otimes I \rightarrow X\otimes X, \\
&I\otimes Z \rightarrow Z\otimes Z, \\
&Z\otimes I \rightarrow Z\otimes I, \\
&I\otimes X \rightarrow I\otimes X.
\end{aligned}
\label{eq:CNOTtrans}
\end{equation}
Where the first operator corresponds to the control qubit and the second operator corresponds to the target.  
Now consider the bit-wise application of seven CNOT gates between logically encoded blocks of 
data [Fig.~\ref{fig:transversal2}].  First the stabilizer set must remain invariant, 
\begin{equation}
\text{CNOT } \mathcal{G}\text{ CNOT} = \{\text{CNOT } (K^{i}\otimes K^{j}) \text{ CNOT}\} = \mathcal{G}.
\end{equation}
i.e. each element in $\mathcal{G}$ must be mapped to another element of $\mathcal{G}$. 
Table~\ref{tab:stabtrans} details the transformation for all the 
stabilizer generators under seven bit-wise CNOT gates, 
demonstrating that this operation is valid on the $[[7,1,3]]$ code.  The transformations in 
Eq.~(\ref{eq:CNOTtrans}) are trivially extended to the logical space, showing that seven 
bit-wise CNOT gates invoke a logical CNOT operation.
\begin{equation}
\begin{aligned}
&\bar{X}\otimes I \rightarrow \bar{X}\otimes \bar{X}, \\
&I\otimes \bar{Z} \rightarrow \bar{Z}\otimes \bar{Z}, \\
&\bar{Z}\otimes I \rightarrow \bar{Z}\otimes I, \\
&I\otimes \bar{X} \rightarrow I\otimes \bar{X}.
\end{aligned}
\label{eq:CNOTtrans2}
\end{equation}
\begin{table*}
\begin{tabular}{|c|c|c|c|c|c|c|}
\hline
$K^i \otimes K^j$ & $K^1$ & $K^2$ &$K^3$ &$K^4$ &$K^5$ &$K^6$  \\
\hline
$K^1$ & $K^1\otimes I$& $K^1\otimes K^1K^2$ & 
$K^1\otimes K^1K^3$ & $K^1K^4 \otimes K^1K^4$ & 
$K^1K^5\otimes K^1K^5$ & $K^1K^6\otimes K^1K^6$\\
\hline
$K^2$ & $K^2\otimes K^1K^2$ & $K^2\otimes I$ & $K^2 \otimes K^2K^3$ 
&$K^2K^4\otimes K^2K^4$ &$K^2K^5\otimes K^2K^5$ & $K^2K^6\otimes K^2K^6$\\
\hline
$K^3$ & $K^3\otimes K^3K^1$ & $K^3\otimes K^3K^2$ &$K^3\otimes I$ & 
$K^3K^4\otimes K^3K^4$ & $K^3K^5\otimes K^3K^5$ &$K^3K^6\otimes K^3K^6$ \\
\hline
$K^4$ & $K^4\otimes K^1$ & $K^4\otimes K^2$ & $K^4\otimes K^3$ &
$I\otimes K^4$ & $K^4K^5\otimes K^5$ &$K^4K^6\otimes K^6$ \\
\hline
$K^5$ & $K^5\otimes K^1$ & $K^5\otimes K^2$ & $K^5\otimes K^3$ &
$K^5K^4\otimes K^4$ & $I\otimes K^5$ &$K^5K^6\otimes K^6$ \\
\hline
$K^6$ & $K^6\otimes K^1$ & $K^6\otimes K^2$ & $K^6\otimes K^3$ &
$K^6K^4\otimes K^4$ & $K^6K^5\otimes K^5$ &$I\otimes K^6$ \\
\hline
\end{tabular}
\caption{Transformations of the $[[7,1,3]]$ stabilizer generators under the gate operation 
$U=$CNOT$^{\otimes 7}$, 
where $\mathcal{G} \rightarrow U^{\dagger}\mathcal{G}U$.  
Note that the transformation does not take any stabilizer outside the group generated by 
$K^i \otimes K^j\; (i,j)\in [1,..,6]$, 
therefore $U=$CNOT$^{\otimes 7}$ represents a valid operation on the codespace.} 
\label{tab:stabtrans}
\end{table*}
The issue of fault-tolerance with these logical operations should be clear.  The $\bar{X}$,$\bar{Z}$,
$\bar{H}$, $\bar{P}^{\dagger}$ and $\bar{P}$ gates are trivially fault-tolerant since the logical operation is performed 
through seven bit-wise single qubit gates.  The logical CNOT is also fault-tolerant since each two-qubit 
gate only operates between counterpart qubits in each logical block.  Hence if any gate is inaccurate 
then, at most, a single error will be introduced in each block.  

In contrast to the [[7,1,3]] code, let us also take a quick look at the [[5,1,3]] code.  
Unlike the 7-qubit code, the full set of Clifford gates cannot be implemented in the 
same transversal manner.  
To see this clearly we can examine how the generators of the 
stabilizer group for the code transforms under a transversal Hadamard operation,
\begin{equation}
\begin{array}{cccccc}
K^1 = & X & Z & Z & X & I \\
K^2 = & I & X & Z & Z & X \\
K^3 = & X & I & X & Z & Z \\
K^4 = & Z & X & I & X & Z \end{array}
\quad \xrightarrow{H^{\otimes 5}} \quad 
\begin{array}{ccccc}
Z & X& X & Z & I \\
I & Z & X & X & Z \\
Z & I & Z & X & X \\
X & Z & I & Z & X \end{array}
\end{equation}
The stabilizer group is not preserved under this transformation and therefore is not a valid logical operation for the [[5,1,3]] code.  One thing to briefly note 
is that there are methods for performing logical Hadamard and phase gates on the [[5,1,3]] code~\cite{G97+}.  However, it essentially involves performing a valid, transversal, 
three qubit gate and then measuring out two of the logical ancilla.  Although it has been demonstrated that a valid set of transversal 
operations for universal computing is incompatible with a quantum code that correct an arbitrary single error~\cite{EK09}, it has not 
been shown that a non-CSS code cannot have a transversal set of operations generating the Clifford group.  This example 
is used to illustrate the transformation for an invalid logical operation on an encoded state.

While these gates can be conveniently implemented on error protected data, they do not represent a universal set 
for quantum computation.  In fact it has been shown that by using the stabilizer formalism, these 
operations can be efficiently simulated on a classical device~\cite{G98,AG04}.  In order to achieve universality 
one of the following gates are generally added to the available set,
\begin{equation}
T = \begin{pmatrix} 1 & 0 \\ 0 & e^{i\pi/4} \end{pmatrix}, 
\end{equation}
or the Toffoli gate~\cite{T81}.  
Applying them in a similar transversal way, for the majority of stabiliser codes, transform the stabilizers 
out the group and is consequently  
not a valid operation.  Circuits implementing these two gates in a fault-tolerant 
manner have been developed~\cite{NC00,GC99,SI05,SFH07}, for various codes (most often the $[[7,1,3]]$ code)
but at this stage the circuits are complicated and resource intensive.  
This has practical implications to encoded operations.  If universality is achieved by adding the 
$T$ gate to the list, arbitrary single qubit rotations require 
long gate sequences (utilizing the Solovay-Kitaev theorem~\cite{K97,DN06}) to approximate arbitrary 
logical qubit rotations and these sequences often require many $T$ gates~\cite{F05+}.   
Finding more efficient methods to achieve universality on encoded data is therefore still an active area 
of research.

\section{Fault-tolerant circuit design for logical state preparation}\label{sec:FTcircuit}
Section~\ref{sec:Fault-tolerance} 
introduced the basic rules for fault-tolerant circuit design and how these rules lead 
to the threshold theorem for concatenated error correction.  However, what does a full fault-tolerant 
quantum circuit look like? Here, we introduce a full fault-tolerant circuit to prepare the $[[7,1,3]]$ logical $\ket{0}$ state.  As the 
$[[7,1,3]]$ code is a single error correcting code, we use the one-to-one definition of fault-tolerance and 
therefore only need to consider the propagation of a single error during the 
preparation (any more that one error during correction represents a higher order effect and is ignored).

As described in Section~\ref{sec:sec:QEC}, 
logical state preparation can be done by initializing an appropriate number 
of physical qubits and measuring each of the $X$ stabilizers that describe the code.  Therefore, 
a circuit which allows the measurement of a Hermitian operator in a fault-tolerant manner needs to be 
constructed.  The general structure of the
circuit used was first developed by Shor~\cite{S96+}, however it should be noted that several more recent 
methods for fault-tolerant state preparation and correction now exist~\cite{S97,S02,DA07,K05}.

\begin{figure*}[ht]
\begin{center}
\includegraphics[width=0.8\textwidth]{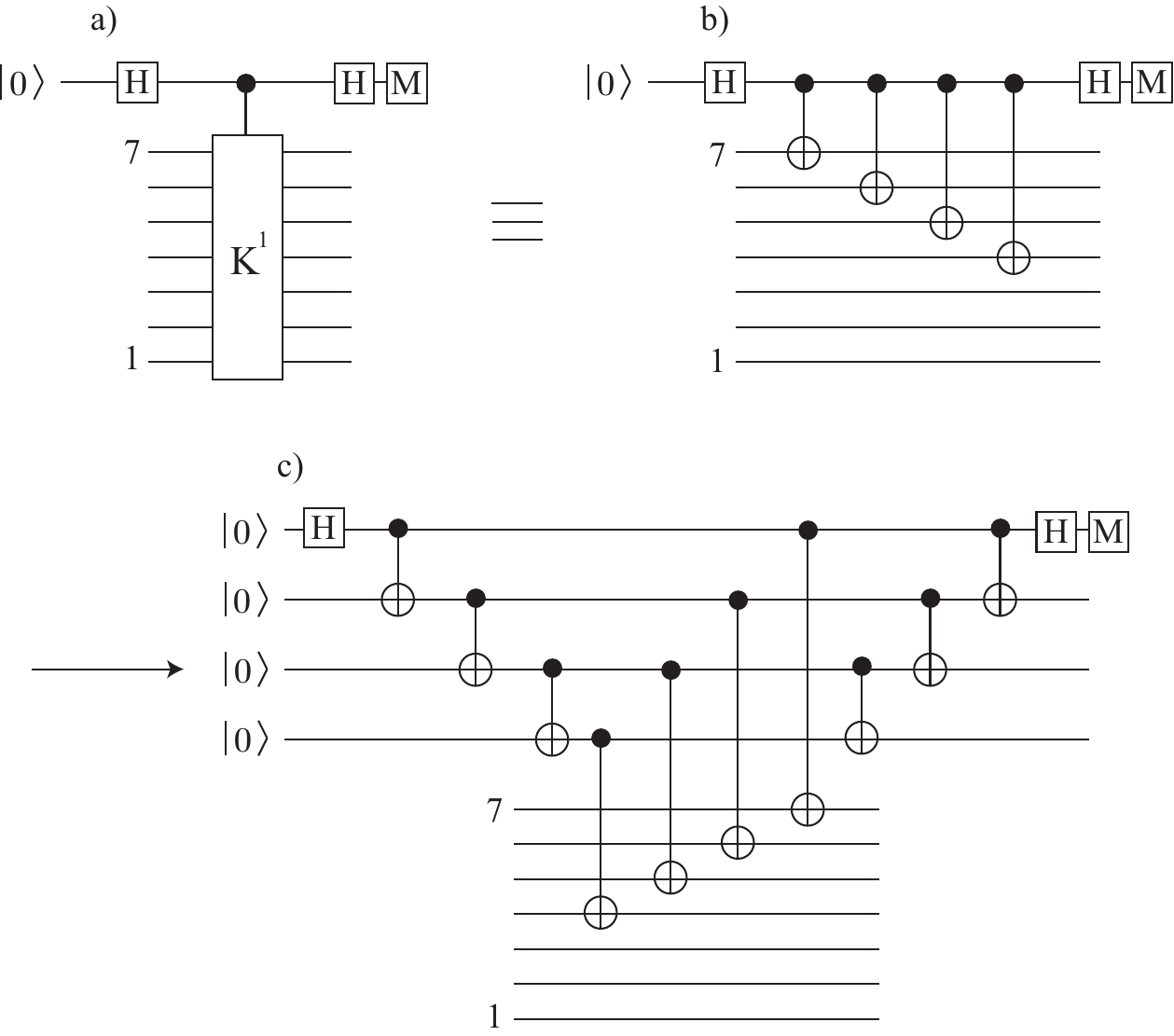}
\caption{Three circuits which measure the stabilizer $K^1$.  Subfigure a) represents a 
generic operator measurement where a multi-qubit controlled gate is available.  Subfigure b) 
decomposes this into single- and two-qubit gates, but in a non-fault-tolerant manner.  Subfigure c) 
introduces four ancilla such that each CNOT is controlled via a separate qubit.  This ensures 
fault-tolerance.}
\label{fig:opmeas2}
\end{center}
\end{figure*}

\begin{figure*}[ht]
\begin{center}
\includegraphics[width=0.8\textwidth]{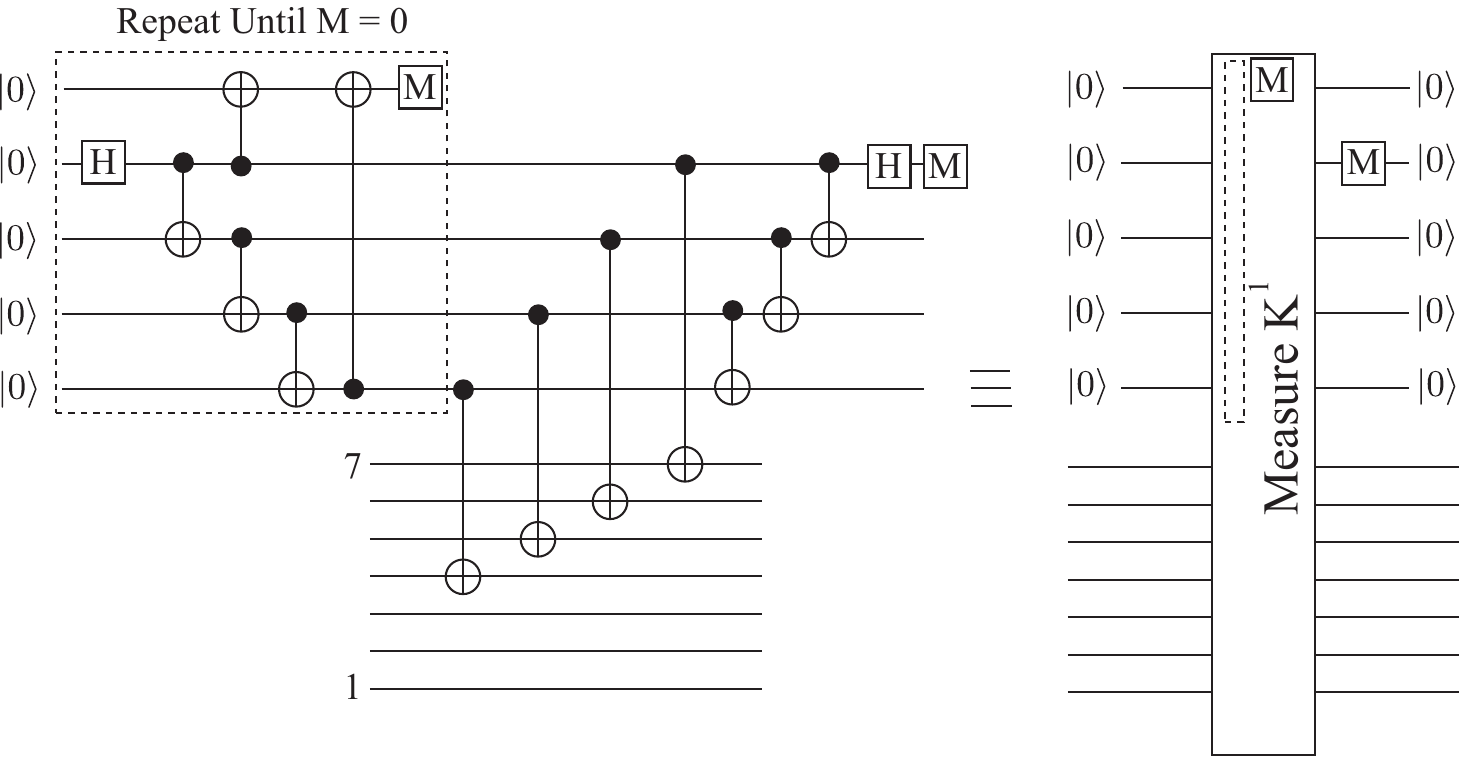}
\caption{Circuit required to measure the stabilizer $K^1$, fault-tolerantly.  A four qubit GHZ 
state is used as ancilla with the state requiring verification against multiple $X$ errors.  After the 
state has passed verification it is coupled to the data block and a syndrome is extracted.}
\label{fig:opmeas3}
\end{center}
\end{figure*}

The circuits shown in Fig.~\ref{fig:opmeas2}a and~\ref{fig:opmeas2}b, 
which measure the stabilizer $K^1 = IIIXXXX$ are 
not fault-tolerant, since a single ancilla is used to control each of the four CNOT gates.  As CNOT gates 
can copy $X$ errors [Fig. \ref{fig:CNOT}], a single $X$ error on this ancilla can be copied to multiple 
qubits in the data block.  Instead, 
four ancilla qubits are used 
which are prepared in the state $\ket{\mathcal{A}} = (\ket{0000}+\ket{1111})/\sqrt{2}$.  
This can be done by initializing four qubits in the $\ket{0}$ state and applying a Hadamard, then a 
sequence of CNOT gates [Fig. \ref{fig:opmeas2}c].  Each of these four ancilla are used to control a separate CNOT gate, after 
which the ancilla state is decoded and measured.  
By ensuring that each CNOT is controlled via a separate ancilla, any $X$ error will only propagate to a single 
qubit in the data block.  However, during the preparation of the ancilla state there is the 
possibility that a single $X$ error can propagate to multiple ancilla, which are then fed forward into the 
data block.  In order to combat this, the ancilla block needs to be verified against possible $X$ errors.  
Tracking through all the possible locations where a single $X$ error can occur during 
ancilla preparation leads to the following unique states.
\begin{equation}
\begin{aligned}
&\ket{\mathcal{A}}_1 = \frac{1}{\sqrt{2}}(\ket{0000}+\ket{1111}),\\
&\ket{\mathcal{A}}_2 = \frac{1}{\sqrt{2}}(\ket{0100} + \ket{1011}),\\
&\ket{\mathcal{A}}_3 = \frac{1}{\sqrt{2}}(\ket{0010} + \ket{1101}),\\
&\ket{\mathcal{A}}_4 = \frac{1}{\sqrt{2}}(\ket{0001} + \ket{1110}),\\ 
&\ket{\mathcal{A}}_5 = \frac{1}{\sqrt{2}}(\ket{0111} + \ket{1000}),\\
&\ket{\mathcal{A}}_6 = \frac{1}{\sqrt{2}}(\ket{0011} + \ket{1100}),\\
\end{aligned}
\end{equation}
From these possibilities, the first four states correspond to no error or at most one bit flip and 
the last two states correspond to multiple errors cased by errors copied by the CNOT operations.
In fact, only the last state, $\ket{\mathcal{A}}_6$, needs to be identified since two bit flip errors exist on the ancilla state.   
To verify this ancilla state, a fifth ancilla is added, initialized and used to perform a parity check on the ancilla block.  This 
fifth ancilla is then measured.  If the result is $\ket{0}$, the ancilla block can be coupled to the data.  
If the ancilla result is $\ket{1}$, then either a single error has occurred on the verification qubit or the ancilla state has been 
prepared in either the  $\ket{\mathcal{A}}_5$ or  $\ket{\mathcal{A}}_6$ state.  
In either case, the entire ancilla block is reinitialized and prepared 
again.  This is continued until the verification qubit is measured to be $\ket{0}$ [Fig.~\ref{fig:opmeas3}].  
\begin{figure*}[ht]
\begin{center}
\includegraphics[width=0.6\textwidth,height=0.9\textheight]{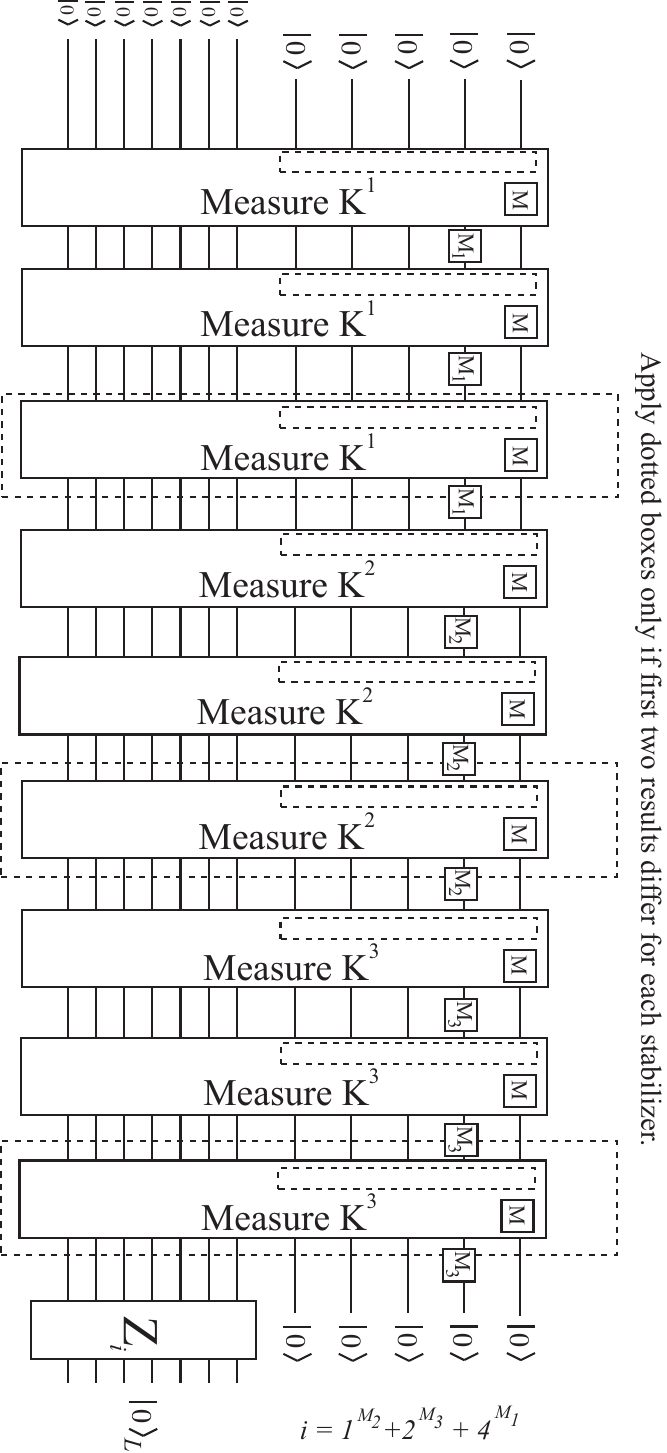}
\caption{ Circuit required to prepare the $[[7,1,3]]$ logical $\ket{0}$ state fault-tolerantly.  
Each of the $X$ stabilizers are sequentially measured using the circuit in Fig.~\ref{fig:opmeas3}.  To 
maintain fault-tolerance, each stabilizer is measured 2-3 times with a majority vote taken.}
\label{fig:stateprep}
\end{center}
\end{figure*}
The re-preparation of the ancilla block protects against multiple $X$ errors, which can propagate forward through the 
CNOT gates.  $Z$ errors propagate in the other direction.  Any $Z$ error 
which occurs in the ancilla block will propagate straight through to the final measurement.  This results in the 
measurement not corresponding to the actual errors which have occurred and can result in 
mis-correction once all stabilizers have been measured.  To protect against this, each stabilizer is 
measured 2-3 times and a majority vote of the measurement results taken.  As any additional error represents 
a second order process, if the first or second measurement has been corrupted by a $Z$ error, then 
the third measurement will only contain additional errors if 
a higher order error process has occurred.  Therefore, we are free to ignore this 
possibility and assume that the third measurement is error free.  
The full circuit for $[[7,1,3]]$ state preparation is shown in Fig.~\ref{fig:stateprep}, where each stabilizer is 
measured 2-3 times.  The total circuit requires a minimum of 12 qubits (7-data qubits and a 5-qubit ancilla block).  

As you can see, the circuit constructions for full fault-tolerant state preparation (and error correction) are not simple 
circuits.  However, they are easy to design in generic ways when employing stabilizer coding.  

\section{Loss protection}
So far we have focused the discussion on correction techniques which assume that error 
processes maintain a qubit structure to the Hilbert space.  
As we noted in section~\ref{sec:lossleakage}, the loss of physical qubits within the computer violates 
this assumption and in general requires additional correction machinery beyond what we 
have already discussed.  
This section examines a basic correction techniques for qubit loss.  
Specifically, we detail one such scheme which was developed with single photon based 
architectures in mind.

Protecting against qubit loss requires a different approach than other general forms of quantum errors such as environmental decoherence 
or systematic control	 imperfections.  The cumbersome aspect related to correcting qubit loss is detecting the presence of a qubit at the 
physical level.  The specific machinery that is required for loss detection is dependent on the underlying physical architecture, but the 
basic principal is that the presence or absence of the physical qubit must be determined without discriminating the actual quantum state.  

Certain systems allow for loss detection is a more convenient way than others.  Electronic spin qubits, for example, can employ Single 
Electron Transistors (SET) to detect the presence or absence of the charge without performing measurement on the spin degree of 
freedom~\cite{DS00,CGJH05,AJWSD01}.  
Optics in contrast requires more complicated non-demolition measurement~\cite{MW84,IHY85,POWBR04,MNBS05}.  This is due to the fact that typical photonic measurement 
is performed via photo-detectors which have the disadvantage of physically destroying the photon.  

Once the detection of the presence of the physical qubit has been performed, a freshly initialized qubit can be injected to 
replace the lost qubit.  Once this has been completed, the standard error correcting procedure can correct for the error.  A freshly initialized 
qubit state, $\ket{0}$ can be represented as projective collapse of a general qubit state, $\ket{\psi} \neq \ket{1}$, as,
\begin{equation}
\ket{0} \propto \ket{\psi}+Z\ket{\psi}.
\end{equation}
If we consider this qubit as part of an encoded block, then the above corresponds to a 50\% error 
probability of experiencing a phase error on this qubit.  Therefore, a loss event that is corrected by non-demolition detection 
and standard QEC, 50\% of the time, will result in a detection event in the QEC cycle and, in the absence of other errors, correction.  
The probability of loss needs to be at a 
comparable rate to standard errors as the correction cycle after a loss detection event will, with high probability, detect an error. 

If a loss event is detected and the qubit replaced, the error detection code shown in  
section~\ref{sec:detection} becomes a single error correction code.  This 
is due to the fact that via the identification of the loss event, these errors subsequently have known locations.  Consequently 
error detection is sufficient to perform full correction, in contrast to 
the error channels we have already considered where in general the location is unknown.  

A second method for loss correction is related to systems that have high loss rates compared to systematic and environmental 
errors, for example in optical systems.  Due to the high mobility of single photons and their relative immunity to 
environmental interactions, loss is a major error channel that generally dominates over other error sources.  The use of error 
detection and correction codes for photon loss is undesirable due to the need for non-demolition detection of the lost qubit.  While techniques for measuring the presence or 
absence of a photon without direct detection have been developed and implemented~\cite{POWBR04}, 
they require multiple ancilla photons and controlled interactions.  
Ultimately it is may be more desirable to redesign the loss correction code such that it can be employed directly with photo-detection rather than 
more complicated non-demolition techniques~\cite{YNM06}.

One such scheme was developed by Ralph, Hayes and Gilchrist in 2005~\cite{RHG05}.  
This scheme was a more efficient extension of an original parity encoding 
method developed by Knill, Laflamme and Milburn~\cite{KLM01}.  The general 
parity encoding for a logical qubit is an $N$ photon GHZ state in the conjugate basis, i.e,
\begin{equation}
\begin{aligned}
&\ket{0}_L^N = \frac{1}{\sqrt{2}}(\ket{+}^{\otimes N} + \ket{-}^{\otimes N}), \\
&\ket{1}_L^N = \frac{1}{\sqrt{2}}(\ket{+}^{\otimes N} - \ket{-}^{\otimes N}),
\end{aligned}
\end{equation}
where $\ket{\pm} = (\ket{0}\pm \ket{1})/\sqrt{2}$.  
The motivation with this type of encoding is that measuring any qubit in the $\ket{0,1}$ basis simply removes it from the state, reducing the 
resulting state by one, 
\begin{equation}
\begin{aligned}
P_{0,N} \ket{0}_L^N &= (I_N + Z_N)\ket{0}_L^N \\
& =  \frac{1}{\sqrt{2}}(\ket{+}^{N-1} + \ket{-}^{N-1})\ket{0}_N = \ket{0}_L^{N-1}\ket{0}_N, \\
P_{1,N} \ket{0}_L^N &= (I_N - Z_N)\ket{0}_L^N \\
&=  \frac{1}{\sqrt{2}}(\ket{+}^{N-1} - \ket{-}^{N-1})\ket{1}_N = \ket{1}_L^{N-1}\ket{1}_N, \\
\end{aligned}
\label{eq:lossenc}
\end{equation}
where $P_{(0,1),N}$ are the projectors corresponding to measurement in the $\ket{0,1}$ basis on the $N$-th qubit (up to normalization).  The 
effect for the $\ket{1}_L$ state is similar. 
Measuring the $N$-th qubit in the $\ket{0}$ state simply removes it from the encoded state, reducing the logical zero 
state by one, while measuring the $N$-th qubit as $\ket{1}$ enacts a logical bit flip at the same time as reducing the size of the logical 
state.   

Instead of introducing the full scheme developed in~\cite{RHG05}, we instead give the general idea of how such encoding allows for 
loss detection without non-demolition measurements.  Photon loss in this model is assumed equivalent to measuring the 
photon in the $\ket{0},\ket{1}$ basis, but not knowing the answer [Sec~\ref{sec:lossleakage}].  
Our ignorance of the measurement result could lead to a logical 
bit-flip error on the encoded state, therefore we require the ability to protect against logical bit-flip errors on the above states.  As already shown, 
the 3-qubit code allows us to achieve such correction.  Therefore the final step in this scheme is encoding the above states into a redundancy code 
(a generalized version of the 3-qubit code), where an arbitrary logical state, $\ket{\psi}_L$ is now given by,
\begin{equation}
\ket{\psi}_L = \alpha\ket{0}_1^N \ket{0}_2^N...\ket{0}_q^N + \beta \ket{1}_1^N \ket{1}_2^N ... \ket{1}_q^N
\end{equation}
where $\ket{0}^N,\ket{1}^N$ are the parity encoded states shown in Eq.~(\ref{eq:lossenc}) 
and the fully encoded state is $q$-blocks of these parity states. 

This form of encoding protects against the loss of qubits by first encoding the system into a code structure that 
allows for the removal of qubits without destroying the computational state and then protecting against logical errors that 
are induced by loss events.  In effect, it maps errors un-correctable by standard QEC to error channels that 
are correctable, in this case qubit loss $\rightarrow$ qubit bit-flip.  Even though this code can be used to convert loss events into 
bit-flips, the size of the code will steadily decrease as qubits are lost.  

This general technique is common with pathological error channels.  If a specific type of error violates the standard ``qubit" assumption of 
QEC, additional correction techniques are always required to map this type of error to a correctable form, 
consequently additional physical resources are usually needed.  

\section{Some modern developments in quantum error correction}
\label{sec:modern}
Up until this stage we have restricted our discussions on error correction to the most basic principals and codes.  
The ideas and methodologies we have detailed represent some of the introductory techniques that were developed when 
error correction was first proposed.  For readers who are only looking for a basic introduction to the field, you 
can quite easily skip the remainder of this paper as we will now examine some of the more modern protocols that 
are utilized when considering the construction of a large scale quantum computer.  

Providing a fair and encompassing review of the more modern and advanced error correction techniques that have been 
developed is far outside our goal for this review.  However, we would be remiss if we did not briefly examine some 
of the more advanced error correction techniques that have been proposed for large scale quantum information processing.  
For the remainder of this discussion we choose two closely related error correction techniques, subsystem coding and 
topological coding which have been receiving 
significant attention in the fields of architecture design and large scale quantum information processing.  While some readers 
may disagree, we review these two modern error correction protocols because they 
are currently two of the most useful correction techniques when discussing the physical construction of a quantum computer.

We again attempt to keep the discussion of these techniques simple and provide specific examples when possible.  However, 
it should be stressed that these error correcting protocols are far more complicated than the basic codes shown earlier.  Topological 
error correction alone has, since its introduction, essentially become its own research topic within the broader error correction field.  
Hence we encourage the reader who is interested to refer to the cited articles below for more rigorous and detailed treatment of these 
techniques.  

\subsection{Subsystem codes: Bacon-Shor codes}
\label{sec:subsystem}

Quantum subsystem codes~\cite{KLP05,KLPL06,B06} are one of the newer and highly flexible techniques to 
implement quantum error correction.  The traditional stabilizer codes that we have reviewed are more formally identified as 
subspace codes, where information is encoded in a relevant coding subspace of a larger multi-qubit system.  In contrast, 
subsystem coding identifies multiple subspaces of the multi-qubit system as equivalent.  
Specifically, multiple states are identified with the logical $\ket{0}_L$ and $\ket{1}_L$ states.  Rather than review the subsystem 
codes in general, we will focus on a subset of these codes known as Bacon-Shor codes \cite{B06}.  

The primary benefit to utilizing Bacon-Shor (BS) codes is the general nature of their construction. The description of arbitrarily large 
error correcting codes is conceptually straightforward, error correction circuits are much simpler to construct~\cite{AC07}, and the 
generality of their construction introduces the ability to perform dynamical code switching in a 
fault-tolerant manner~\cite{SEDH07}.  This final property gives BS coding significant flexibility as the strength of error correction within a 
quantum computer can be changed fault-tolerantly during operation of the computer.

As with the other codes presented in this review, BS codes are stabilizer codes but now defined over a square lattice.  The lattice dimensions 
represent the $X$ and $Z$ error correction properties and the size of the lattice in either of these two dimensions dictates the total 
number of errors the code can correct.  In general, a $\mathcal{C}$($n_1$,$n_2$) BS code is defined over a 
$n_1\times n_2$ square lattice which encodes one logical qubit into $n_1n_2$ physical qubits with the ability to correct 
at least $\lfloor\frac{n_1-1}{2}\rfloor$ $Z$ errors and at least $\lfloor\frac{n_2-1}{2}\rfloor$ $X$ errors.
Again, keeping with the spirit of this review, we focus on a specific example, the $\mathcal{C}$(3,3) BS code.  
This code, encoding one logical qubit with 9 physical qubits can correct for one $X$ and one $Z$ error.  In order to define 
the code structure we begin with a $3\times 3$ lattice of qubits, where qubits are identified with the vertices of the lattice.  
Note that this 2D structure represents the structure of the code, it does not imply that a physical array of qubits {\em must} be 
arranged into a 2D lattice.  
\begin{figure}[ht!]
\begin{center}
\includegraphics[width=0.4\textwidth]{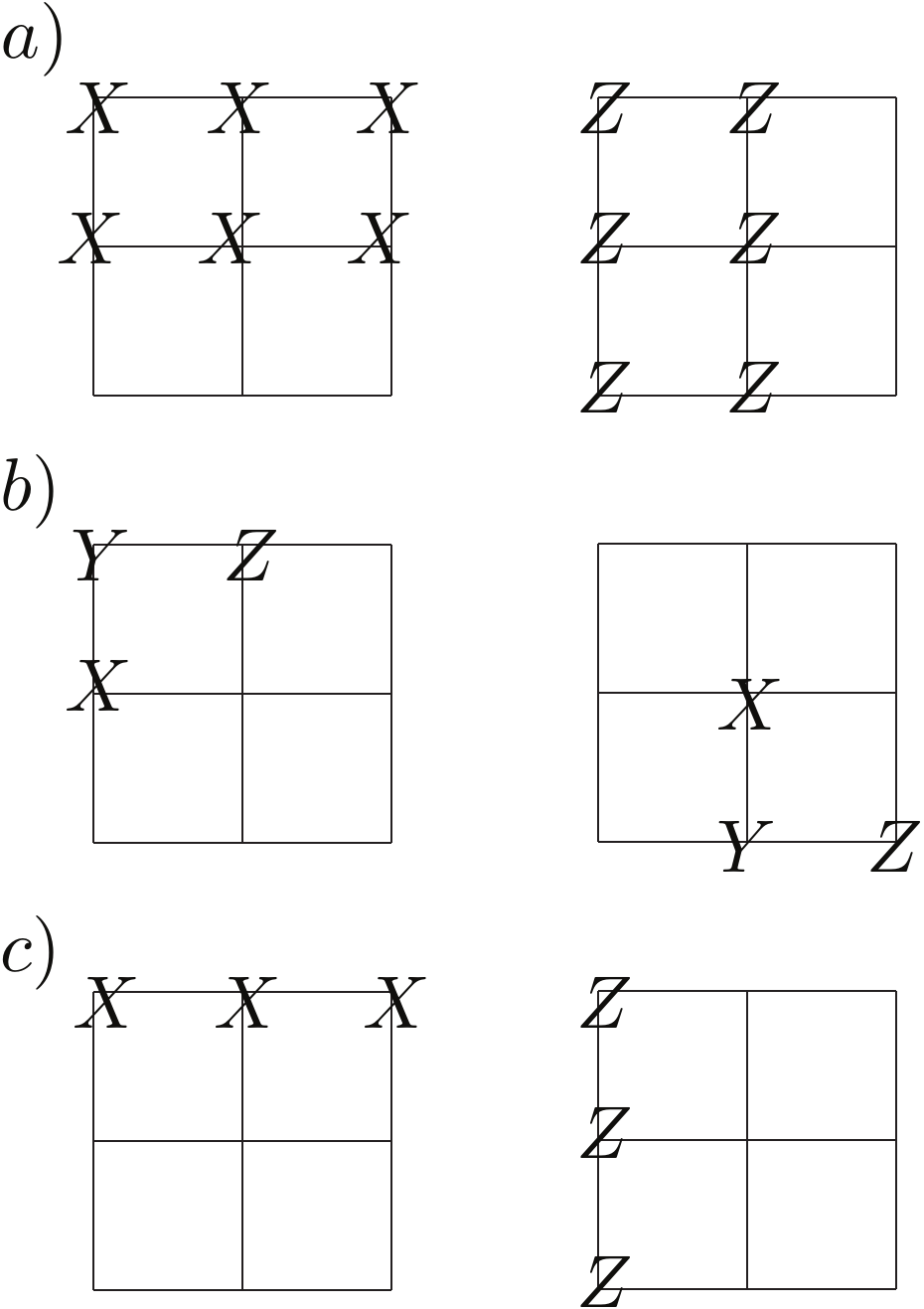}
\caption{Stabilizer structure for the $\mathcal{C}$(3,3) code.  Subfigure a) gives two of the four stabilizers from the group $\mathcal{S}$.  Subfigure b)  
illustrates one of the four encoded sets of Pauli operators from each subsystem defined with the Gauge group, $\mathcal{T}$.  Subfigure c) gives the 
two logical operators from the group $\mathcal{L}$ which enact valid operations on the encoded qubit. }
\label{fig:stab}
\end{center}
\end{figure}
Fig.~\ref{fig:stab} illustrate three sets of stabilizer operators which are defined over the lattice.  The first group, illustrated in Fig.~\ref{fig:stab}a. 
is the stabilizer group, $\mathcal{S}$, which is generated by the operators,
\begin{equation}
\begin{aligned}
\mathcal{S} = \langle X_{i,*} X_{i+1,*} ; Z_{*,j}Z_{*,j+1} \ | \ i \in \Z_{2} ; j \in \Z_{2} \rangle,
\end{aligned}
\label{stabilizers:bs}
\end{equation}
where we retain the notation utilized in~\cite{AC07,SEDH07}. $U_{i,*}$ and $U_{*,j}$ represent an operator, $U$, acting on all qubits in a given row, $i$, or column, $j$, 
respectively, and $\Z_2=\{1,2\}$.  The second relevant subsystem is known as the gauge group~[Fig.~\ref{fig:stab}b.], 
$\mathcal{T}$, and is described via the non-Abelian group generated by the pairwise operators
\begin{equation}
\begin{aligned}
\mathcal{T} =    \langle X_{i,j}X_{i+1,j} ; Z_{j,i}Z_{j,i+1} \ | \ i \in \Z_{2} ; j \in \Z_{3} \rangle, 
\end{aligned}
\end{equation}
The third relevant subsystem is the logical space~[Fig.~\ref{fig:stab}c], $\mathcal{L}$, which can be defined through the logical Pauli operators
\begin{equation}
\mathcal{L} = \langle  Z_{*,1} ; X_{1,*} \rangle.
\end{equation}

The stabilizer group $\mathcal{S}$, defines all relevant code states, i.e. {\em every} valid logical space is a $+1$ eigenvalue of this set.  
For the $\mathcal{C}$(3,3) code, there are a total of nine physical qubits and a total of four independent stabilizers in $\mathcal{S}$, 
hence there are five degrees of freedom left in the system which can house $2^5$ logical states which are simultaneous eigenstates of 
$\mathcal{S}$.  This is where the gauge group, $\mathcal{T}$, becomes relevant.  As the gauge group is non-Abelian, there 
is no valid code state which is a simultaneous eigenstate of all operators in $\mathcal{T}$.  However, if you examine closely there 
are a total of four encoded sets of Pauli operations within $\mathcal{T}$.  Fig~\ref{fig:stab}b. illustrates two such sets.  
As all elements of $\mathcal{T}$ commute with all elements of $\mathcal{S}$ we can 
identify each of these four sets of valid ``logical" qubits to be equivalent, i.e. we define $\{\ket{0}_L,\ket{1}_L\}$ pairs 
which are eigenstates of $\mathcal{S}$ and one abelian subgroup of $\mathcal{T}$ and then ignore exactly what 
gauge group we are in\footnote{Provided we are in eigenstates of $\mathcal{S}$ we can, in principal, be in 
superpositions of eigenstates of the operators in $\mathcal{T}$.  However, the error correction procedure for the BS code projects 
us back to eigenstates of the gauge operators.}.  
Therefore, each of these gauge states represent a subsystem of the code, with each subsystem logically equivalent.  

The final group we consider is the logical group $\mathcal{L}$.  This is the set of two Pauli operators which 
enact a logical $X$ or $Z$ gate on the encoded qubit {\em regardless} of the gauge choice and consequently represent true 
logical operations to our encoded space.  

In a more formal sense, the definition of these three group structures allows 
us to decompose the Hilbert space of the system.  
If we let $\mathcal{H}$ denote the Hilbert space of the physical system, $\mathcal{S}$ forms an Abelian group and hence can act as a stabilizer set defining subspaces of $\mathcal{H}$.  If we describe each of these subspaces by the binary vector, $\vec{e}$, formed from the eigenvalues of the stabilizers, $\mathcal{S}$, then each subspace splits into a tensor product structure
\begin{equation}
\mathcal{H} = \bigoplus_{\vec{e}} \mathcal{H}_{\mathcal{T}} \otimes \mathcal{H}_{\mathcal{L}},
\end{equation}
where elements of $\mathcal{T}$ act only on the subsystem $\mathcal{H}_{\mathcal{T}}$ and the operators $\mathcal{L}$ act only on the subsystem $\mathcal{H}_{\mathcal{L}}$.  In the context of storing qubit information, a logical qubit is encoded into the two 
dimensional subsystem $\mathcal{H}_{\mathcal{L}}$.  As the system is already stabilized by operators in $\mathcal{S}$ and the operators 
in $\mathcal{T}$ act only on the space $\mathcal{H}_{\mathcal{T}}$, qubit information is only altered when operators in the group 
$\mathcal{L}$ act on the system.  

This formal definition of how BS coding works may be more complicated than the standard stabilizer codes shown earlier, but this 
slightly more complicated coding structure has significant benefits when we consider how error correction is performed.  

In general, to perform 
error correction, each of the stabilizers of the codespace must be checked to determine which eigenvalue changes have occurred due to 
errors.  The stabilizer group, $\mathcal{S}$, consist of qubit operators 
that scale with the size of the code.   In general, for 
a $n_1\times n_2$ lattice, the $X$ stabilizers are $2n_1$ dimensional and the $Z$ stabilizers are $2n_2$ dimensional.  If 
techniques such as Shor's method~[Section~\ref{sec:FTcircuit}] were used, we would need to prepare a large ancilla state to perform 
fault-tolerant correction,  this is clearly undesirable.  This problem can be mitigated due to the 
gauge structure of these codes~\cite{AC07}.  

Each of the stabilizers in $\mathcal{S}$ are simply the product of certain elements from $\mathcal{T}$, for example,
\begin{equation}
\begin{aligned}
&X_{1,1}X_{1,2}X_{1,3}X_{2,1}X_{2,2}X_{2,3} \in \mathcal{S} \\
= &( X_{1,1}X_{2,1} ) . (X_{1,2}X_{2,2}).(X_{1,3}X_{2,3}) \in \mathcal{T}.
\end{aligned}
\end{equation}
Therefore if we check the eigenvalues of the three, 2-qubit operators from $\mathcal{T}$ we are able to calculate what the 
eigenvalue is for the 6-dimensional stabilizer.  This decomposition of the stabilizer set for the code can only occur since the 
decomposition is in terms of operators from $\mathcal{T}$ which, when measured, has no effect on the logical information encoded within the system.  
In fact, when error correction is performed, the gauge state of the system will almost always change based on the order in which 
the eigenvalues of the gauge operators are checked. 

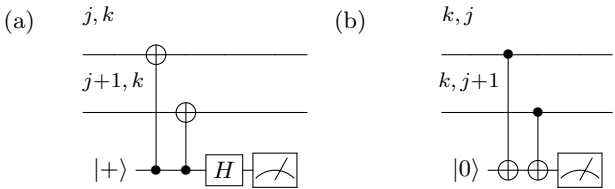
\begin{figure}[ht]
\begin{center}
\begin{tabular}{ccc}
\put(-30,10){(a)}
\put(95,10){(b)}
\Qcircuit @C=1ex @R=2.3ex @!R {  \put(0.1,15){\footnotesize{$j,k$}}
   & \qw                             & \targ      & \qw       & \qw     & \qw & \qw \\
   & \qw  \put(-12,12){\footnotesize{$j{+}1,k$}}  & \qw        & \targ     & \qw     & \qw & \qw\\
   & \push{|+\rangle \hspace{0.1cm}} & \ctrl{-2}  & \ctrl{-1} & \gate{H} & \meter 
                                }    
   &\hspace{1.2cm} & \hspace{0.2cm}
\Qcircuit @C=1ex @R=2.3ex @!R {  \put(0.1,15){\footnotesize{$k,j$}}
   & \qw                             & \ctrl{+2}  & \qw       & \qw     & \qw \\
   & \qw  \put(-12,12){\footnotesize{$k,j{+}1$}}  & \qw        & \ctrl{+1} & \qw     & \qw \\
   & \push{|0\rangle \hspace{0.1cm}} & \targ      & \targ     & \meter &
                                }    
 \vspace{-0.2cm}
\end{tabular}
\end{center}
\caption{\label{fig:BScheck} (From Ref.~\cite{AC07}) Circuits for measuring the gauge operators and hence performing error correction for subsystem 
codes.  Subfigure a) measures, fault-tolerantly, the operator $X_{j,k}X_{j+1,k}$ with only one ancilla. Subfigure b) measures $Z_{k,j}Z_{k,j+1}$.  The results of these 
two qubit parity checks can be used to calculate the parity of the higher dimensional stabilizer operators of the code. }
\end{figure}

This exploitation of the gauge properties of subsystem coding is extremely beneficial for the design of fault-tolerant correction circuits.  
As the stabilizer operators can now be decomposed into multiple 2-dimensional operators, fault-tolerant circuits for error correction 
do not require any encoded ancilla states.  Furthermore, even if we decide to scale the code-space to correct more errors (increasing the 
lattice size representing the code), we do not require measuring operators with higher dimensionality.  Fig~\ref{fig:BScheck} taken from Ref.~\cite{AC07} 
illustrates the fault-tolerant circuit constructions for Bacon-Shor subsystem codes.
As each ancilla qubit is only coupled to two data qubits, no further circuit constructions are required to ensure fault-tolerance\footnote{It 
should be noted that for higher distance BS codes, specifically for $d\geq 5$, multiple measurements for each syndrome are 
required to ensure fault-tolerance}.  
The measurement results from these 2-dimensional parity checks are then combined to calculate the parity of the higher dimensional 
stabilizer of the subsystem code.  

A second benefit to utilizing subsystem codes is the ability to construct fault-tolerant circuits to perform dynamical code switching.  Dynamical code switching is a technique that could be 
utilized when noise sources are highly biased.  In many systems the physical mechanisms 
which lead to errors are not symmetric in $X$ and $Z$.  For example, simple Markovian 
dephasing introduces only $Z$ errors into the system and in all circumstances is the more 
dominant error channel.  

One technique is to use concatenation in a clever way.  
The QEC codes we have largely examined in this review are symmetric in the $X$ and $Z$ 
sector.  The 5-, 7- and 9-qubit codes correct for {\em one} $X$ error and {\em one} $Z$ error. 
Aliferis and Preskill considered a 
concatenated code, where the lower level code is a simple $n$ qubit 
repetition code [Section~\ref{sec:sec:3qubit}] that only corrected for $Z$ errors~\cite{AP08}.  By utilizing this 
lower level code, they symmeterize the $Z$ and $X$ noise (the size of the $n$ qubit code 
is determined by the level of asymmetry in the physical noise).  The subsequent levels are 
standard QEC codes, but it now operates at the next encoded level which will operate with symmetrical rates for logical 
$X$ and $Z$ errors.  By utilising the simpler $n$-qubit repetition code at the 
lower level, qubit resources are reduced, compared to using a symmetric code at every level.
This work was extended and adapted to a realistic model in superconducting systems~\cite{ABD09}.

Asymmetric coding may also be required if the noise acting on a quantum computer 
changes over the course of its operation.  In this case it may be require to switch from 
a symmetric code to a asymmetric code, dynamically.  Naively, this can be done be 
simply decoding all qubits and then re-encoding them with the new code.  However, 
this procedure would not be fault-tolerant, as any error that occurs to the system 
when it is decoded will propagate through to the new encoded state.  A second 
technique would be to encode the system with an asymmetric code at the next level of 
concatenation and then decode the lower level.  However, when utilizing the BS codes, 
we can perform this code switching in a different way.
As noted before, the $\mathcal{C}(n_1,n_2)$ code corrects for $\lfloor \frac{n_1-1}{2}\rfloor$ $X$ 
errors and $\lfloor \frac{n_1-1}{2}\rfloor$ $Z$ errors.  
The ability to convert between two codes, $\mathcal{C}(n_1,n_2)$ and $\mathcal{C}(n_1',n_2')$, 
in a fault-tolerant manner 
would allow for us to dynamically change the strength (or asymmetry) of the error correction 
whenever noise rates fluctuate in the computer.  It was shown in Ref.~\cite{SEDH07} how such 
switching can be achieved, allowing for the fault-tolerant conversion between 
arbitrary sized QEC codes. 

\subsection{Topological codes}
\label{sec:topological}
A conceptually similar (but technically distinct) coding technique to the Bacon-Shor subsystem codes is the idea of topological error correction, first introduced with the Toric code of Kitaev in 1997~\cite{K97}.  Topological coding is similar to subsystem codes in that the code structure is defined on a lattice (which, in general, can be of dimension $\ge 2$) and the scaling 
of the code to correct more errors is conceptually straightforward.  However, in topological coding schemes the protection afforded to 
logical information relies on the unlikely application of error chains which define non-trivial topological paths over the code surface.  

Topological error correction is a complicated area of QEC and fault-tolerance and any attempt to fairly summarize the 
field is not possible within this review.  In brief, there are two ways of approaching these schemes.  
The first is simply to treat 
topological codes as a class of stabilizer codes over a qubit system.  This approach is more amenable to current information technologies 
and is being adapted to methods in cluster state computing~\cite{RHG07,FG08}, optics~\cite{DFSG08,DMN08}, ion-traps~\cite{SJ08} and superconducting systems~\cite{IFIITB02}.  
The second approach is to construct a physical Hamiltonian model based on the structure of the topological code or choose 
systems which appear to exhibit topological order.  This leads to the more 
complicated field on anyonic quantum computation~\cite{K97}.  For example, the coding structure of the Toric code \cite{K97} can 
be treated as a physical Hamiltonian system, with the
ground state corresponding to the logical states of the code (in the case of the Torus it is a 4-fold degenerate ground state 
corresponding to two encoded qubits).  Excitations from the 
ground state of this Hamiltonian correspond to the creation of anyonic quasi-particles and 
are energetically unfavourable (since their physical Hamiltonian symmetries reflect the 
coding structure imposed).  

This and other models of anyonic computation utilise quasi-particles that exhibit fractional quantum statistics 
(they acquire fractional phase shifts when their positions are exchanged twice with other anyons, in contrast to Bosons or Fermions which 
always acquire $\pm 1$ phase shifts).  The unique properties of anyonic systems 
therefore allow for natural, robust, error protection.  However, the major issue with this model is that it relies on quasi-particle 
excitations that do not, in general, arise naturally.  Although certain physical systems have been shown to exhibit anyonic properties, 
most notably in the 
fractional quantum hall effect~\cite{NSSFS08}.  However, it is a daunting task to both manufacture a reliable anyonic system and to reliably design and construct a 
large scale computing system.  Note:  Recently it has 
been shown that anyonic models based on the 2D codes illustrated in this section do 
not exhibit self-correcting properties~\cite{KC08,BT09}.  Higher dimensional topological coding 
models, currently a minimum of 4D, are actually required to implement self-correcting quantum memories~\cite{AHHH08}. 

As there are several extremely good discussions of both anyonic~\cite{NSSFS08} and non-anyonic topological computing~\cite{DKLP02,FSG08,FG08}, we will not 
review any of the anyonic methods for topological computing and simply provide 
a brief example of one topological coding scheme, namely the surface code~\cite{BK01, DKLP02,FSG08}.  
The surface code for QEC is a desirable error 
correction model for several reasons.  As it is defined over a 2-dimensional lattice of qubits it can be implemented on architectures that 
only allow for the coupling of nearest neighbour qubits (rather than the arbitrary long distance coupling of qubits in separate regions of the 
computer).  The surface code also exhibits one of the highest fault-tolerant thresholds of any QEC scheme, recent simulations 
estimate a threshold approaching 1\%~\cite{RHG07,WFSH09}.  
Finally, the surface code has been analysed with respect to loss errors, showing a high tolerance when such errors 
are heralded~\cite{SBD09}.  This subfield of error correction has been heavily researched in recent years.  Methods in statistical 
physics are now routinely utilised to help calculate fault-tolerant thresholds \cite{BD08,KBM09,BAO12,ABKM12}, more 
advanced coding models and methods for computing with these models are under investigation \cite{BD07,BD+07,KBAD10,B11,F12}.  
This section will present a basic introduction, hopefully readers will feel more comfortable studying these advanced techniques afterwards.

\begin{figure}[ht!]
\begin{center}
\includegraphics[width=0.4\textwidth]{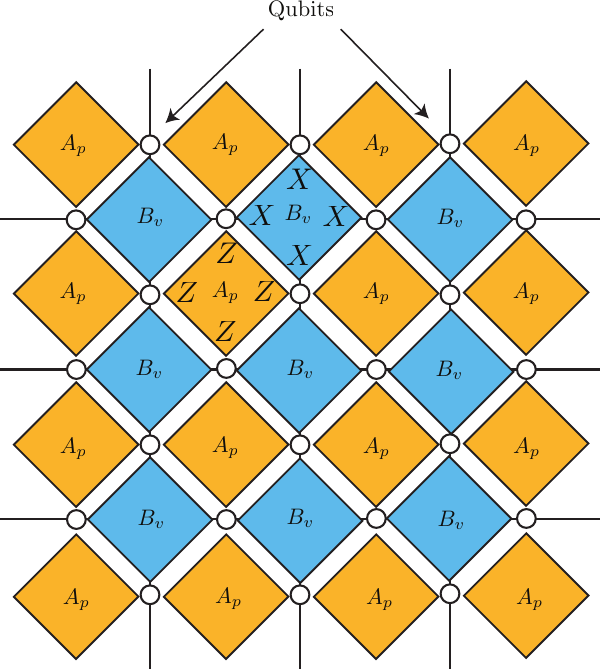}
\caption{General structure of the surface code.  The edges of the lattice correspond to 
physical qubits.  The four qubits surrounding each face (or plaquette) are +1 eigenstates 
of the operators $A_p$ while the four qubits surrounding each vertex are +1 eigenstates of the 
operators $B_v$.  If all eigenstate conditions are met, a unique multi-qubit state is defined 
as a ``clean" surface.}
\label{fig:surface1}
\end{center}
\end{figure}

The surface code, as with subsystem codes, is a stabilizer code defined over a 2-dimensional qubit lattice, as Fig.~\ref{fig:surface1} illustrates.  We 
identify each edge of the 2D lattice with a physical qubit.  The stabilizer set consists of two types of operators.  The first is the 
set of $Z^{\otimes 4}$ operators which circle every lattice face (or plaquette).  The second is the set of $X^{\otimes 4}$ operators which 
encircle every vertex of the lattice.  The stabilizer set is consequently generated by the operators,
\begin{equation}
A_p = \bigotimes_{j \in b(p)} Z_j, \quad B_v = \bigotimes_{j \in s(v)} X_j,
\end{equation}
where $b(p)$ is the four qubits surrounding a plaquette and $s(v)$ is the four qubits surrounding each vertex in the lattice and 
identity operators on the other qubits are implied.  First note that 
all of these operators commute as any plaquette and vertex stabilizer will share either zero or two qubits.  If the lattice is not periodic in either dimension, this stabilizer 
set completely specifies one unique state, i.e. for a $N\times N$ lattice there are $2N^2$ qubits and $2N^2$ stabilizer generators.  
Hence this stabilizer set defines a unique multi-qubit entangled state which is generally referred to as a ``clean" surface.  
\begin{figure*}[ht!]
\begin{center}
\includegraphics[width=0.8\textwidth]{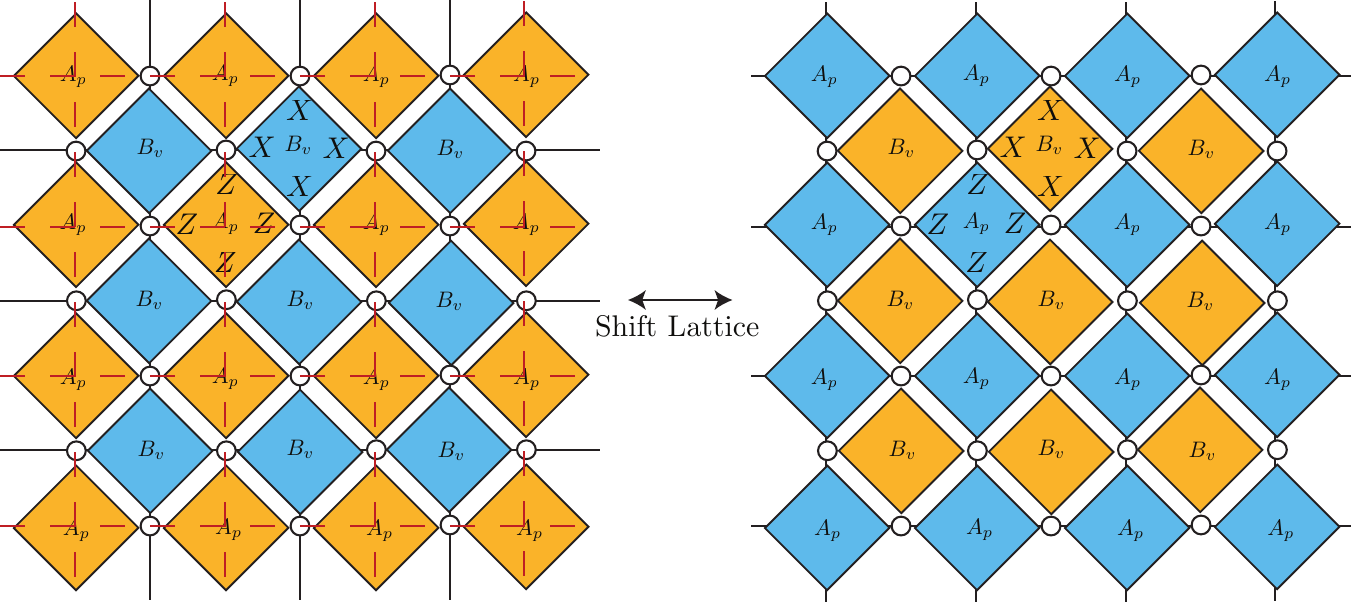}
\caption{The surface code imbeds two self similar lattices that are interlaced, generally 
referred to as the primal and dual lattice.  Subfigure a) illustrates one lattice where plaquettes are 
defined with the stabilizers $A_p$.  Subfigure b) illustrates the dual structure where plaquettes 
are now defined by the stabilizer set $B_v$.  The two lattice structures are interlaced 
and are related by shifting along the diagonal by half a lattice cell.  Each of these 
equivalent lattices are independently responsible for $X$ and $Z$ error correction.}
\label{fig:surface2}
\end{center}
\end{figure*}
\begin{figure*}[ht!]
\begin{center}
\includegraphics[width=\textwidth]{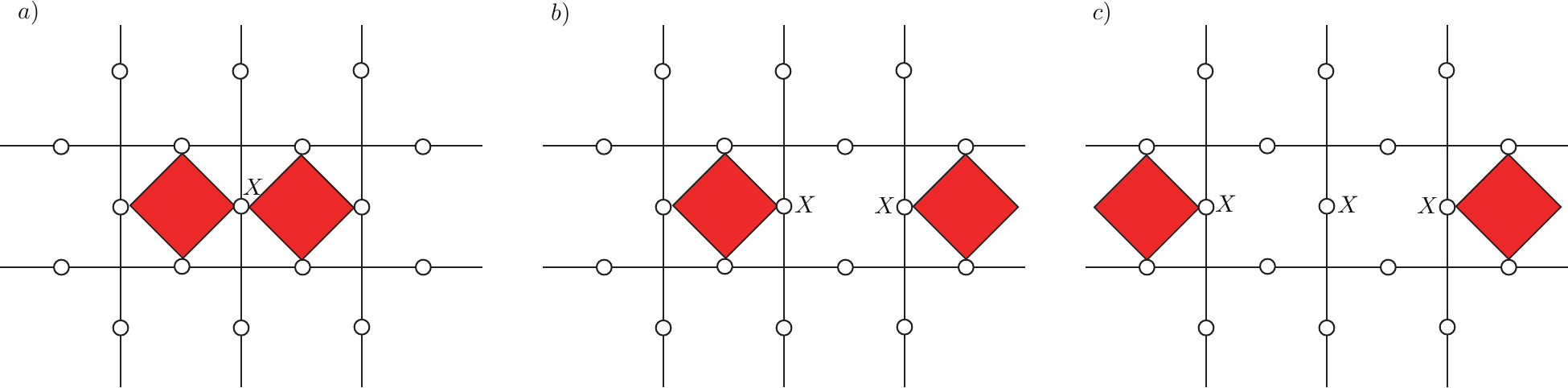}
\caption{Examples of error chains and their effect on the eigenvalues for each plaquette stabilizer.  Subfigure a). A single $X$ error causes the parity of two 
adjacent cells to flip. Subfigures b) and c).  Longer chains of errors only cause the end cells to flip eigenvalue as each intermediate cell will have two 
$X$ errors and hence the eigenvalue for the stabilizer will flip twice.}
\label{fig:surface3}
\end{center}
\end{figure*}
Detailing exactly how this surface can be utilized to perform robust quantum computation is far outside the scope of this review and there 
are several papers to which such a discussion can be referred~\cite{RH07,RHG07,FSG08,FG08}.  
Instead, we can quite adequately show how robust error correction 
is possible by simply examining how a ``clean" surface can be maintained in the presence of errors.  
The $X$ and $Z$ stabilizer sets, $A_p$ and $B_v$ define two equivalent 2D lattices which are interlaced, as 
Fig.~\ref{fig:surface2}, illustrates.  
If the total 2D lattice is shifted along the diagonal by half a cell then the operators $B_v$ are now arranged around a plaquette and 
the operators $A_p$ are arranged around a lattice vertex.  Since protection against $X$ errors are achieved by detecting eigenvalue flips 
of $Z$ stabilizers and visa-versa, these two interlaced lattices correspond to error correction against $X$ and $Z$ errors respectively.  
Therefore we can quite happily restrict our discussion to one possible error channel, for example correcting $X$ errors (since the correction for 
$Z$ errors proceeds identically when considering the stabilizers $B_v$ instead of $A_p$).  

Fig~\ref{fig:surface3}a. illustrates the effect a single $X$ error has on a pair of adjacent plaquettes.  
Since $X$ and $Z$ anti-commute, a single 
bit-flip error on one qubit in the surface will flip the eigenvalue of the $Z^{\otimes 4}$ stabilizers on the two plaquettes adjacent to the 
respective qubit.  As single qubit errors act to flip the eigenvalue of adjacent plaquette stabilizers we examine how chains of errors 
affect the surface.  Figs~\ref{fig:surface3}b. and Fig.~\ref{fig:surface3}c.  examine two longer chains of errors.  
As you can see, if multiple errors occur, only the 
eigenvalues of the stabilizers associated with the ends of the error chains flip.  Each plaquette along the chain will always have two 
$X$ errors occurring on different boundaries and consequently the eigenvalue of the $Z^{\otimes 4}$ stabilizer around these plaquettes will 
flip twice.  

\begin{figure}[bt]
\begin{center}
\includegraphics[width=0.3\textwidth]{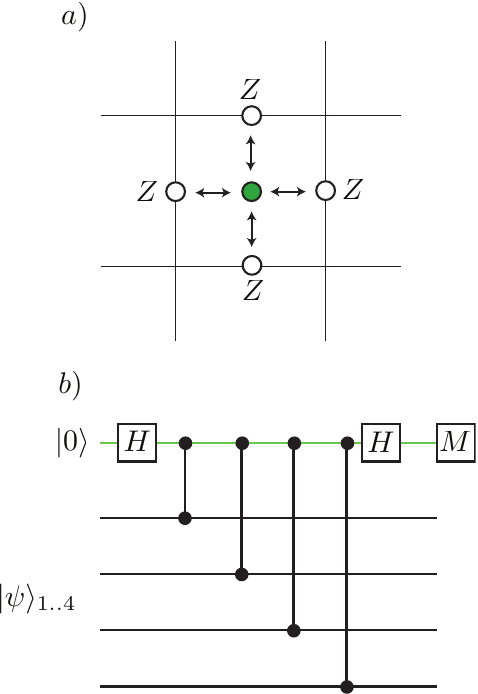}
\caption{Subfigure a).  Lattice structure to check the parity of a surface plaquette.  An additional ancilla qubit is coupled to the four neighboring qubits that 
comprise each plaquette.  Subfigure b).  Quantum circuit to check the parity of the $Z^{\otimes 4}$ stabilizer for each surface plaquette.}
\label{fig:surface4}
\end{center}
\end{figure}

If we now consider an additional ancilla qubit which sits in the center of each plaquette and can couple to the four surrounding qubits, we can 
check the parity by running the simple circuit shown in Fig~\ref{fig:surface4}.  If we assume that we initially prepare a 
perfect ``clean" surface we then, at some later time, check the parity of every plaquette over the surface.  
If $X$ errors have occurred on a certain subset 
of qubits, the parity associated with the endpoints of error chains will have flipped.  We now take this 2-dimensional {\em classical} data 
tree of eigenvalue flips and pair them up into the most likely set of error chains.  Since it is assumed that the probability of error on any individual 
qubit is low, the most likely set of errors which reflects the eigenvalue changes observed is the minimum weight set (i.e. connect up all 
plaquettes where eigenvalues have changed into pairs such that the total length of all connections is minimized).  This classical 
data processing is quite common in computer science and minimum weight matching algorithms such as the Blossom package~\cite{CR99,K08} have 
a running time polynomial in the total number of data points in the classical set.  
Once this minimal matching is achieved, we can identify the likely error chains corresponding to the end points and correction can be applied 
accordingly.  

\begin{figure}[ht!]
\begin{center}
\includegraphics[width=0.45\textwidth]{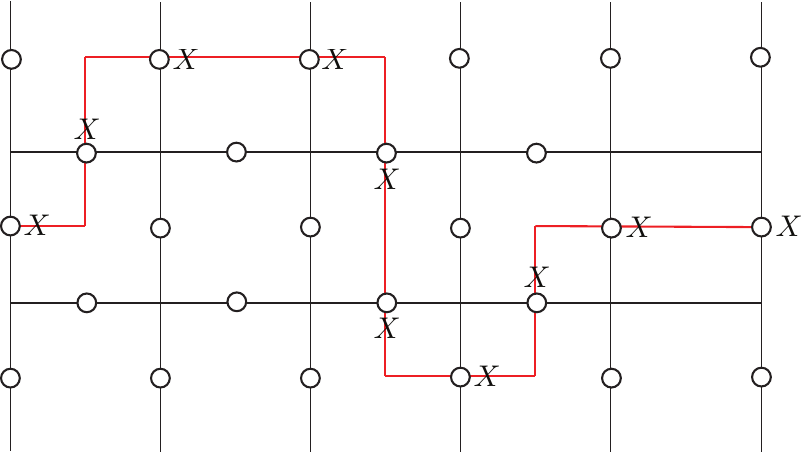}
\caption{Example of a chain of errors which do not cause any eigenvalue changes in the surface.  If errors connect boundaries to other 
boundaries, the error correction protocol will not detect them.  In the case of a ``clean" surface, these error chains are invariants of the 
surface code.  When computation is considered, qubit information are artificial boundaries within the surface.  Hence if error chains connect 
these information qubits to other boundaries, logical errors occur.}
\label{fig:surface5}
\end{center}
\end{figure}

The failure of this code is therefore dictated by error chains that cannot be detected through changes in plaquette eigenvalues.  
If you examine Fig~\ref{fig:surface5}, we consider an 
error chain that connects one edge of the surface lattice to another.  In this case every plaquette has two associated qubits that 
have experienced a bit flip and no eigenvalues in the surface have changed.  Since we have assumed that we are only wishing to 
maintain a ``clean" surface, these error chains have no effect, but when one considers the case of storing information in the lattice, these 
types of error chains correspond to logical errors on the qubit~\cite{BK98,FSG08}.  Hence undetectable errors are chains which connect boundaries of the 
surface to other boundaries (in the case of information processing, qubits are artificial boundaries within the larger lattice surface).
It should be stressed that this is a simplified description of the full protocol, but it does encapsulate the basic idea.  The important 
thing to realize is that the failure rate of the error correction procedure is suppressed, exponentially with the size of the lattice.  If we consider an error 
model where each qubit experiences a bit flip, independently, with probability $p$, then an error chain of one occurs with probability $p$, 
error chains of weight two occur with probability $O(p^2)$, chains of three $O(p^3)$ etc.  If we have an $N \times N$ lattice and we extend 
the surface by {\em one} plaquette in each dimension, then the probability of having an error chain connecting two boundaries will drop 
by a factor of $p$ (one extra qubit has to experience an error)\footnote{The reduction is not exactly $p$ as there is a combinatorial factor that 
accounts for all the possible extra error chains that are now possible.}.  Extending an $N\times N$ lattice by one 
plaquette in each dimension requires $O(N)$ extra qubits, hence this type of error correcting code suppresses the probability of having undetectable 
errors exponentially with a qubit resource cost which grows linearly.  

As we showed in Section~\ref{sec:Fault-tolerance}, 
standard concatenated coding techniques allow for a error rate suppression which scales with the 
concatenation level as a double exponential while the resource increase scales exponentially.  For the surface code, the error rate suppression 
scales exponentially while the resource increase scales linearly.  While these scaling relations might be mathematically equivalent, the 
surface code offers much more flexibility at the architectural level.  Being able to increase the error protection in the computer with only a 
linear change in the number of physical qubits is far more beneficial than using an exponential increase in resources when utilizing 
concatenated correction.  Specifically, consider the case where a error protected computer is operating at a logical error rate which is 
just above what is required for an algorithm.  If concatenated error correction is employed, then adding another later of correction will not 
only increase the number of qubits by an exponential amount, but it will also drop the effective logical error rate far below what is actually required.  
In contrast, if surface codes are employed, we increase the qubit resources by a linear factor and drop the logical error rate sufficiently for 
successful application of the algorithm.  

We now leave the discussion regarding topological correction models.  We emphasize again that this was a {\em very} broad overview 
of the general concept of topological codes.  There are many details and subtleties that we have deliberately left out of this discussion and 
we urge the reader, if interested, to refer to the referenced articles for a more thorough treatment of this topic.   

\section{Conclusions and future outlook}

This review has hopefully provided a basic introduction to some of 
the most important theoretical aspects of QEC and 
fault-tolerant quantum computation.  The ultimate goal of this discussion was not 
to provide a rigorous theoretical framework for QEC and fault-tolerance, but 
instead attempted to illustrate most of the important rules, results and techniques that have evolved out of this field.  Hopefully this introduction will serve as a starting point for 
individuals introducing themselves to this topic.  For those wishing to continue 
research into the QEC field, we highly recommend consulting the papers 
referenced throughout this review and especially 
Refs.~\cite{G97+,NC00,G00,KLABVZ02,S03+,G09} 
which provide a more mathematically rigorous review of QEC and fault-tolerance.

We not only covered the basic aspects of QEC through specific examples, 
but also we have briefly discussed how physical errors influence quantum 
computation and how these processes are interpreted within the context of 
QEC.   One of the more important aspects of this review is the discussion related to 
the stabilizer formalism, circuit synthesis and fault-tolerant circuit construction.  Stabilizers are 
arguably the most useful theoretical formalism in QEC
as once it is sufficiently understood, most of the 
important properties of error correcting codes can be investigated and understood 
largely by inspection.

The study of QEC and fault-tolerance is still and active area of 
QIP research.  
Although the library of quantum codes and error correction techniques are vast there is 
still a significant disconnect between the abstract framework of quantum coding and 
the more physically realistic implementation of error correction for large-scale quantum 
information processing.  

There are several future possibilities for the direction of quantum information processing.  
Even with the development of many of these advanced techniques, the physical 
construction and accuracy of current qubit fabrication is still insufficient to obtain any 
benefit from QEC.  Many in the field now acknowledge that the future development of 
quantum computation will most likely split into two broad categories.   The first is arguably the 
more physically realistic, namely small qubit applications in quantum simulation.  
Beyond these smaller qubit applications, we move to truly large scale quantum computation, i.e. 
implementing large algorithms such as Shor on qubit arrays well beyond
1000 physical qubits.  This would undoubtably require active techniques in 
error correction.  Future work needs to focus on adapting the many codes and 
fault-tolerant techniques to the architectural level.  As we noted in section~\ref{sec:threshold}, 
the implementation of QEC at the design level largely influences the fault-tolerant 
threshold exhibited by the code itself.  Being able to efficiently incorporate 
both the actual quantum code and the error correction procedures at the physical level 
is extremely important when developing an experimentally viable, large-scale quantum computer.  

There are many differing opinions within the quantum computing community as to the 
future prospects for quantum information processing.  Many remain pessimistic regarding 
the development of a million qubit device and instead look towards quantum simulation in 
the absence of active error correction as the realistic goal of quantum information.  However, 
in the past few years, the theoretical advances in error correction and 
the fantastic speed in the experimental development of few qubit devices continues to 
offer hope for the near-term construction of a large scale computer, incorporating many of the ideas 
presented within this review.  While we could never 
foresee the possible successes or failures in quantum information science, we remain hopeful 
that a large scale quantum computer is still a goal worth pursuing.

\section{Acknowledgments}
The authors wish to thank 
A. M. Stephens, R. Van Meter, A.G. Fowler, L.C.L. Hollenberg, A. D. Greentree for helpful comments and
acknowledge the support of MEXT, JST and FIRST projects.
\bibliographystyle{alpha}
\bibliography{QEC}

\end{document}